\documentclass[printer]{aa} 

\usepackage{natbib}
\usepackage{graphicx}
\usepackage{txfonts}
\usepackage{amsfonts}
\usepackage{xcolor}
\usepackage{longtable}
\usepackage{afterpage}
\usepackage{changepage}
\usepackage{geometry}
\usepackage{hhline}
\usepackage{tablefootnote}
\usepackage{graphicx}
\usepackage{lscape}
\usepackage{comment}
\usepackage{booktabs}
\usepackage[squaren,Gray]{SIunits}
\usepackage{hyperref}
\bibpunct{(}{)}{;}{a}{}{,}

\newcommand{\Mjup}{M$_{\rm Jup}$}

\newcommand{\vsini}{$v\sin{i}$}
\newcommand{\Msun}{M$_{\sun}$}

\newcommand{\aastroruwe}{$\alpha _{\rm ruwe }$}
\newcommand{\aastropma}{$\alpha _{\rm PMa }$}
\newcommand{\aastroaen}{{\bf $\alpha _{\rm AEN }$}}
\newcommand{\ruwe}{ruwe}
\newcommand{\nsigmaruwe}{{\bf $N$-$\sigma_{\rm ruwe}$}}
\newcommand{\nsigmapma}{{\bf $N$-$\sigma_{\rm PMa}$}}
\newcommand{\nsigmaaen}{{\bf $N$-$\sigma_{\rm AEN}$}}

\begin{document}

\title{Searching for substellar companion candidates with Gaia.}
\subtitle{III. Search for companions to members of young associations}
\author{A.-M. Lagrange\inst{1,2} \and  F. Kiefer\inst{1} \and P. Rubini\inst{3} \and V. Squicciarini\inst{1,4} \and A. Chomez \inst{1,2}  \and J. Milli\inst{2} \and A. Zurlo\inst{5,6} \and J. Bouvier\inst{2} \and P.  Delorme\inst{2} \and H.  Beust\inst{2} \and J. Mazoyer\inst{1} \and O. Flasseur\inst{7} 
\and N. Meunier\inst{2} \and L. Mignon\inst{2} 
\and G. Chauvin\inst{8}\and P. Palma-Bifani\inst{8,1}}
\institute{
\label{inst:1}LIRA, CNRS, Observatoire de Paris, Université PSL, 92190 Meudon, France\thanks{Please send any request to anne-marie.lagrange@obspm.fr} \and \label{inst:2}Univ. Grenoble Alpes, CNRS, IPAG, F-38000 Grenoble, France  \and
\label{inst:3}Pixyl, 5 Avenue du Grand Sablon, 38700 La Tronche, France \and
\label{inst:4}INAF -- Osservatorio Astronomico di Padova, Vicolo dell'Osservatorio 5, I-35122, Padova, Italy \and \label{inst:5}Instituto de Estudios Astrof\'isicos, Facultad de Ingenier\'ia y Ciencias, Universidad Diego Portales, Av. Ej\'ercito Libertador 441, Santiago, Chile  \and
\label{inst:6}Millennium Nucleus on Young Exoplanets and their Moons (YEMS), Chile \and
\label{inst:7}Univ. Lyon, Univ. Lyon1, ENS de Lyon, CNRS, Centre de Recherche Astrophysique de Lyon (CRAL) UMR5574, F-69230 Saint-Genis-Laval, France \and
\label{inst:8}Université Côte d’Azur, OCA, CNRS, Lagrange, 96 Bd de l'Observatoire, 06300 Nice, France
}
        
   \date{Received ; accepted }

  
  \abstract
{Absolute astrometry with Gaia is expected to detect and characterize the orbits of thousands of exoplanets in the coming years. A  tool, GaiaPMEX, was recently developed to characterize multiple systems based on two binarity indicators derived from the DR3 astrometric solution: the astrometric signature \aastroruwe{} in linear motion residuals in Gaia-only data, and, when the sources were also observed by Hipparcos, the astrometric signature \aastropma{} in the Gaia-Hipparcos proper motion anomaly (PMa).}
{Our aim is to identify close (less than typ. 20 au)  (sub)stellar companion candidates to star members of close-by young associations
previously monitored in radial velocity (RV) surveys. We wish to compare the  detection capabilities of absolute astrometry and spectroscopy, and to characterize planetary-mass companions, combining the astrometric data with direct imaging and RV data.}
{We use GaiaPMEX to identify binary stars members of close young associations and constrain the mass and semi-major axes (sma) of possible companions. 
For companion masses possibly in the planetary range, we use direct imaging and when possible, RV data as well, to further constrain their nature and orbital properties.}
{For each of our targets, we provide a diagnosis on its binarity based on absolute astrometry.  
When no binary is detected, GaiaPMEX provides detection limits in the (sma, mass) space. 
We identify several companions with possible masses down to the brown dwarfs (BD; 50+) or planetary masses (13). 
Around the M-type star G80-21, we detected a new companion orbiting at less than 1-2 au. 
Adding RV and high contrast imaging data shows this companion is a giant planet. In other two cases, AB Pic and HD 14082 B, we confirm the presence of substellar companions, and determine the first robust solutions for their mass and orbital properties. 
We further identified 9 potentially interesting candidates for planetary mass companions, which remain to be studied. Finally, a detailed treatment of noises in Gaia astrometric measurements shows that there are no evidence at a 2--$\sigma$ level of two exoplanet detections that were previously announced based on the same set of data.}
{Our approach allows to detect all stellar mass companions with sma in the range 0.1-10 au. For separations below 0.1 au, however, spectroscopy outperforms absolute astrometry. Combining GaiaPMEX and RV data is therefore perfectly adapted for a full exploration of the 0.01-10  au sma range when searching for stellar companions, and increases the expected rate of detections derived from RV surveys. Moreover, in the 0.5 to 5 au domain,  GaiaPMEX has an excellent sensitivity to BDs, and a good sensitivity to planetary mass planers as well for this sample.  
}


   \maketitle
%

\section{Introduction}
Planet formation is one of the most challenging questions in today's astronomy. From an observational point of view, the search for low-mass companions (planets, BDs) of stars has driven a lot  of efforts in the last decades, to contain statistical information on planets' properties or to characterize exoplanets individually. In both cases, the results are compared to model outputs: population synthesis outputs for surveys, and interiors or atmospheric models for individual objects, which in turn allow further constraining the models.

The conditions in which planetary systems form are key information that motivates, in particular, studies of stellar multiplicity. Indeed, binarity or multiplicity plays a major role in the formation and evolution of planetary systems through protoplanetary disk truncation and shape, disk lifetime, star-planet interactions, etc. The identification and characterization of multiple systems in star-forming regions or in young stellar associations have then motivated many observational efforts for decades. We refer to \citet{Offner22} for a recent review.


So far, most estimates of stellar binary occurrence rates for small semi-major axis ($\lesssim $ 1-10 au) have been obtained using ground based spectroscopic surveys \cite[see results in  ][]{Offner22}. However, such surveys are biased towards nearly equal mass binaries (companion-to-star mass ratio -- hereafter, $q$ -- close to 1), towards  periods less than a few years, and towards inclined systems, as the "projected" radial velocity (RV) amplitude, and therefore the survey sensitivity, increases as $\sin i$,
with $i$ the inclination of the companion's orbital momentum with respect to the line of sight. Interferometric surveys can also detect close binaries \cite[see ref. in][]{Offner22}. They are also biased towards a $q$-ratio close to 1. High contrast imaging allows identifying companion stars with a low $q$-ratio,  typically at separations down to 10-20 au (depending on the $q$-ratio). Absolute astrometry with Gaia is well suited to identify and characterize stellar multiplicity in the intermediate region (1-10 au), down to brown dwarfs companions (small $q$-ratio), as well as planetary mass companions~\citep{Perryman2014,Sahlmann2015,Holl2022,Arenou2023,Holl2023}. \footnote{Gaia spectroscopy also revealed very close companions with periods in the 1-1000d (see ref. above), but no statistical results have been derived yet.}


Using proper motion anomalies (PMa) estimated from both Gaia and Hipparcos \cite[hereafter PMa; see][]{Kervella2019,Brandt2021,Kervella2022} data has allowed to identify previously unknown substellar companions. Gaia data, on its own, provide information that allows both identifying companions that the PMa alone cannot detect, and lifting degeneracies in the solutions found using the PMa only. Recently, we developed GaiaExplore, a tool designed to identify binaries by analyzing the observed excess astrometric Gaia signal. Additionally, it leverages available data on PMa from both Gaia and Hipparcos data. Complementing GaiaExplore is GaiaPMEX, that allows to constrain the candidate companion mass and sma axis using the same information. GaiaPMEX is described in \citet{Kiefer2024a} (hereafter K24). It has been used already to identify planetary companion candidates around solar-type stars in \citet{Kiefer2024b}.

In this paper, we use GaiaExplore and GaiaPMEX to search for  companions to star members of young and close-by associations that were already surveyed in RV. Such associations have been studied in detail in the last decades as they are ideal sites to study planetary systems formation and early evolution. The young age of these associations allows us to make the best use of direct imaging to search for planetary mass companions at separations typically larger than 10 au. We identify new stellar binary systems, compare the respective sensitivities of GaiaPMEX and spectroscopic surveys, and identify possible BD and planetary mass candidates. We further vet the planetary mass candidates.

We present our approach in Section \ref{sec:approach}. We analyze the output of GaiaPMEX in Section \ref{sec:results}, and present the results related to binarity in Section \ref{sec:binaries_main}. We analyze the 13 systems identified with possible planetary mass companions based on Gaia and PMa data in Section \ref{sec:vetting_cc}, and further characterize three "Gaia" planets in Sections \ref{sec:AB Pic} and \ref{sec:HD14082 B}. Grounded on our method, we discuss in Sect. \ref{sec:noplanets} the status of some candidate planets around stars from the present survey,  recently detected from PMa measurements. A brief conclusion is provided in Section \ref{sec:conclusion}. 

\section{Approach}\label{sec:approach}

\subsection{Sample} \label{sec:sample}
We first considered the sample gathered by ~\citet{Zuniga21} (hereafter ZF21) for their census of spectroscopic binary systems among close associations and moving groups (MG, see Table~\ref{tab:YMG}): $\beta$ Pictoris (BPC), Octans (OCT), $\epsilon$ Chamaeleontis (ECH), TW Hya (TWA), Argus (ARG), AB Doradus (ABDor), Columba (COL) and Tucana-Horologium (THA). Their primary membership assessment relies on the SACY approach, described in ~\citet{Torres06} and in ~\citet{Torres08}. They also use the BANYAN $\Sigma$ classification\footnote{\url{https://github.com/jgagneastro/banyan_sigma}} (hereafter BANYAN) based on Gaia DR2 \citep{Gagne18}. The SACY and BANYAN classifications do not always agree; in particular, 60 targets have a BANYAN-based membership but no SACY-based membership.\footnote{Whether these conclusions would be changed once using Gaia DR3 instead of Gaia DR2 for the BANYAN classification is certainly to be tested, but is beyond the scope of this paper, as it would not allow us to compare our results with theirs any longer.} 
ZF21 considered both classifications in their analysis, and so did we. An additional criterion in their selection process was that at least one spectrum or one radial velocity (either coming from ground-based surveys or from Gaia DR2) with an uncertainty of less than 3 km/s had to be available. This ensured a first data point to their spectroscopic survey, aiming at identifying companions.

\begin{table}
    \centering
    \caption{Young Moving Groups (YMG) considered in the present study, with approximate ages and distances ($d$).}.
    \label{tab:YMG}
    \begin{tabular}{cccccc}
        \toprule 
YMG	&	Age &	$d$ & $N_{bf}$ & $N$ & ref.	\\
        \midrule
&	Myr	& pc & & \\
        \midrule
ECH 	&	$\sim$5 &	100	& 36 & 17 & (1),(1),(2)\\
TWA	    &	$10\pm3$ &	50 & 67 & 17 & (3),(1),(4)\\
BPC	&	$24\pm3$ &	50 & $\sim$200 & 42 & (3),(1),(5)\\
THA	&	$45\pm4$ & 50 & 142 & 76& (3),(6),(7)\\
COL	    &	$42^{+6}_{-4}$ & 50 & 50 &  43 & (3),(6),(3)\\
OCT 	&	$35\pm5$	&	130 & 48 &16  & (8),(6),(8)\\
ARG	&	$45\pm5$	&	70 & 40 & 30  & (3),(9),(3)\\
ABDor	&	$150^{+50}_{-19}$ &	30 & 89 & 46 & (3),(6),(3) \\
        \bottomrule
    \end{tabular}
    \tablefoot{
    $N_{bf}$ indicates the number of bona-fide members of these groups, and $N$ indicates the number of stars found in ZF21.  
    References, respectively, for distance, age and $N_{bf}$ are provided as well. (1):~\citet{Kastner22},
    (2):~\citet{Dickson-Vandervelde21}, (3):~\citet{Bell15}, (4):~\citet{luhman23}, (5):~\citet{Shkolnik17},    
    (6):~\citet{Gagne18},
    (7):~\citet{kraus14},
    (8):~\citet{murphy15},
    (9):~\citet{Zuckerman19}.
    }
\end{table}

The initial ZF21 list contains 410 targets. Many targets are, however,  not members of one of the above-mentioned associations, according to either SACY or BANYAN classification. We do not consider them in the present study, and we keep only the 328 members of these associations according to SACY or BANYAN. They are listed in Table B.1 \footnote{Table B.1 is only available in electronic form at the CDS via anonymous ftp to cdsarc.u-strasbg.fr}. For each star, we provide the SACY and/or BANYAN parent association, together with relevant information.

\begin{figure}[hbt]
    \centering
    \includegraphics[width=40.mm,clip=True]{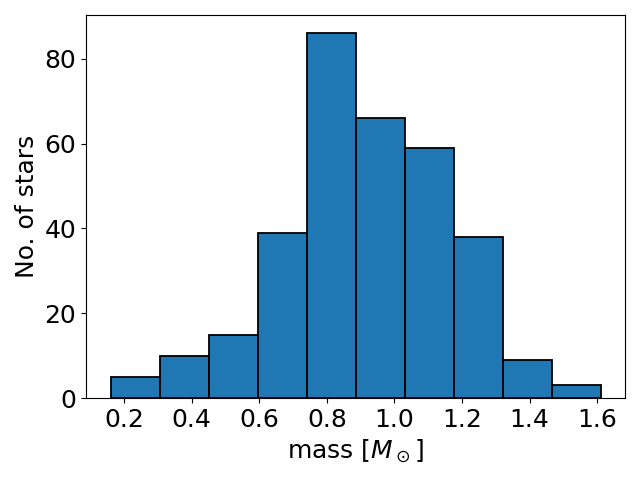}
    \includegraphics[width=40.mm,clip=True]{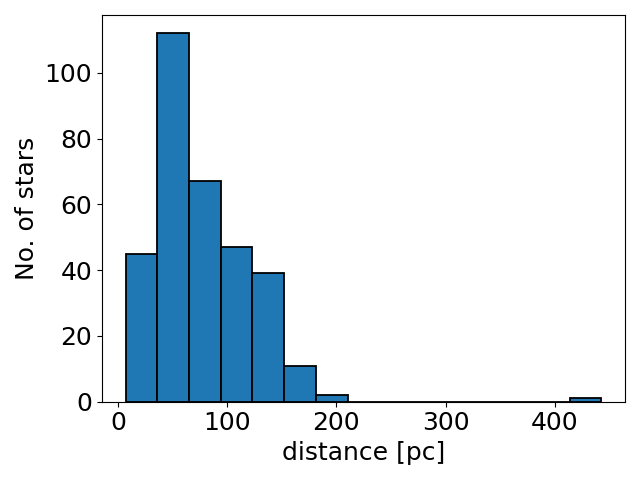}
    \caption{Histogram of the sample stars masses and distances.}
    \label{fig:histo_sample}
\end{figure}

The moving group (MG) ages range between 3 Myr and more than 100 Myr. Each MG contains between 16 and 75 members for the SACY classification and between 12 and 58 targets for the BANYAN classification. Most of the stars have masses (as provided by ZF21) between 0.7 and 1.2 \Msun~(median 0.9 \Msun) and distances as estimated by ZF21 between 10 and 200 pc (median 67 pc), see Figure~\ref{fig:histo_sample}\footnote{Note that CD-621197, classified an ARG member according to SACY ($p=0.8$), and as a field star according to BANYAN, has a distance of more than 400 pc; it is probably an outlier.}. Finally, we note that, due to the selection process adopted by ZF21, the present sample gathers only a fraction of the known members of the considered associations.

\subsection{Absolute astrometry analysis}\label{sec:PMEX}

GaiaPMEX (for Gaia DR3 Proper Motion anomaly and astrometric noise EXcess) uses the Gaia DR3 database to characterize the properties of companions around any star that was observed with Gaia (K24). Starting from the Renormalised Unit Weight Error (ruwe), it calculates the astrometric signature \aastroruwe{}, that is the angular wandering of the photocenter beyond proper motion, parallax and noise\footnote{\samepage Noise levels in the data are estimated for any source of a given $G$, $B_p-R_p$, ra and dec, from a statistical study of the whole Gaia database of sources with $G$\,\!$<$16. This is thoroughly demonstrated in K24.}. When available, it also calculates the astrometric signature \aastropma{}, that is the part of proper motion anomaly (PMa) in excess of the expected PMa of a single star with astrometric noise. The PMa is taken from~\citet{Kervella2022}. Both astrometric information may provide constraints on a possible companion orbiting the target star. A large number of astrometric models of \ruwe{} -- and PMa if relevant -- are calculated at each node of a grid of mass and sma of a hypothetical companion. At any of those nodes, the companion and orbital parameters, as well as parallax and stellar mass, are taken randomly, following the prior distributions shown in Table~\ref{tab:distro}. The resulting distributions of \ruwe{} -- and PMa -- are then used to calculate the likelihood of the corresponding observed data given the model -- i.e. mass and sma of the companion. Through Bayesian inversion, this likelihood is used to calculate the confidence region of the companion's mass and sma given the data. When the PMa is available, we also combine its constraints with those from the \ruwe{}, leading to narrower confidence regions on mass and sma. 

Moreover, comparing the observed value of \aastroruwe{} -- and \aastropma{} -- to the distributions they should follow if a star were single, led us to define the significance of these data. Formally, we calculate the significance of \aastroruwe{} or \aastropma{} from their $p$-value in the distributions based on the null hypothesis of the star being single. We translate this $p$-value into an $N$--$\sigma$ significance level by using the normal law, such that $p$=0.046 is equivalent to a 2--$\sigma$ significance. In the following, we will note the significance as \nsigmaruwe{} and \nsigmapma{} for the various related quantities, and we will use the 2--$\sigma$ significance as the thresholds for identification of binaries.

Note that a similar analysis can be made using the AEN instead of the RUWE. In exceptional cases where we use the AEN, the astrometric signature relative to the AEN  will be noted \aastroaen{}. Similarly, we will note the significance as \nsigmaaen{} when relative to the AEN.

\begin{table}[hbt]
    \centering
    \caption{Distribution of parameters sampled at each tested bin of the mass-sma grid.}
    \label{tab:distro}
    \begin{tabular}{lcc}
        \toprule
        parameter &  type & bounds or law \\
        \midrule 
        $\log M_comp$ & uniform &  $\log M_c$ $\pm$ $\Delta\log M_c$  \\
        $\log \text{sma}$ & uniform &  $\log \text{sma}$ $\pm$ $\Delta\log \text{sma}$ \\ 
        $e$ & uniform &  $0$--$0.9$ \\
        $\omega$ & uniform &  $0$--$\pi$ \\
        $\Omega$ & uniform &  $0$--$2\pi$ \\
        $\phi$ & uniform &  $0$--$1$ \\
        $I_c$ & uniform or $\sin i$ &  $0$--$\pi/2$ \\
        $\varpi$ & normal & $\mathcal N$(PLX,\,$\sigma_{\rm PLX}^2$)  \\
        $M_\star$ & normal & $\mathcal N$($M_\star$,\, $\sigma_{M_\star}^2$)  \\
        \bottomrule
    \end{tabular}
\end{table}

In addition to GaiaPMEX, which  aims  at characterizing the mass and sma of the binaries, we  also built GaiaExplore, that just aims at  quickly identifying sources with a significance beyond the 2--$\sigma$ threshold adopted above. This allows splitting split the faster detection phase and the slower characterization phase of our analysis. For any list of sources, it searches online archives to gather Simbad, Gaia DR3, and, if available, Hipparcos-2 data \citep{vanleeuwen2007} and PMa  data from \citep{Kervella2022}. Then it calculates, as explained above, the astrometric signature from the RUWE and its significance, as well as, when Hipparcos-2 data are available, the significance of the PMa.  We note that a 5-parameters fit is necessary on a given source's astrometry in the Gaia DR3 (\verb+astrometric_params_solved+ = 31 or 95, see Z24). For some sources,  only a 2-parameters (sky RA and DEC coordinates) fit solution is released in the Gaia DR3. In such cases, GaiaPMEX cannot be used, since it requires that parallax and proper motion be also fitted.  

\subsection{Limitations of GaiaPMEX}

One of the most stringent limitations of GaiaPMEX is the source magnitude, as the calibration and AL-scan measurement noises were only determined for sources with $G$\,\!$<$16. This is nonetheless with a weak impact on studies, such as this work, based on RV surveys that are strictly limited to $V$-mag $<$ 15.

For the sources with $G$\,\!$<$16, the determination of calibration and AL-scan measurements noises are certainly only approximative, assuming constancy of noise levels over time, and cannot model systematic non-Gaussian wanderings of astrometric signals, if any, beyond single Keplerian motion. We thus assume that "in average" any source behaves well over time, that is within the typical behavior of single sources of the same category in $G$ and $Bp-Rp$ magnitudes, which consists in calibration noise and AL-scan measurement error. 

Although this tool can apply, within the above-mentioned limitations, to most sources, the interpretation of the AEN, \ruwe\ and PMa and their respective GaiaPMEX curves, is limited by the possible presence of additional unmodeled signals, beyond single Keplerian orbit, including for instance stellar activity, disk accretion, and multiplicity -- as discussed in Section~\ref{sec:assumptions} below -- as well as blended PSF due to visual binary companions or background sources. 
The presence of a binary stellar companion with a separation typically 200 mas or more (up to 1000 mas) may impact the Gaia PSF, depending on the brightness ratio between the two components  \citep{Holl2023}. In such a case, the (sma, mass) solutions found by GaiaPMEX can be impacted, and should be taken with care. In practice, the $ IPD{\_ \-}frac{\_ \-}multi{\_ \-}peak $ (hereafter Frac) and $ IPD{\_ \-}gof{\_ \-}harmonic{\_ \-}amplitude $ (hereafter GoF) are two useful indicators of such a situation determined by the Image parameter determination (IPD) of Gaia. The first one indicates that multiple peaks are occasionally detected in the PSF and removed from the aperture window, and the second one measures the amplitude of the goodness-of-fit variations when fitting the PSF by a single source model, as a function of the position angle of the scan direction. Note, however, that  multiple peaks can both be due to background stars, or physical companions, and thus is not a strong indicator multiplicity.

\subsection{Note on the origin of the GaiaPMEX signal} \label{sec:assumptions}
In our approach, we implicitly assume that \nsigmaruwe{}  $\geq$  2 or  \nsigmapma{}  $ \geq$  2  indicates the presence of one companion. 
The rationale for assuming a single companion instead of several is that such rudimentary astrometric data, basically two points -- \ruwe\  and PMa -- cannot, on its own, disentangle signals from different companions.
Assuming that both the Gaia astrometric excess and PMa are due to a single companion is an additional hypothesis, also taken for the sake of parsimony. 

Finally, we identified three other effects, besides a companion, that could a priori produce an astrometric signal: stellar activity, and in the case of young systems, accretion and disks. They are discussed below.

\subsubsection{Spots}\label{sec:spots}
Low-mass to solar-type stars show various types of activity ~\cite[see e.g.][for a review]{Meunier21} that can, a priori, induce shifts in the star's photocenter. This is the case, in particular, for dark spots and bright faculae. The astrometric jitter (AJ, defined as the rms of the induced shifts of the photocenter) expected from the Sun seen at 10 pc is less than 0.1 $\micro$as for spots only ~\citep{Lagrange11}. This translates into an astrometric jitter of 0.04 $\micro$as for a  Sun-twin at 50 pc. For such a star, the relation between the RV semi-amplitude (RVSA) and the astrometric jitter (AJ) was found to be: $\text{AJ} (\micro \text{as}) = 0.2 \times \text{RVSA}$ (m/s). For close-by FGK-type stars, simulations yield astrometric jitters less than 0.5 $\micro$as ~\citep{Meunier22}. Such values are orders of magnitudes smaller than the values (typ. 0.1 mas) associated  to binaries.

The activity of young stars is, however, much more intense than the Sun's one, as reflected by their log(R'HK), down to $-4.2$ to be compared to that of the Sun, which varies between $-4.93$ and $-5.03$ ~\citep{Hall07}. It is dominated by a smaller amount of dark spots or bright structures  with sizes larger than those of their older counterparts ~\citep{Berdyugina05,Lockwood07}. The temperature difference between the structures and the star's surface is up to a few hundred of degrees.
In the case of the nearby (9.7 pc) M1-type star AU Mic, known for decades to show intense magnetic activity clearly seen in photometry and in spectroscopy ~\citep{Lannier17, Zicher22,Mignon23},  the amplitude of these long-term photometric variations is of the order of 5$\%$. These variations are best reproduced by one dominating, long-lived structure that induces RV variations due to stellar rotation modulation, with an amplitude of 600 m/s, plus a smaller one  ~\citep{Martioli21}\footnote{Note that we do not consider the impact of the flares that produce additional, high amplitude photometric variations because they are spurious and very short-lived.}.  

A rough estimate of the impact of spots/faculae on the astrometric jitter can be made as follows. We assume a single, large and fully dark spot that produces a flux variation of 5 $\%$  of the star flux (as in the extreme case of AU Mic). Note that multiple structures spread over the star longitudes would decrease the astrometric jitter. The maximum shift of the photocenter is obtained when the spot is located at the edge of the stellar disk and is $\simeq  0.05 \times R_*$. We computed the induced photometric shift for all our targets, taking into account their estimated radius, their distance, and assuming such a spot. To estimate the stellar radii, we employed the \textsc{madys} tool \citep{squicciarini22} for stellar parameter determination based on isochrone fitting: the conversion between observed Gaia DR3 \citep{GaiaDR3} and 2MASS \citep{2mass} photometry and stellar parameter was mediated by PARSEC isochrones \citep{parsec} with solar metallicity, under the assumption of moving groups' ages.

For all targets, the induced shifts are found to lead to \aastroruwe{} and \aastropma{} significances much smaller than the thresholds used for companion detection: indeed, the median value of the induced shifts is about  0.003 mas, while all \aastroruwe{} are larger than 0.05 mas. 
Hence, we conclude that the jitters due to stellar activity are negligible in the context of the present study. 

\subsubsection{Accretion and circumstellar  extinction}\label{sec:accretion}
In the case of very young (a few Myr) stars, the photocenter shifts can have other origins than a companion. This is potentially the case when stars are still accreting material from their  disks. For classical T Tauri members of Taurus, Lupus, or Chameleon, the  accretion luminosity may be as high as the star luminosity itself \citep{Alcala21,Gangi22}. The accretion luminosity is dominated by the accretion spot at the stellar surface but regions within au from the stars may also contribute to this luminosity. Also, 
 scattered light from dust in the non homogeneous disk, variable in time, can also contribute to the photometric budget. In such cases, attributing the astrometric variations to a companion may then be incorrect.

Episodic extinction by the inner disk \cite[dippers, see e.g. ][]{McGinnis2015} may also occur in the case of accreting stars seen with a high inclination (greater than typically 55 degrees). Such effects cannot be estimated without a dedicated modeling of the systems. Hence, in the case of accreting stars, the Gaia signal cannot be straightforwardly attributed only to companions. 

Most of the stars considered in the present paper are older than 10 Myr, i.e. much older than classical T Tauri stars. Yet, some members of the youngest associations, eps Cha and TWA do show signs of accretion or of being still embedded in massive disks.

With an age of 3-5 Myr, eps Cha is the youngest nearby moving group in our sample. Three members of Eps Cha in our our sample are actually known to be surrounded by accretion disks:  DX Cha (HD 104237A), a Herbig AeBe star which forms, together with the T Tauri-type companion HD 104237B a close binary system, surrounded by a CO-rich circumbinary disk \citet{Hales14}, and two CTT stars, MP Mus (PDS 66, CPD-681894) and T Cha  \citep{Kastner10,Sacco14}. We will therefore not consider these targets in our study. Six do not show IR excesses indicative of accretion disks \citep{Dickson-Vandervelde21}. Among the remaining ones, RXJ 10053-7749, and [K2001c] 17 are reported as weak-line T Tauri stars (WTTs) by \citet{Wahhaj10}; we therefore do not expect accretion to be active for these stars.  

None of our TWA targets, except TWA 3A, belongs to the list of the 14 confirmed or putative TWA members showing evidence of dense circumstellar material from mid-infrared excess in Spitzer/IRAC or WISE bands \citep{Venuti19}. We furthermore note that for those 14 accretors, the accretion luminosity is 100 to 1000 times lower than stellar luminosity. Hence, we can assume that the astrometric signal will probably not be impacted by accretion for our TWA targets, but the results have to be taken with caution because of this risk. 

For the remaining stars, we collected AllWISE \citep{wise} photometry to look for evidence of infrared excess. Assuming ages corresponding to their respective moving group, we computed expected AllWISE magnitudes based on PARSEC isochrones, and then compared the synthetic $G-Wi,\, \forall i \in [1,4]$ colors with the observed ones. Whereas the above-mentioned T Cha and DX Cha show a significant ($\Delta(G-Wi) > 2$) IR excess in all bands, other stars only show infrared disks in W3 and W4 bands, indicative of a more evolved disk: in addition to the already mentioned TWA 3A(TWA according to SACY; ? according to Banyan), to this list belong TW Hya (TWA), DZ Cha (ECHA), HD 98800 (TWA), HD 319139 (BPC) and CPD-68 1894 (ECHA according to SACY; LCC according to BANYAN classifications). For the sake of precaution, in the following, we will not consider these stars for the GaiaPMEX analysis, except HD 319139 given its older age.

\section{GaiaPMEX overall results \label{sec:results}}
\subsection{GaiaPMEX maps}\label{sec:outputs}

\begin{figure*}[hbt]
    \centering
    \includegraphics[width=60.mm,clip=True]{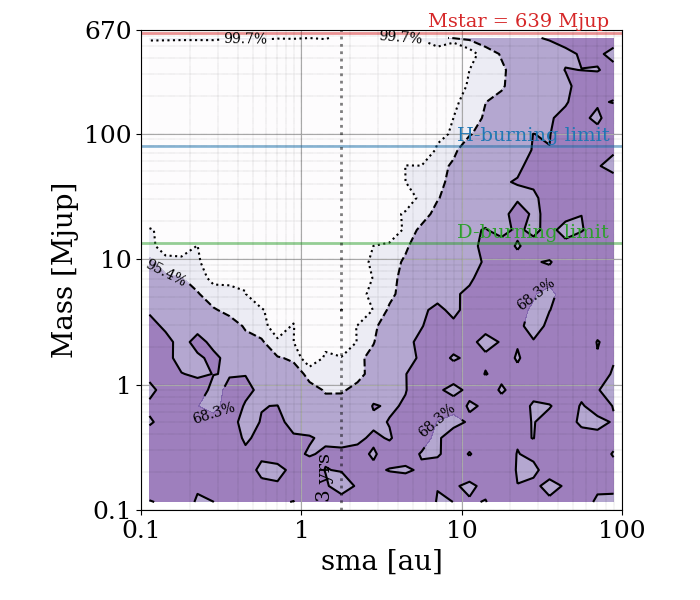} 
    \includegraphics[width=60.mm,clip=True]{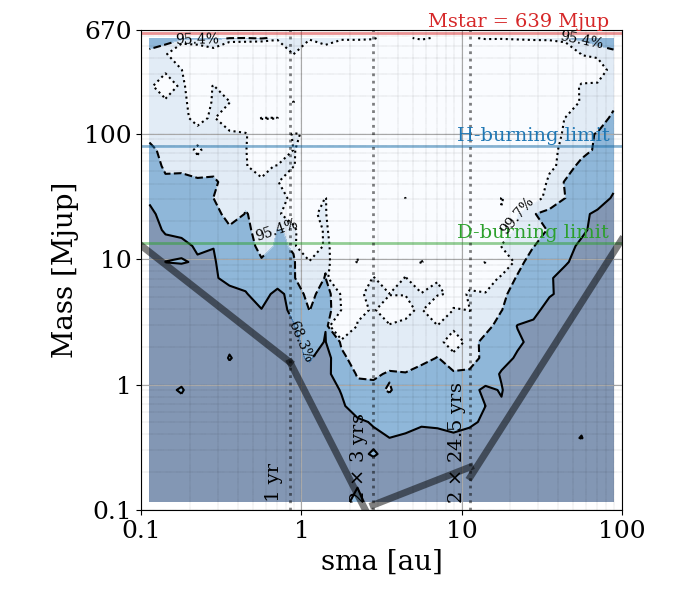} 
    \includegraphics[width=60.mm,clip=True]{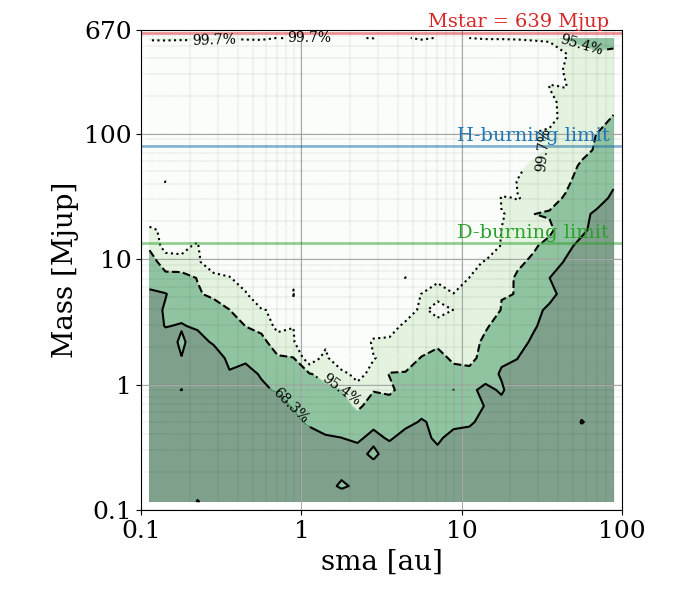} \\
    \includegraphics[width=60.mm,clip=True]{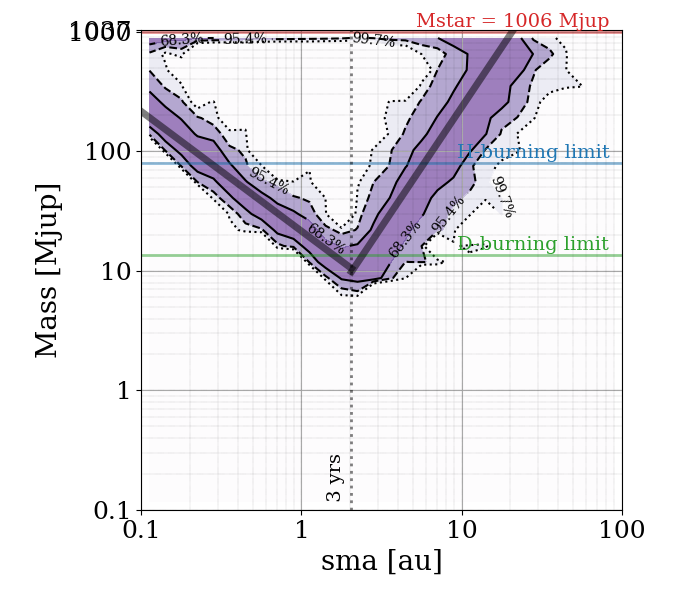} 
    \includegraphics[width=60.mm,clip=True]{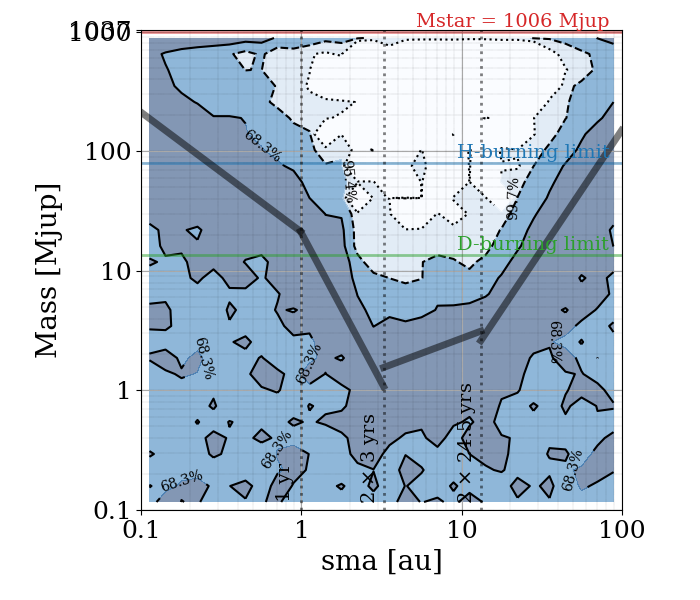} 
\includegraphics[width=60.mm,clip=True]{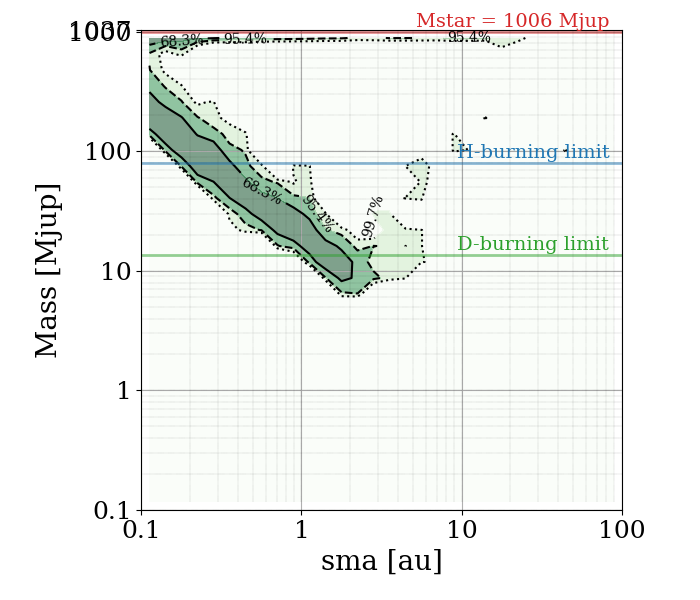} \\
    \includegraphics[width=60.mm,clip=True] {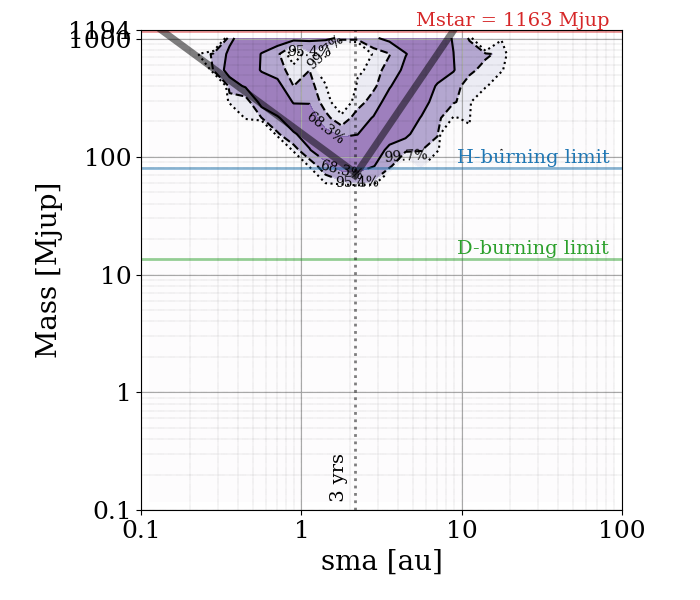} 
    \includegraphics[width=60.mm,clip=True] {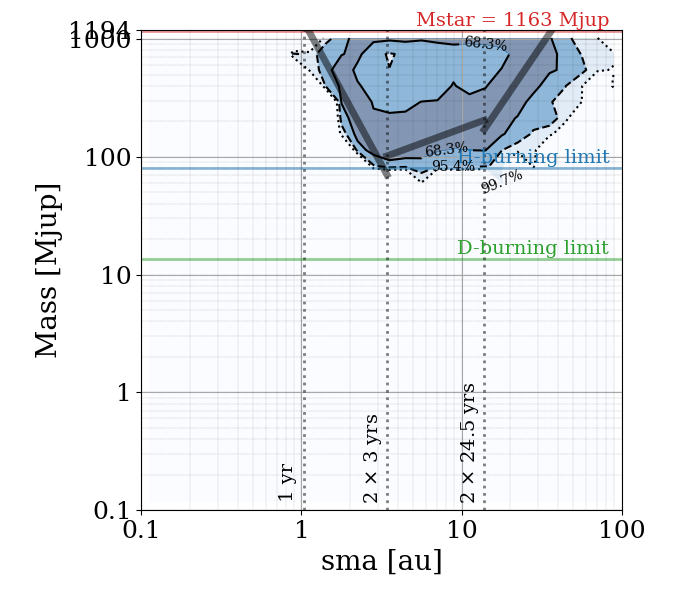} 
    \includegraphics[width=60.mm,clip=True]{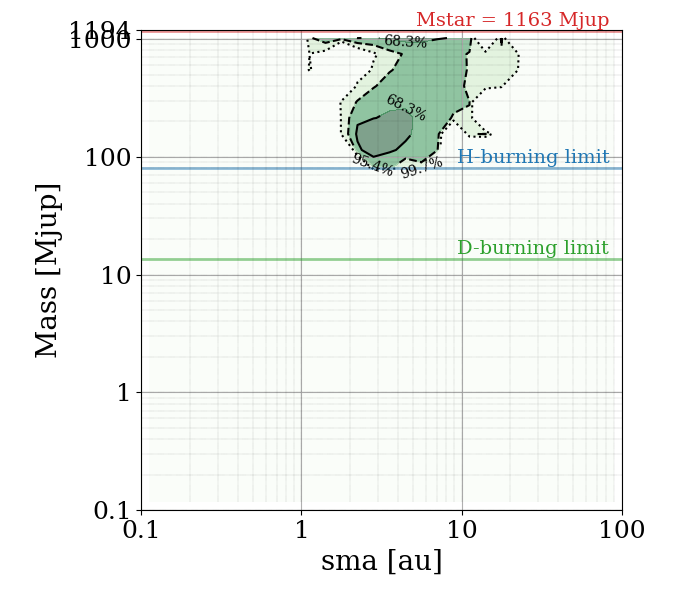}\\ 
\includegraphics[width=60.mm,clip=True]{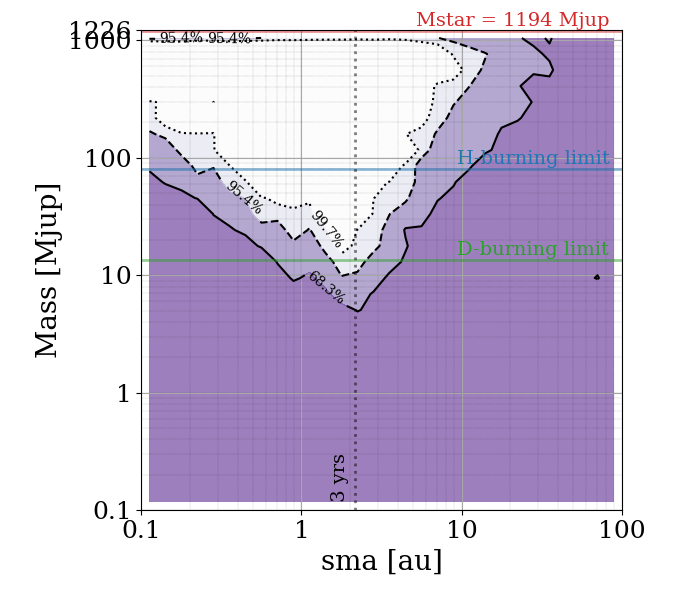} 
    \includegraphics[width=60.mm,clip=True]{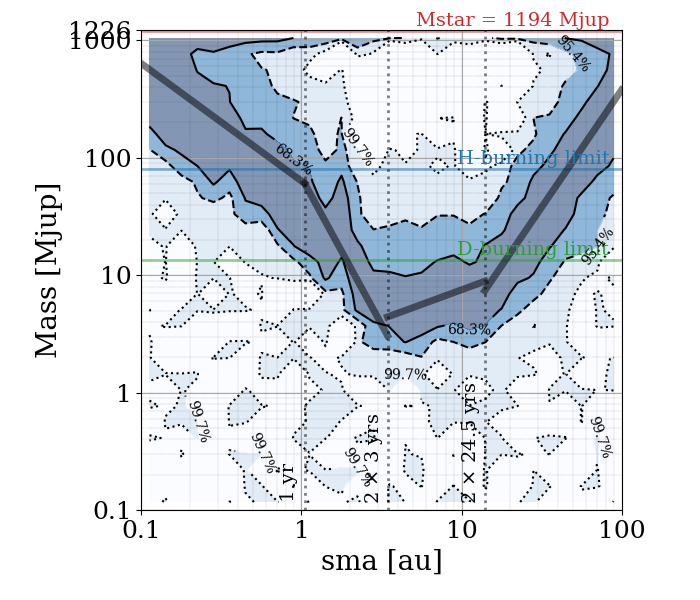} 
    \includegraphics[width=60.mm,clip=True]{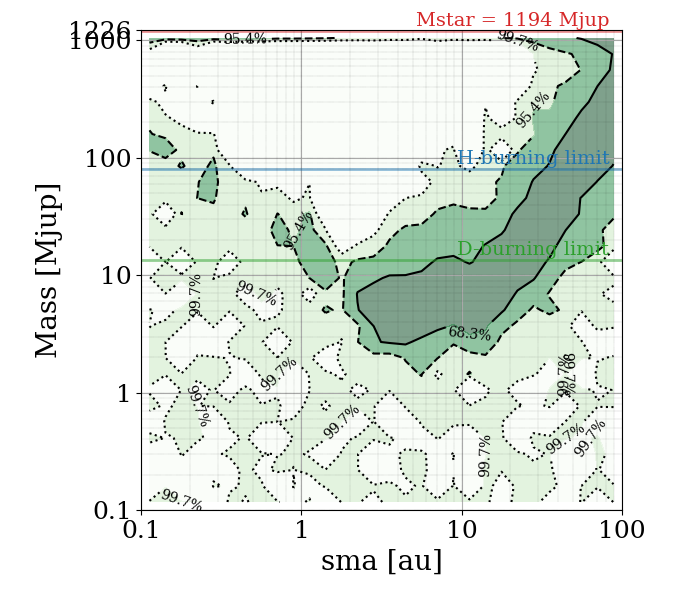} 
  \includegraphics[width=60.mm,clip=True]{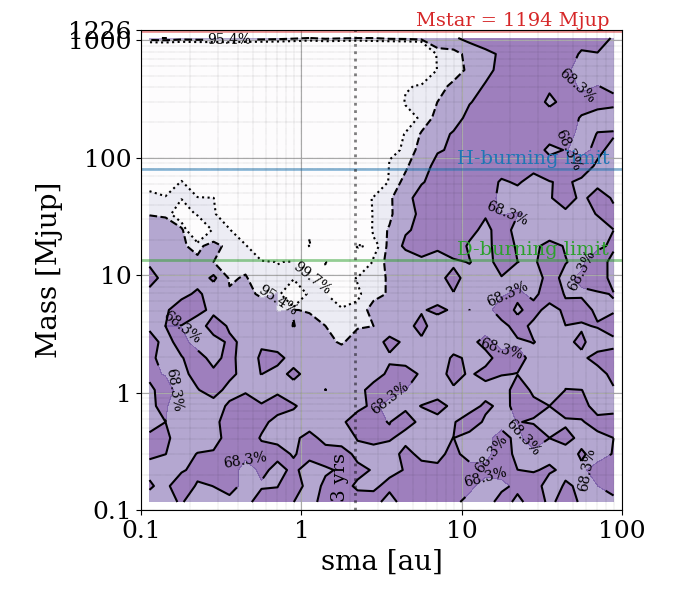} 
    \includegraphics[width=60.mm,clip=True]{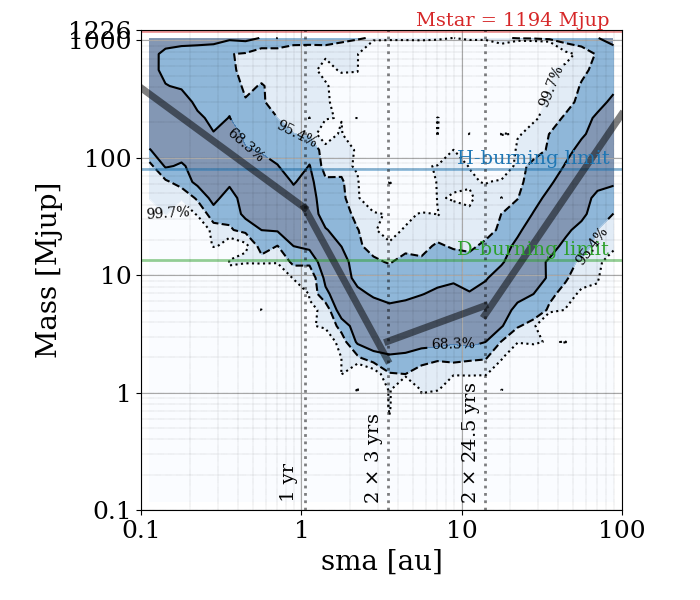} 
    \includegraphics[width=60.mm,clip=True]{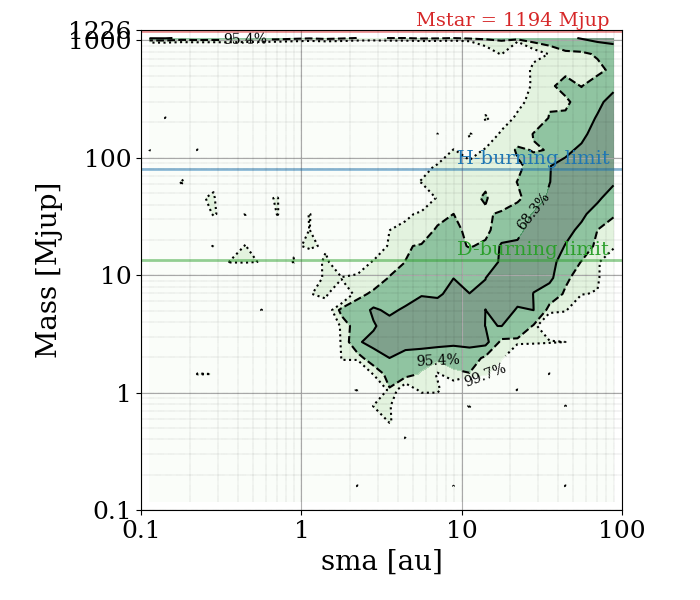} 
    \caption{GaiaPMEX (sma, mass) solutions for potential companions.  Constraints from \ruwe{}  only  (Left), constraints from PMa only (Middle), and  constraints from \ruwe{}   + PMa (Right). Each row, from Top to Bottom: 1/ No detection: AU Mic. 2/ Star with a "small" sma  or BD companion:  HD 3221. 3/ Star with a well constrained sma companion:  HD 139084. 4/ Star with a "large" sma  or BD companion: PZ Tel. 5/ Star with  a "large" sma  or BD or Planet companion: AF Lep.
    The  contours show the 68\%, 95\%, and 99.7\% confidence intervals. The black lines show the  model relationships developed in K24. }
    \label{fig:examples}
\end{figure*}

Figure~\ref{fig:examples} shows examples of  GaiaPMEX results on five stars members of the BPC according to SACY and BANYAN classifications, when using our \ruwe{}  analysis (left column), when using the PMa  only (middle column), and  when combining both constraints from \ruwe{}  and PMa (right column). 
Five  different cases are considered: no  detection of companions (AU Mic, 1st row), detection of a less-than-3-au-sma companion, which, on the basis of the present astrometric data only, can be either a star or a BD (HD 3221, 2nd row), detection of a very-well-constrained-sma  companion (HD 139084, 3rd row), detection of a "large"-sma (larger than typically 2-3 au)  companion which, on the basis of the present astrometric data only, can be either a star or a BD (PZ Tel, 4th row), and detection of a large sma companion which, on the basis of the present astrometric data only, can be either a star, or a BD, or a planet (AF Lep, 5th row). Note that the companion of HD 3221 is actually known to be a star, thanks to RV data ~\citep{vogt15} and that HD 139084 is classified as a SB1 spectroscopic binary in ZF21. Also, PZ Tel has a BD companion directly imaged ~\citep{Biller10,Mugrauer10}; its dynamical mass and sma parameters found by ~\cite{Franson23} are compatible with the GaiaPMEX solutions. AF Lep has a planetary mass companion also directly imaged with high contrast imaging ~\citep{Mesa23,deRosa23,Franson2023_aflep}, and the mass and sma found for this planet are compatible with GaiaPMEX solutions. 

As expected, when a companion is detected, the (mass, sma) solutions are, in most cases, degenerate when considering \ruwe{}  only or PMa only. Combining the constraints from \ruwe{}   and PMa  significantly reduces the (sma, mass) solution ranges in case of detection, and improves the detection limits in case of non-detection. In very few cases like HD 139084, the (sma, mass) domain is very well constrained with \ruwe{}  and PMa constraints. Most of the time, though, the (mass, sma) domain of the companion is still degenerate, and additional data such as RV or high contrast imaging data are  needed to fully constrain the companions' mass and orbital properties. This was the case for PZ Tel and AF Lep (see above).

When a companion is detected, we can  distinguish different types of degeneracies, as described in K24:
\begin{itemize}
    \item the solutions are on both parts of the V-shape locus, with possible sma ranging from a fraction of au to a several (sometimes dozens of) au. This may happen happens  when Gaia data only  are  considered or when only PMa are considered or when they are both considered. Examples can be seen in Figure~\ref{fig:examples}: HD 138084 (left) and PZ Tel or AF Lep (middle). 
    \item all solutions are in the left part of the V-shape locus of solutions only, forming thus a L-shape. All have sma below typ. 1-3 au. The companion mass is often  above 13 \Mjup. This can happen only when both Gaia and Hipparcos data are available. An example can be seen in Figure~\ref{fig:examples}: HD 3221 (right).
    \item all solutions are located in the right part of the V-shape locus of solutions. All have sma greater than typ. 1-3 au. Depending on the stars, the sma can extend to dozens of au. The companion mass can be as low as 2-3 \Mjup. This can happen only when both \ruwe{} and PMa  are available.  Examples can be seen in Figure~\ref{fig:examples}: HD 138084, PZ Tel, or AF Lep (right). 

\end{itemize}

\subsection{GaiaPMEX sensitivity}\label{sec:sensitivity}
When no companion is detected, the (mass, sma) maps (either based from the \ruwe{}  analysis or PMa only or considering both constraints from \ruwe{}  and PMa) obtained  provide valuable detection limits. An example can be seen in Fig. \ref{fig:examples} in the case of AU Mic, and other examples are provided in subsequent figures. The sensitivity domain corresponds to the white/clear upper parts of the maps on the left column (\ruwe{}  analysis only) or on the middle (PMa analysis only) on the right column (\ruwe{}  + PMa). These areas give the (sma, mass) of companions that can be excluded as they would have been detected if present around the stars. Of course, the sensitivity domain depends on the star's properties (distance and mass, magnitudes, and colors).  

\begin{figure}[hbt]
    \centering
    \includegraphics[width=40.mm,clip=True]{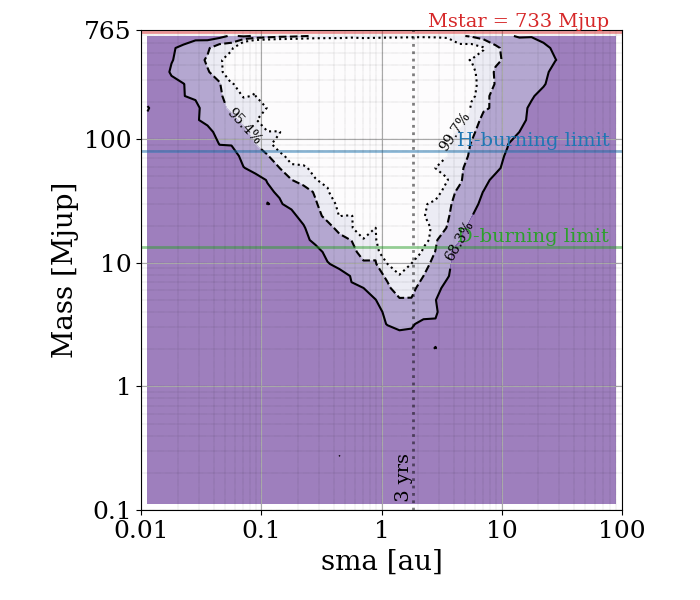}
    \includegraphics[width=40.mm,clip=True] {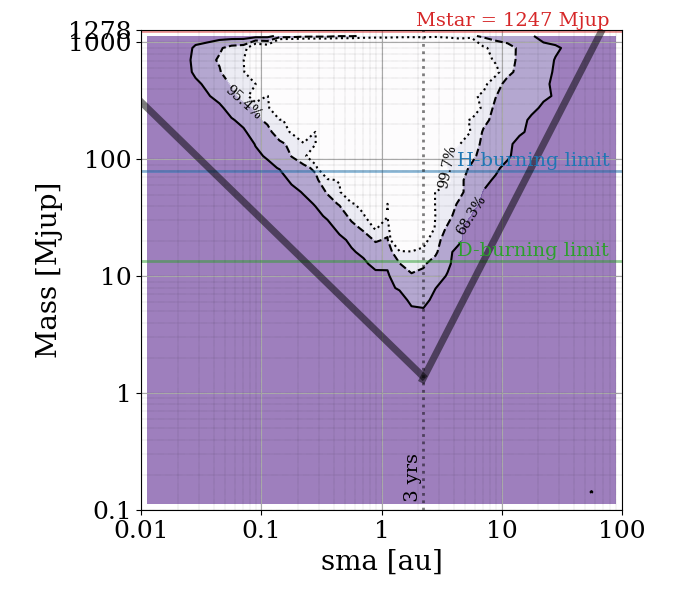} 
    \caption{Impact of the star masses on the Gaia alone sensitivity domain. Left: CD-351167 (0.7 \Msun) and Right: HD 13246 (1.2 \Msun). Both stars have parallaxes of $\simeq$ 22 mas. Note that the y-axes cover different mass ranges.}
    \label{fig:impact_mass}
\end{figure}

At a given distance, the sensitivity domain increases with decreasing star mass. The effect is, yet, rather limited for most of our targets, given the limited range of masses considered in the present analysis (see Figure~\ref{fig:histo_sample}). This is illustrated in Figure~\ref{fig:impact_mass} where we show the maps obtained  
for CD-351167 (0.7 \Msun) and HD 13246 (1.2 \Msun), both members of THA (according to both classifications), and with similar parallaxes ($\simeq$ 22 mas). 

\begin{figure}[hbt]
    \centering
    \includegraphics[width=40.mm,clip=True]{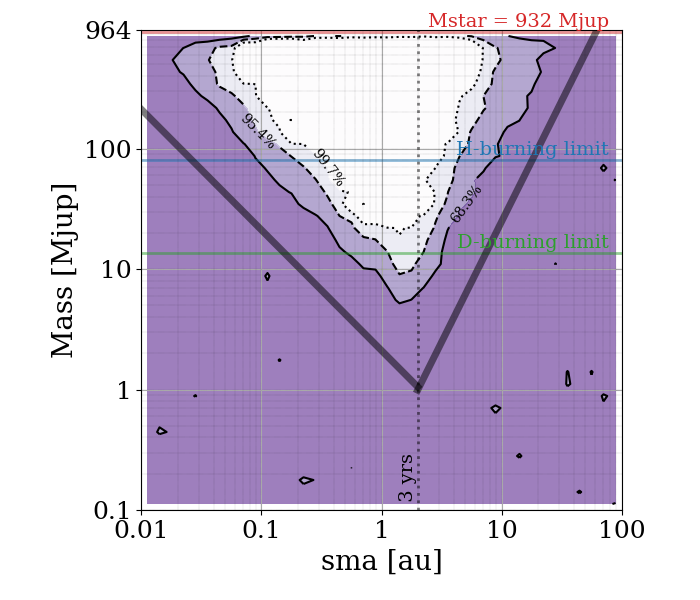} 
    \includegraphics[width=40.mm,clip=True]{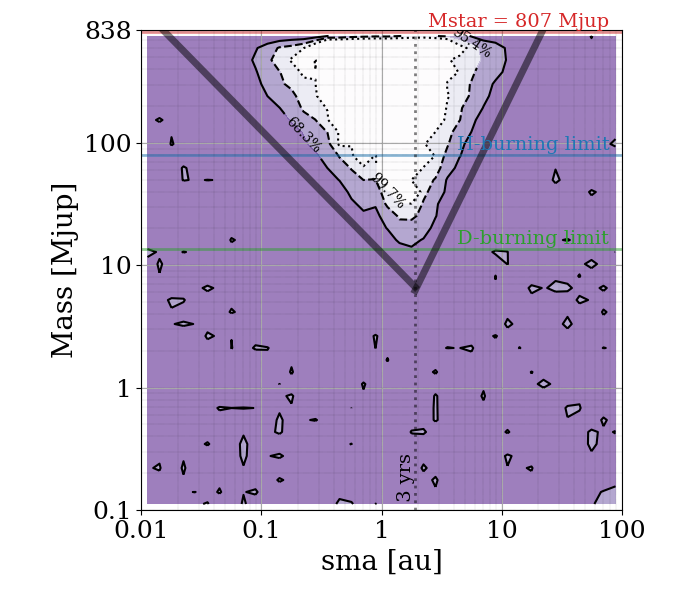} \\ 
    \includegraphics[width=40.mm,clip=True]{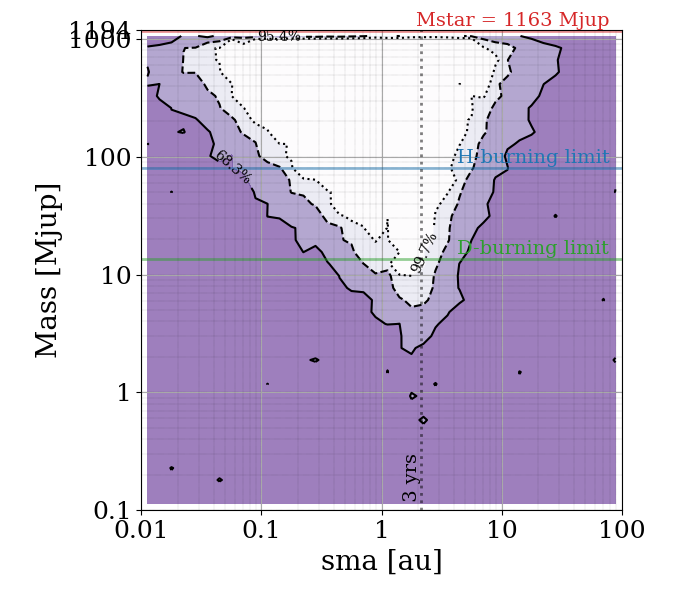}
  \includegraphics[width=40.mm,clip=True]{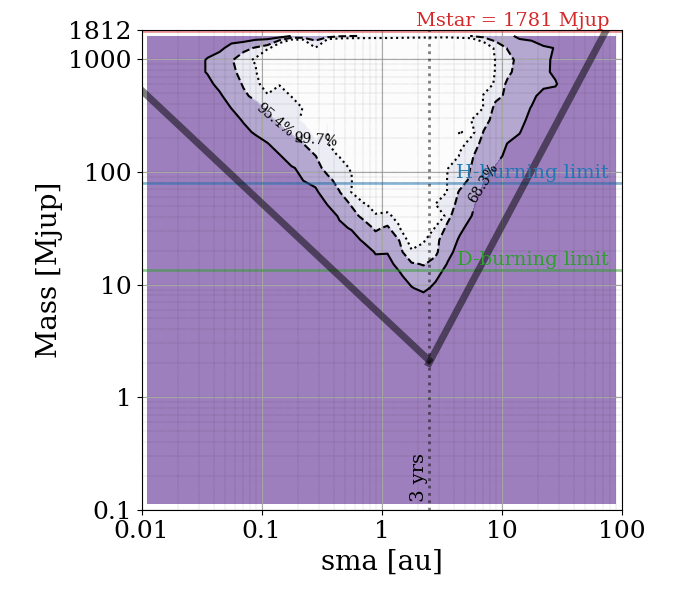} 
    \caption{Impact of the distance on the Gaia alone sensitivity domain. Top, Left: HD 23208 (0.9 \Msun,  $\pi$ =  17.6 mas) and Right: TYC 7066-1037-1 (0.8 \Msun, $\pi$ = 7.4 mas). Bottom, Left:  HD 105 (1.1 \Msun, $\pi$ = 25 mas) and Right: HD 47875 (1.1 \Msun, $\pi$ = 14 mas). Note that the y-axes cover different mass ranges.}
    \label{fig:impact_distance}
\end{figure}

For a star with a given mass, the sensitivity domain increases with decreasing distance. This is illustrated in Figure~\ref{fig:impact_distance} where we show the obtained maps 
for HD 23208 (0.9 \Msun, $\pi$ = 17.6 mas), TYC 7066-1037-1 (0.8 \Msun, $\pi$ = 7.4 mas), two members of OCT according to both SACY and BANYAN classifications, and HD 105 (1.1 \Msun, $\pi$ = 25 mas) and HD 47875 (1.1 \Msun, p$\pi$ = 14 mas), two members of THA according to both SACY and BANYAN. 

From the inspection of all the GaiaPMEX maps obtained, we conclude that the survey is  sensitive to all stellar mass companions in the $ \simeq $ 0.1 to $ \simeq $ 10 au range for all but 6 (resp. 2, 2, 1, 5, 5, 1, 0) stars members of COL (resp. THA, BPC, OCT, ABD, ARG, TWA, ECH). The sensitivity to BDs is also remarkably good in the $ \simeq $ 1 to $ \simeq $ 3 au range.  

For sma below 0.1 au, and above 10-20 au, the sensitivity is, as expected, limited. The $\leq $ 0.1 au domain can be  well explored by spectroscopy, while the $\geq $ 10 au domain can be well covered by high contrast imaging, for stellar and massive BDs companions. Finally, this survey is also sensitive to planetary-mass companions, as detailed in Section \ref{sec:vetting_cc}.

\section{Astrometric binaries}\label{sec:binaries_main}
\subsection{GaiaPMEX detections (all masses)}\label{sec:binaries}

Table C.1\footnote{Table C.1 is only available in electronic form at the CDS via anony-mous ftp to cdsarc.u-strasbg.fr} provides the binary status found for each target of our sample. For each target, we indicate its \nsigmaruwe{} and \nsigmapma{}. As mentioned above, a binary status is attributed when one of these parameters is larger than 2.  For comparison purposes, we also indicate the conclusions on their binary status based on spectroscopic diagnostics using ZF21, as well as other indicators of binarity from the literature: $WDS flag$, $SB9 flag$, and $NSS flag$ (see Appendix C). We also provide the Frac given by Gaia. 
In case of a 5 or 6 parameters fit, Frac  greater than typically 20-30 indicates a stellar pair (companion or background) with a projected separation of about 400-1000 mas ~\citep{Holl2023}. In case of a 2-parameter solution, a Frac greater than typically 20 indicates a stellar pair with a projected separation of about 200 mas, while a Frac greater of 30-60  indicates a stellar pair with a projected separation of about 200-400 mas. Finally, a Frac of about 0 associated with a  GoF greater than 0.2 is indicative of a galaxy (such a situation is not present in our sample) or a bright binary companion with separation $<$200 mas.  The GoF is below 0.2 for all our targets except two targets, CD-431395 and HD161460, which have also high Frac, and have 3 parameters-solved so no GaiaPMEX computation. Therefore, we do not report the individual values of this indicator.
 We finally flag, for information, the targets for which companions were detected in direct imaging by ~\citet{Bonavita22}. 
 
 Table C.1 also reports the  sma at mimimum mass value, and the maximum masses of the GaiaPMEX solutions for each target. The minimum masses vary significantly (between stellar masses to planetary masses) from one target to the other. The maximum sma is often smaller than about 20 au, but may in some cases be as high as to a few dozen of au. 

In the SACY classification, 126 targets are found to be binaries. A similar number is found in the BANYAN classification. When removing the two youngest associations, eps Cha and TWA according to SACY classification, GaiaPMEX finds 60 binaries based on the \ruwe{} only, 13 on PMa only, and 10 on \ruwe{}  + PMa. Note we also find 4 targets: CD-471999 (OCT), HD 55279 (COL), VXRPSPC16a (ARG), and HD217379A (ABDor) with \nsigmaruwe{} lower than but close to 2 (1.8, 1.7, 1.8, 1.7, respectively), and \nsigmaaen{} (see K24) greater than 2. We consider them as interesting candidates. 

Finally, Table~\ref{tab:stats} summarizes the number of targets for these associations, as well as the number of binaries found using  the \ruwe{}  only, PMa only, and \ruwe{}  or PMa. These numbers are given for both the SACY and the BANYAN classifications. We note that most detections are with \nsigmaruwe{} or \nsigmapma{}$ > 3$. In addition, we note that many of the targets detected with \nsigmaruwe{} or \nsigmapma{} between 2 and 3, are known as binaries based on other criteria.  

\begin{table}[]
\centering
\caption{Summary of detections and associated binary rates. SACY classification. }
\label{tab:stats}
\resizebox{\linewidth}{!}{%
\begin{tabular}{lcccccc}

 \toprule
	&	THA	&	OCT	&	COL	&	BPC	&	ARG	&	ABDor	\\
 \midrule

Sample	&	76	&	16	&	43	&	42	&	30	&	46	\\
 \midrule
\nsigmaruwe{}$ \ge 2$	&	20	&	8	&	8	&	14	&	6	&	15	\\
 2 $\le  $\nsigmaruwe{}$ \le  $  3	&	5	&	0	&	1 (1 known)	&	5 (4 known)	&	2	&	1	\\
\nsigmapma{}$ \ge 2$	&	5	&	1	&	4	&	6	&	0	&	7	\\
2 $\le  $\nsigmaruwe{}$ \le  $  3	&	2 (2 known)	&	0	&	0	&	0	&	0	&	0	\\
Astrom. bin. (total)	&	22	&	9	&	12	&	18	&	6	&	17	\\
2 $\le  N$-$\sigma  \le 3$	&	6 (3 known)	&	1	&	1 (1 known)	&	5 (4 known)	&	2	&	1	\\
 \midrule
Spec. bin. ZF21	&	13	&	2	&	1	&	11	&	1	&	4	\\
Spec. bin. ZF21 only	&	6	&	0	&	1	&	4	&	1	&	2	\\
 \midrule
Binary (total)	&	28	&	9	&	13	&	22	&	7	&	19	\\
\midrule
Astrom. bin. rate&	29 $\pm $ 6	&		&	28 $\pm $8	&	43 $\pm $10	&		&	37 $\pm $9	\\
Binary rate 	&	37 $\pm $ 6	&		&	30 $\pm $ 8	&	52 $\pm $10	&		&	41 $\pm $9	\\
 \midrule
Nbr Planets/light BD	&		&		&	1$^a$	&	3$^b$	&		&	1$^c$	\\
\bottomrule
\end{tabular}%
}
\tablefoot{
Note that binaries and binary rates include stars, BD, and planets.  1st row: number of astrometric binaries with \nsigmaruwe{}$ \ge 2$; 2nd row: number of astrometric binaries with  2 $\le  $\nsigmaruwe{}$ \le  $  3; 3rd row: number of astrometric binaries with \nsigmapma{}$ \ge 2$; 4th row : number of astrometric binaries with  2 $\le  $\nsigmapma{}$ \le  $  3; 5th row: total number of astrom. binaries (this work); 6th row: number of astrom. binaries with 2 $\le  N$-$\sigma \le  $  3. In parenthesis, the number of already known binaries, independently of the astrometric data. The following lines provide respectively the number of SB or SB? listed by ZF21, the number of spec. binaries not found as astrometric binaries among the spectroscopic  binaries listed in the previous row, the rate of astrom. binaries, the rate of astrom+spec. binaries, and the number of known planets/light BDs among the sample. (a): AB Pic, (b) : PZ Tel, AF Lep, HD 14082B b, (c) G80-21.
}
\end{table}


\subsection{Comparison with ZF21 survey }\label{sec:Analysis}

 ZF21 lists 
 0 binaries in OCT (among 15 targets)\footnote{HD 155177 is flagged as a SB2 on the basis of three RV by ZF21, though.}, 8 binaries in BPC (40 targets), 10 in THA (75 targets), 0 in OCT, 0 in ARG (29 targets), 1 in COL (44 targets), and 3 in AB Dor (40 targets). The 22 binaries are flagged as SB or  SB1, SB2, SB?, SB1?, SB2?, SB3 when classifications can be applied, or as ? when no conclusion was drawn from their analysis. These numbers can be compared to the numbers of binaries  with sma unambiguously  less than 10 au found by GaiaPMEX: 6, 7, 6, 3, 7, and 5 in the same associations (total of 28) (Table~\ref{tab:stats}). 
Hence, GaiaPMEX finds more binaries (28) with sma unambiguously less than 10 au than ZF21 (22). Note that additional binaries identified by GaiaPMEX with sma up to a few dozens of au could actually have their sma in this range as well. The larger sensitivity to binary can be explained by the fact that GaiaPMEX is sensitive to binaries with larger sma, and with lower mass ratio than spectroscopy. Also, spectroscopic detection relies on favorable inclinations of the companion orbital plane, while absolute astrometry does not. 

\begin{figure*}[hbt]
    \centering
    \includegraphics[width=40.mm,clip=True]{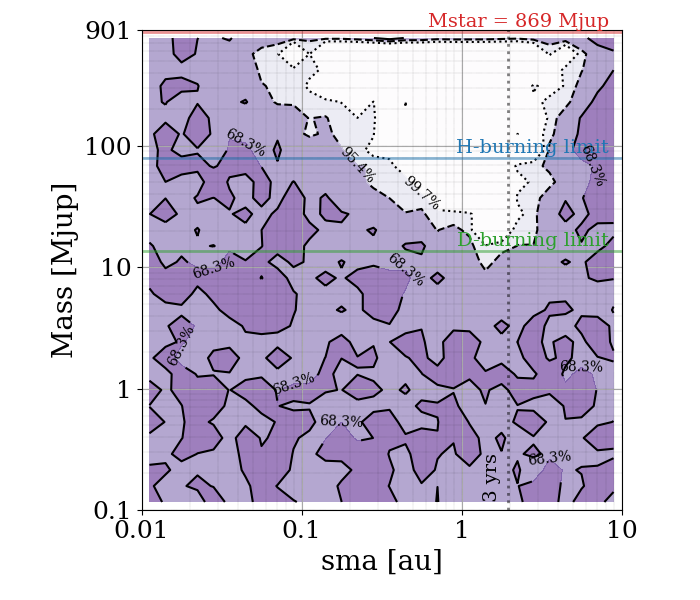} 
    \includegraphics[width=40.mm,clip=True]{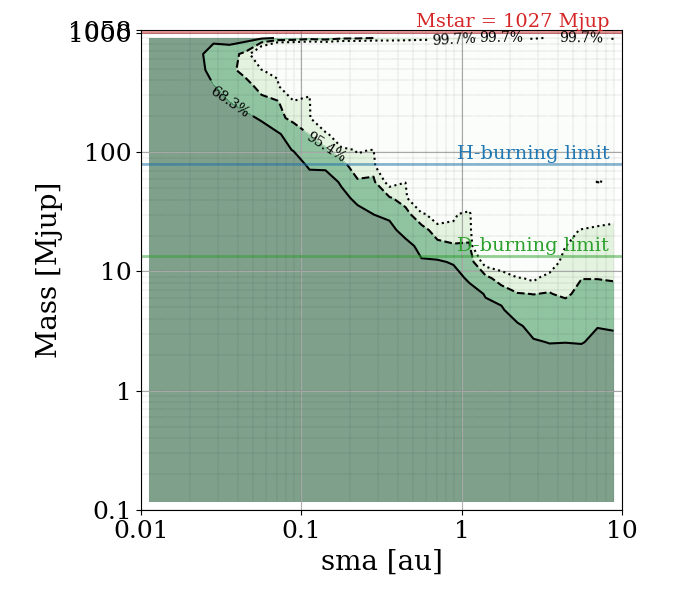} 
    \includegraphics[width=40.mm,clip=True]{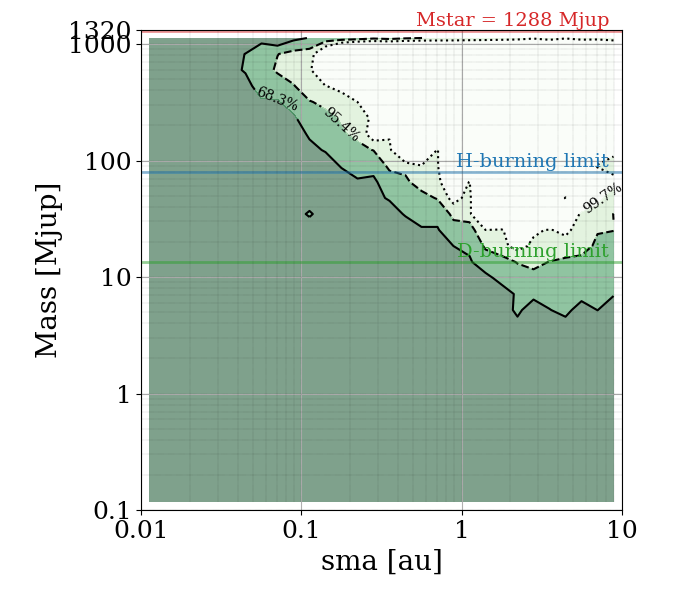} 
\includegraphics[width=40.mm,clip=True]{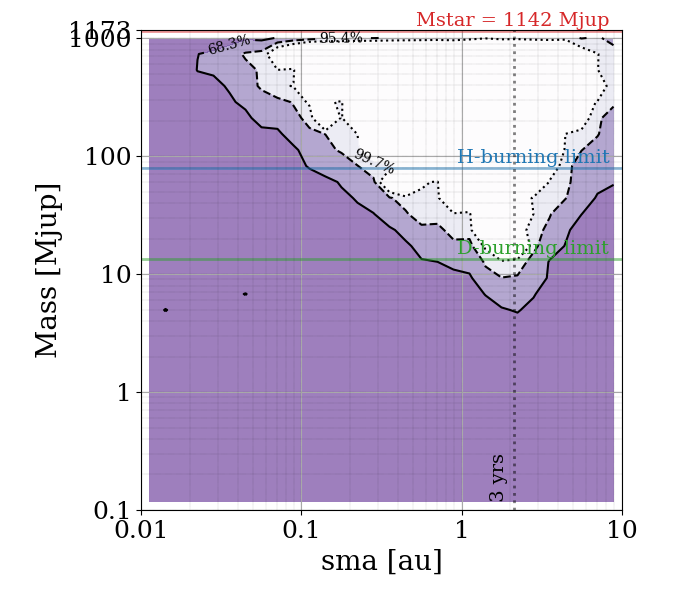}  
   \\
 \includegraphics[width=40.mm,clip=True]{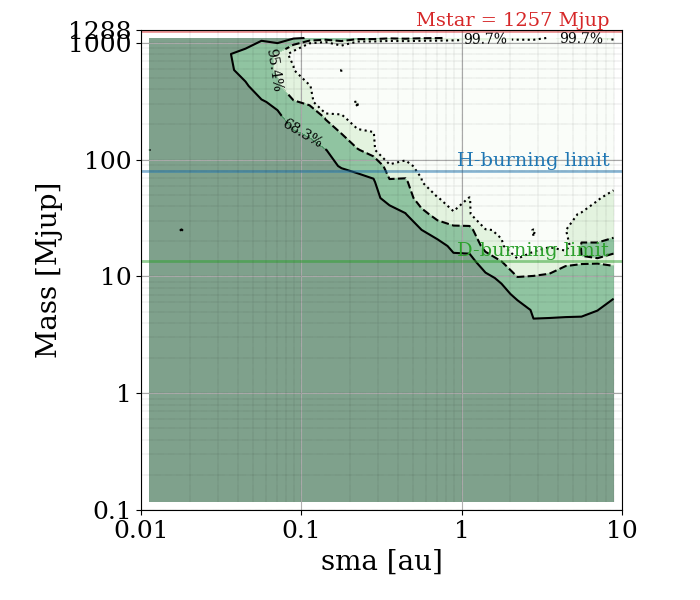}  
\includegraphics[width=40.mm,clip=True]{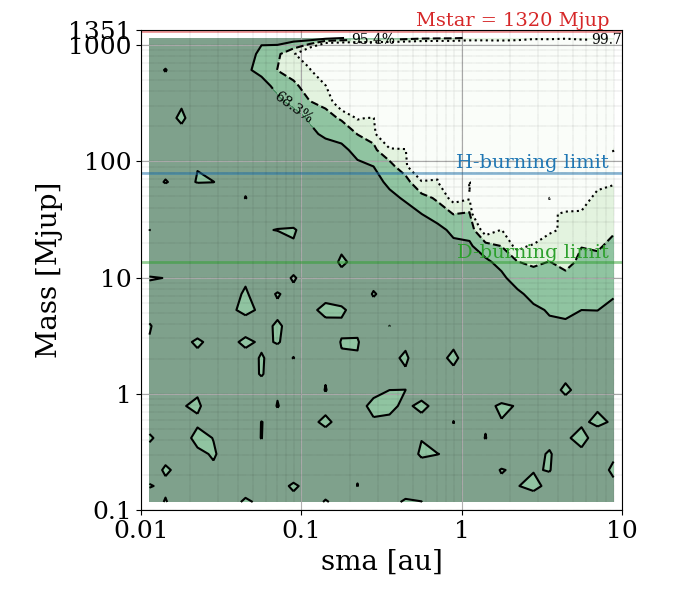}  
\includegraphics[width=40.mm,clip=True]{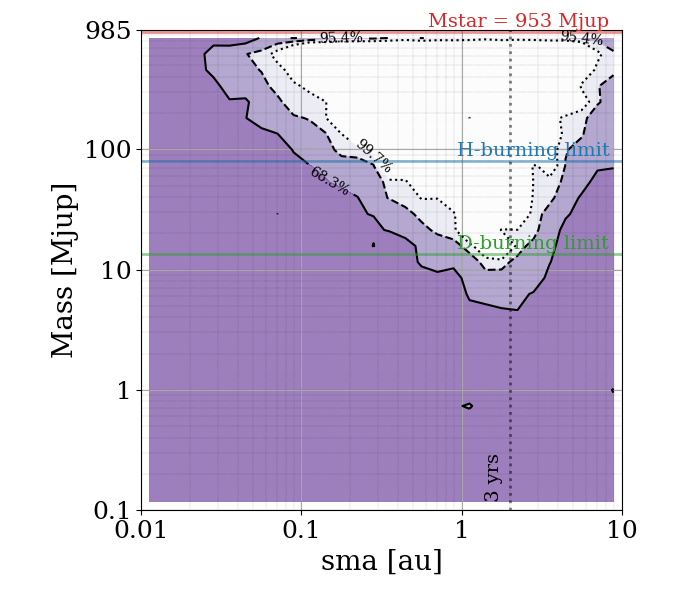}  
\includegraphics[width=40.mm,clip=True]{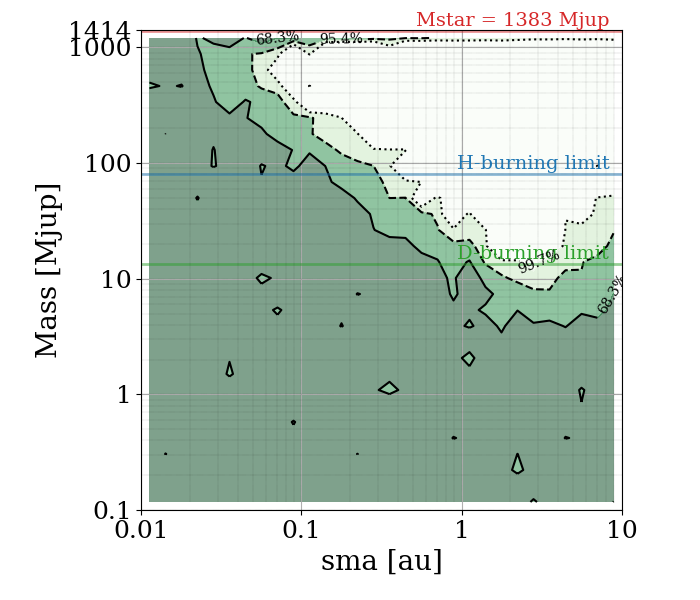}  
\\
  \includegraphics[width=40.mm,clip=True]{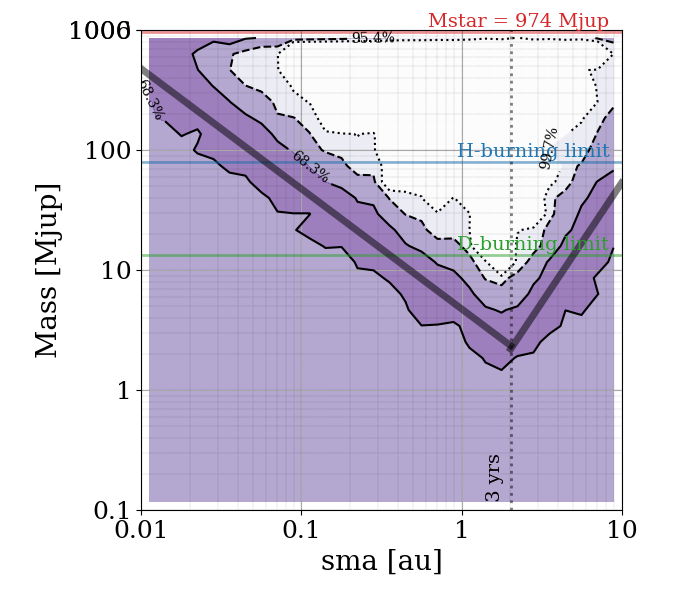}  
\includegraphics[width=40.mm,clip=True]{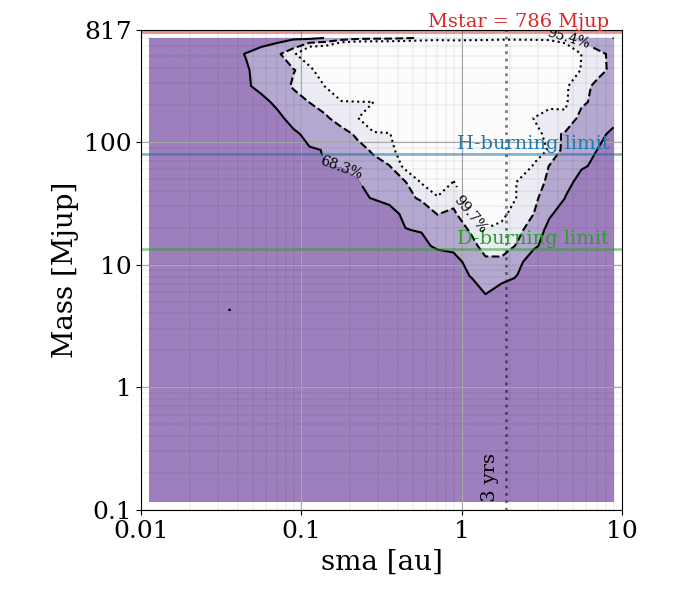}  
\includegraphics[width=40.mm,clip=True]{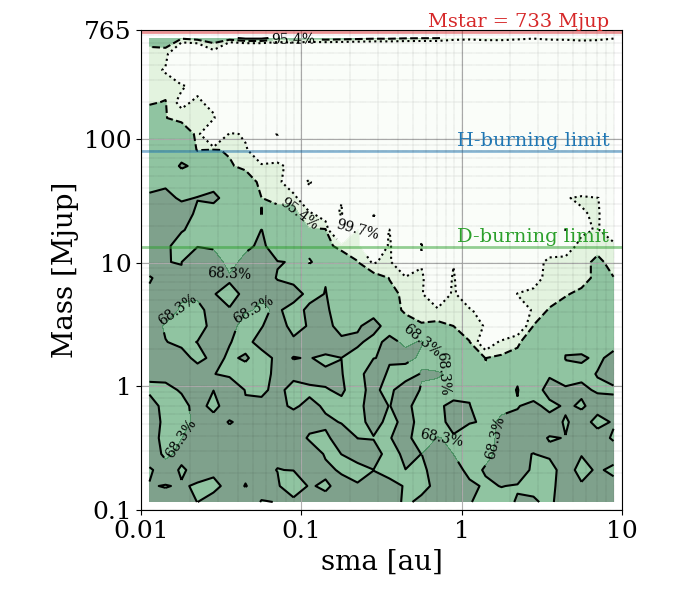}  
    \caption{GaiaPMEX maps for stars listed as SB by ZF21 and not classified as binaries according to  GaiaPMEX results (see text). }
    \label{fig:SBnotdetected}
\end{figure*}

Fourteen of the stars identified by ZF21 as binaries or possible binaries are not found as binaries by GaiaPMEX given our detection criteria using \ruwe{} or PMa. One of them, HD 217379A, reported as an SB3 by ZF21 on the basis of ~\citet{Elliott14} results, and characterized by ~\citet{Tokovinin16a} is just below the  $2$-$\sigma_{\rm \texttt{RUWE}}$ detection limit and just above the $2$-$\sigma_{\rm AEN}$ one. 
 We discuss these thirteen stars below, and show in Fig. \ref{fig:SBnotdetected} their GaiaPMEX maps, when available.

\begin{itemize}
   \item HD 86356 (THA) is reported as an SB2 by ZF21. It was classified as a THA member by ~\citet{Torres06} but was later classified as a PMS member of Chameleon ~\citep{Lopez-Marti13}, and then newly rediscovered as a THA member ~\citep{Elliott16}. An SB2? status was tentatively proposed by ~\citet{James06} on the basis of CCF analysis, but according to these authors, new data were needed for confirmation. Yet, ~\citet{James16} did not confirm its SB2 status. No new data were available to ZF21. The Frac is 0. GaiaPMEX excludes all stellar companions with sma in the range of 0.2 to 4 au, and all BDs in with sma in the range of 1 to 2 au. Depending on their masses, stellar or BD companions can also be excluded outside these boundaries (see Fig. \ref{fig:SBnotdetected}).
 We conclude that the binary status of this star is not assessed. 

   \item CD -42 3328 (THA) is reported as an SB2 by ZF21, but no RV could be found in their paper or in the literature to assess this classification. The Frac is 0. 
 GaiaPMEX excludes all stellar companions with sma in the range 0.2 to 3 au, and all BDs in with sma in the range 1 to 3 au. Depending on their masses, stellar or BD companions can also be excluded outside these boundaries  (see Fig. \ref{fig:SBnotdetected}).  We conclude that the binary status of this star is not assessed.

   \item HD 13183 (THA)  is mentioned as a SB1 by ZF21 but not reported in their Table 3. This target was flagged as an SB1 by ~\citet{Cutispoto02} on the basis of RV variations. Yet, ZF21 do not confirm the RV variations but find an asymmetrical CCF profile.  Note that it is classified as a WDS with a large separation (700+ as), but the wide companion cannot explain a SB1 status. The Frac is 0. GaiaPMEX excludes all stellar companions with sma in the range 0.1 to more than 10 au, and all BDs  in with sma in the range 1 to 10 au.  Depending on their masses, stellar or BD companions can also be excluded outside these boundaries  (see Fig. \ref{fig:SBnotdetected}). Notably, no companion around this star is reported in \citet{Gratton24}. We conclude that the binary status of this star is not assessed.

   \item HD 22213 (THA)  is flagged as SB1 by ZF21 on the basis of 2 UVES RV at 8 and 14 km/s, while the target is a relatively fast rotator (\vsini~ about 41 km/s). A visual binary (0.9 and 0.5 \Msun) with a projected separation of 1.7" ($ \simeq $ 85 au) was reported by  ~\citet{Hagelberg20}. The Frac is low. GaiaPMEX is not sensitive to this visual binary.  GaiaPMEX excludes all stellar companions with sma in the range 0.1 to 4 au, and all BDs  in with sma in the range 1 to 2 au. Depending on their masses, stellar or BD companions can also be excluded outside these boundaries (see Fig. \ref{fig:SBnotdetected}). 

   \item HD 32195 (THA) is classified as an SB? by ZF21, but it has a high \vsini, more than 45 km/s. UVES data reveal variations with 1 km/s amplitude. The Frac is 0. This star was reported as a PMa binary by ~\citet{Kervella19}, but GaiaPMEX does not confirm this status. GaiaPMEX excludes all stellar companions with sma in the range 0.3 to  more than 10 au, and all BDs  in with sma in the range 2 to 10 au. Depending on their masses, stellar or BD companions can also be excluded outside these boundaries  (see Fig. \ref{fig:SBnotdetected}). No companion around this star is reported in \citet{Gratton24}. We conclude that the binary status of this star is not assessed.

   \item HD 17250 (THA) is classified as a SB by ZF21. Three UVES data separated by 5 and 14 days show variations of 0.3 and 0.5 km/s but it has a high vsin(i) of 42 km/s. SOPHIE data also show variations on days timescales which can be compatible either with a companion or with pulsations. It is classified as a less than 2" (120 au) binary in the WDS catalogue.   ~\citet{Tokovinin16b} report a SB for this star member of a quadruple system but without references. Frac is 0. GaiaPMEX excludes all stellar companions with sma in the range of 0.3 to  more than 10 au, and all BDs  in with sma in the range of 2 to 4 au.  Additional stellar or BD companions can be excluded from these boundaries, depending on their masses (see Fig.\ref{fig:SBnotdetected}). Additional data are needed to assess the SB status.

   \item HD 36329 (COL) is classified by Torres et al. (2006) as an SB2. Frac is 0. GaiaPMEX excludes all stellar companions with sma in the range 0.3 to more than 10 au, and all BDs  in with sma in the range 2 to  4 au. Depending on their masses, stellar or BD companions can also be excluded outside these boundaries (see Fig. \ref{fig:SBnotdetected}). It is classified as a non single star by Gaia.

        \item HD 161460 (BPC) is reported as a  SB2 \citep{Torres06}. HD 161460 is a SB2 (K0IV + K1IV), observed only twice, one near conjunction and in the other spectrum the lines are not very well resolved, but both Li lines are strong. An almost equal mass ($q \approx 0.9$) stellar companion at $\sim 8$ au is reported in \citet{Gratton24}. GaiaPMEX cannot find a solution because the Gaia DR3 archive only reports a 2-parameters fit (\verb+astrometric_params_solved+=3). Frac is 35, indicating a possible pair.  The absence of a 5-parameters fit solution in the DR3 leads to suspect a possibly unidentified issue with the 5-parameters fit previously performed in the DR2.

   \item HD 191089 is classified as SB1? by ZF21. Yet, our HARPS data recorded over more than 1000 days do not reveal any hint of binarity and show rather pulsations ~\citep{Grandjean20}. It is classified as a visual binary in the WDS catalog, but all companions are unbound. Frac is 0. GaiaPMEX excludes all stellar companions with sma in the range 0.3 to more than 10 au, and all BDs  in with sma in the range 1.5 to 6 au. Depending on their masses, stellar or BD companions can also be excluded outside these boundaries (see Fig.\ref{fig:SBnotdetected}). Furthermore, no companion is reported in the analysis by \citet{Gratton24}. 
 
       \item CD -27 11535 is classified as an SB1 by ZF21 and was recently found as a member of a very close resolved system by ~\citet{Thomas23}. The latter, however, cannot reconcile the orbital fit results and the luminosity constraints based on the Gaia DR3 data, maybe due to a discrepant parallax measurement from Gaia attributed by the authors to variability, or to the presence of an additional unresolved companion to one of the two components. No companion around this star is reported in \citet{Gratton24}. GaiaPMEX cannot find a solution because the Gaia DR3 archive only reports a 2-parameters fit (\verb+astrometric_params_solved+=3). Frac is 19, indicating possibly a possible pair. The absence of a 5-parameters fit solution in the DR3 leads to suspect a possibly unidentified issue with the 5-parameters fit previously performed in the DR2. It is thus likely that the DR2 reported an anomalous parallax.  

   \item V* V1005 Ori (BPC) is an M-type star classified as a SB by ZF21, from the publication of ~\citet{Elliott14} based on "literature RV". However, ZF21 do not confirm  RV variations with their data. Our HARPS data, spread over 3000 days, do reveal RV variations (amplitude 300 m/s), but also significant (100 m/s) bissector variations indicative of  stellar activity, well correlated with the RV variations, indicative of stellar activity. They do not show direct indication of a stellar binary.  V* V1005 Ori is classified as a very wide visual binary, a fact that cannot account for an SB status. Frac is 0. GaiaPMEX excludes all stellar companions with sma in the range 0.03 to more than 10 au, and all BDs in with sma in the range 0.2  to more than 10 au. Depending on their masses, stellar or BD companions can also be excluded outside these boundaries (see Fig. \ref{fig:SBnotdetected}). Such limits leave very little room for a close stellar companion, except at very short separations, but there are today no evidence for such a companion in the available RV data.

   \item V*V379 Vel (ARG) is classified as a SB? by ZF21 on the basis of two epochs available with two instruments, and a variation of 3.2 km/s. Its \vsini~ is about 14 km/s. Frac is 0. GaiaPMEX excludes most stellar companions with sma between 0.1 and 8 au, and most BD companions with sma between 0.5 and 3 au as well. GaiaPMEX excludes all stellar companions with sma in the range 0.3 to 4 au, and all BD in with sma in the range 1.5 to 2 au. Depending on their masses, stellar or BD companions can also be excluded outside these boundaries (see Fig. \ref{fig:SBnotdetected}). Additional data are needed to assess the SB status. 

        \item V*PX Vir (AB Dor) is classified as a SB1 by ZF21, and mentioned as a single line binary characterized by \citet{Griffin10} with a period of 216 d. Recently, both components were characterized using RV, absolute and relative astrometry, revealing a mass ratio of about 0.5 \citep{Wang23-PXVir}. Frac is 3. GaiaPMEX cannot find a solution because the Gaia DR3 archive only reports a 2-parameters fit (\verb+astrometric_params_solved+=3).
   
\end{itemize}

In summary, among the 14 SB reported by ZF21 in BPC, THA, OCT, ARG, COL and ABDor which  are not found by GaiaPMEX as binaries based on the \ruwe{}  or PMa criteria, eight need to be confirmed as spectroscopic binaries. In three cases, GaiaPMEX failed at finding solutions, but two have high Frac, possible indicating a pair, and one has a Frac of 3. One is detected using the AEN excess. For the remaining cases, the non-detection by Gaia indicates that the companions must orbit closer than 0.1-0.3 au.

Globally, we conclude that GaiaPMEX is very efficient at identifying  stellar binaries for non-accreting stars members of young and close associations in the range of about 0.1 to 10-20 au. Moreover, it is yet only partly sensitive to very close binaries (sma less than typically 0.1 au), which consequently represents the sweet spot region for spectroscopic detection of binaries in the Gaia context.  For separations greater than 10-20 au, high contrast imaging is best suited to the detection of binaries down to BD masses for these young systems. This demonstrates the complementarity of the various detection methods.



\subsection{Binary rates }\label{sec:binary_rates}
Determining  accurate binary rates is beyond the scope of this paper, because the number of targets considered for each association is small and the sample is probably biased due to selection effects (stars suitable for RV measurements), and because, as seen above,  there are for a number of sources doubts about their  classifications among the different associations. Furthermore, computing  binary rates corrected from completeness effects requires taking several assumptions on the distribution of the $q$-ratio, from very low values (planets) to 1 (equal-mass binaries) and on the radial distribution of the binaries that are not justified (these are information that we would actually like to know). We note that such assumptions were taken by ZF21 to estimate their corrected binary rates and because of the associated uncertainties, they caution about the obtained corrected rates. 
 
Nonetheless, we provide hereafter very tentative estimates of {\it observed } binary rates  
for BPC ($\simeq $ 24 Myr, see Table 1), COL ($\simeq $ 42 Myr), THA ($\simeq $ 45 Myr), and AB Dor (120-200 Myr), which have more than 40 targets in the ZF21 sample. Using the numbers reported in Table~\ref{tab:stats}, we compute the binary rates as Ndet/Ntotal, where Ndet is the number of binaries detected with Gaia (either from the \ruwe{}  and/or PMa). We do not correct Ndet from any false positive (FP) effect related to the detections with 
\nsigmaruwe{} or \nsigmapma{} between 2 and 3, because they are very few (see Table~\ref{tab:stats}), because among them, several are known as binaries from other technics, and also because 
in a second step, we also consider the spectroscopic detections, for which we do not know the FP. The uncertainties were computed as  sqrt(Ndet)/Ntotal. We obtained rates of astrometric binaries of 43 $\pm $ 10, 28 $\pm $ 8, 29 $\pm $ 6 and 37 $\pm $ 9 $\% $ for  BPC, COL, THA, and AB Dor, respectively. 
 
In a second step, we added the number of astrometric binaries found with GaiaPMEX and the number of spectroscopic binaries (SB or SB?) that were not already identified as astrometric binaries. We assume thus that when an SB is reported and a GaiaPMEX companion is found, they correspond to the same companion (which may actually not be always the case, and may reduce the actual number of binaries). We obtained rates of 52 $\pm $ 11, 30 $\pm $ 8, 37 $\pm $ 6 and 41 $\pm $ 9 $\% $. We see that in most cases, the rates are of about 30-40 $\% $, except for BPC, for which the rate is higher. We note, however, that the BPC  hosts three planetary mass companions, that are included in the estimation, and contribute to the estimated rate. 

Finally, considering the BANYAN classification leads to  rates of 
41 $\pm $ 10, 37 $\pm $ 10, 28 $\pm $ 8, and 43  $\pm $ 9 $\% $ 
for  BPC, COL, THA, and AB Dor, respectively.
When adding the spectrosopic binaries, we get rates of 51 $\pm $ 10, 40 $\pm $ 10, 39   $\pm $8, 43  $\pm $ 9 $\% $. This highlights the importance of securing the members of these associations.

\section{Vetting planetary candidate companions }
\label{sec:vetting_cc}

\subsection{Planetary candidate companions derived from \ruwe{} and from combining \ruwe{} and PMa}
\begin{figure*}[hbt]
    \centering
    \includegraphics[width=40.mm,clip=True]{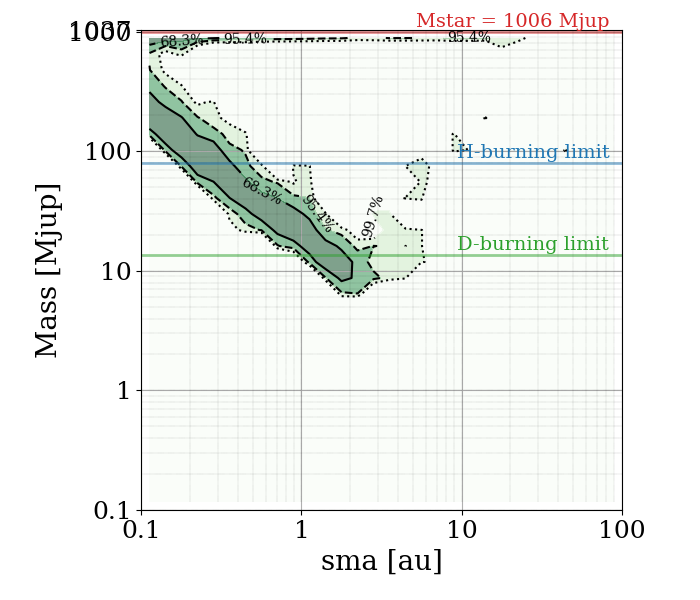} 
    \includegraphics[width=40.mm,clip=True]{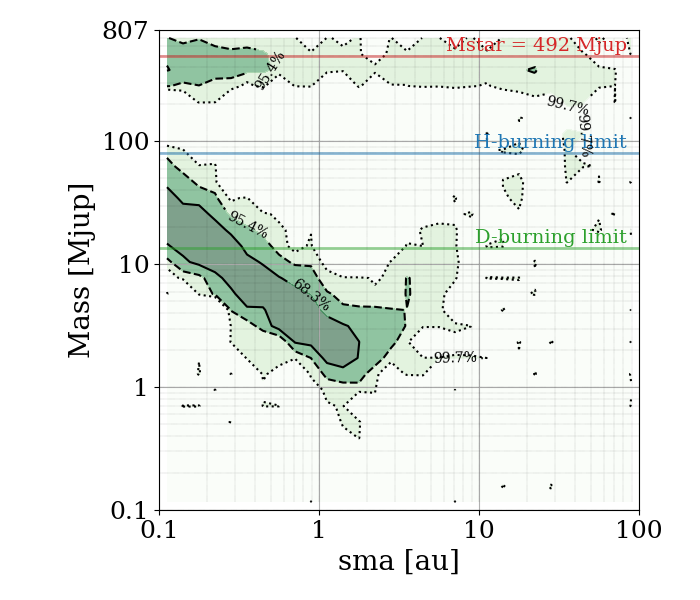} 
    \includegraphics[width=40.mm,clip=True]{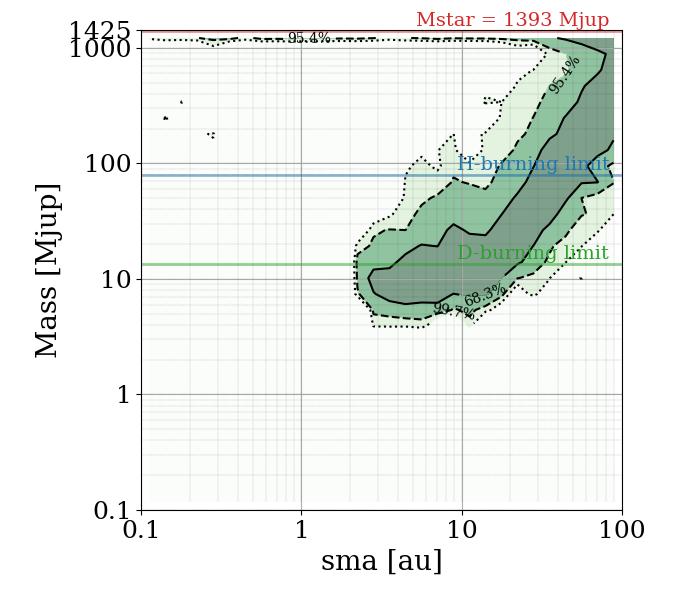} 
    \includegraphics[width=40.mm,clip=True]{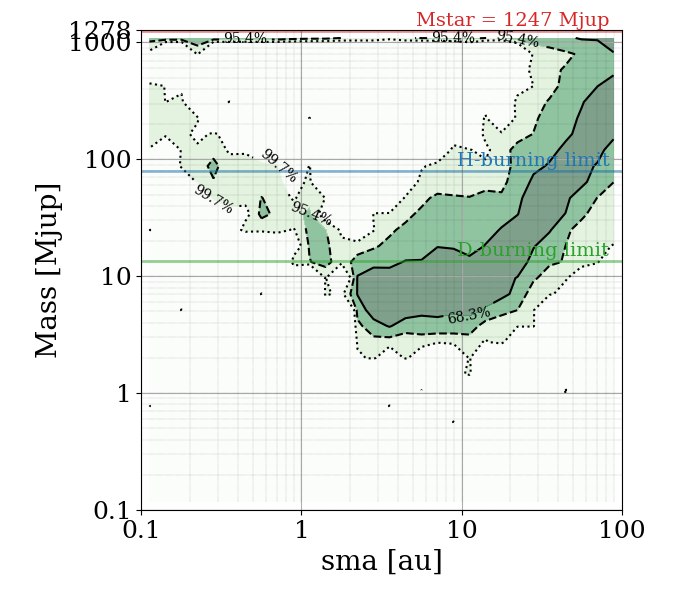} 
\includegraphics[width=40.mm,clip=True]{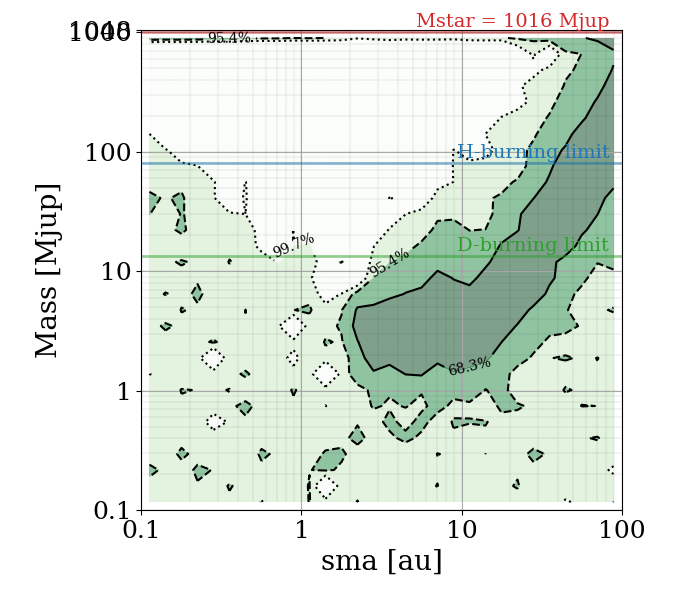} 
\includegraphics[width=40.mm,clip=True]{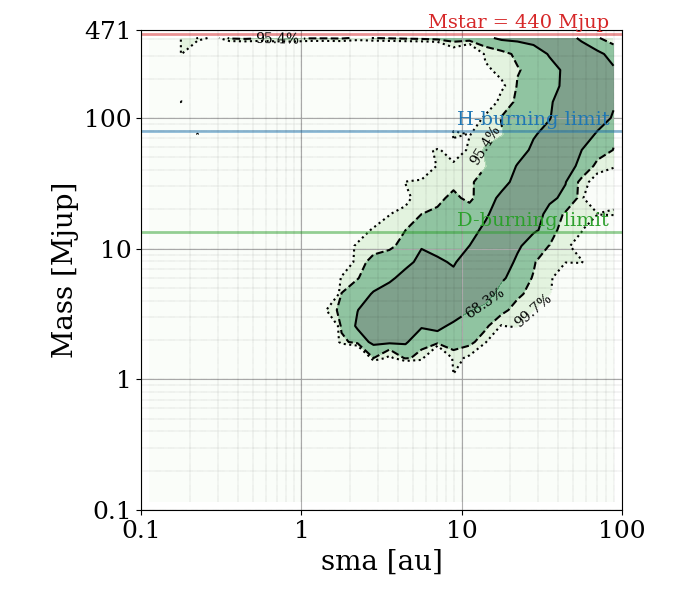} 
\includegraphics[width=40.mm,clip=True]{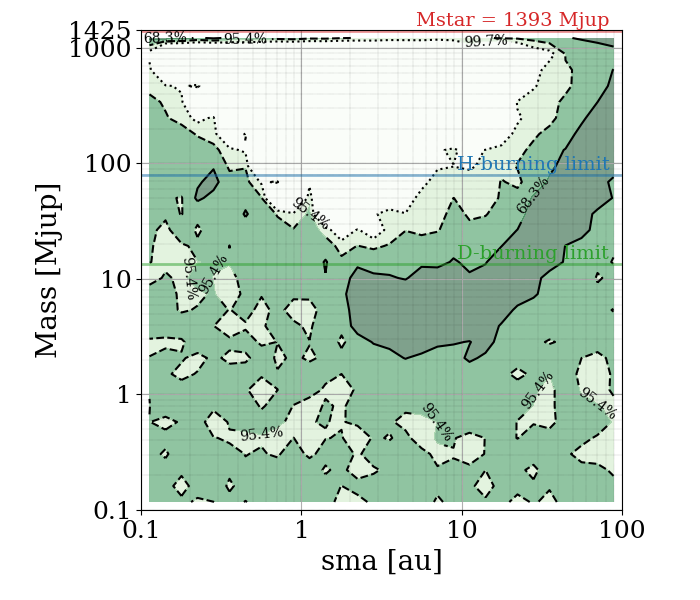} 
\includegraphics[width=40.mm,clip=True]{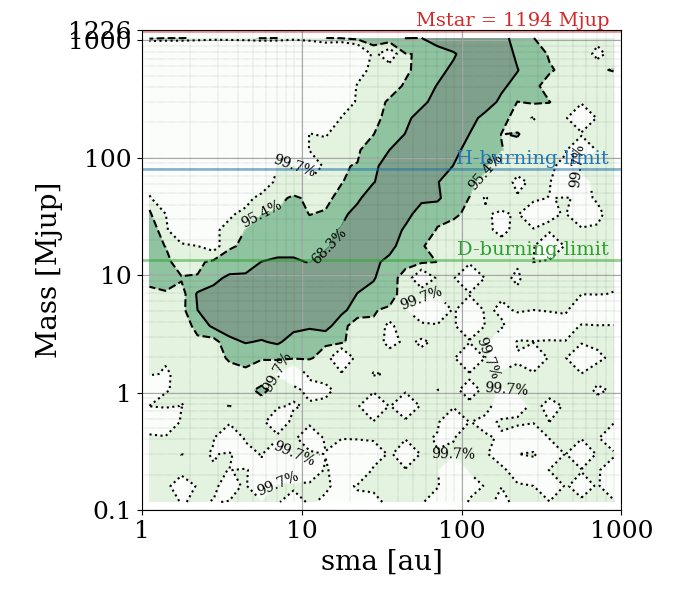} 
\includegraphics[width=40.mm,clip=True]{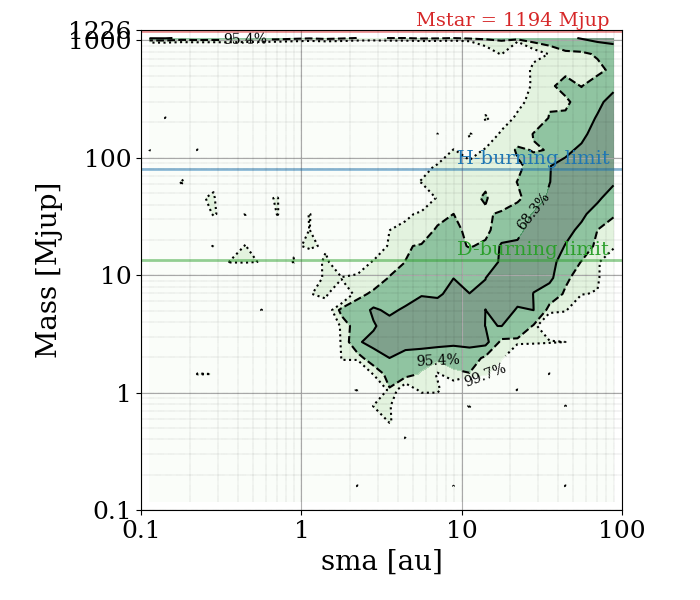} 
\includegraphics[width=40.mm,clip=True]{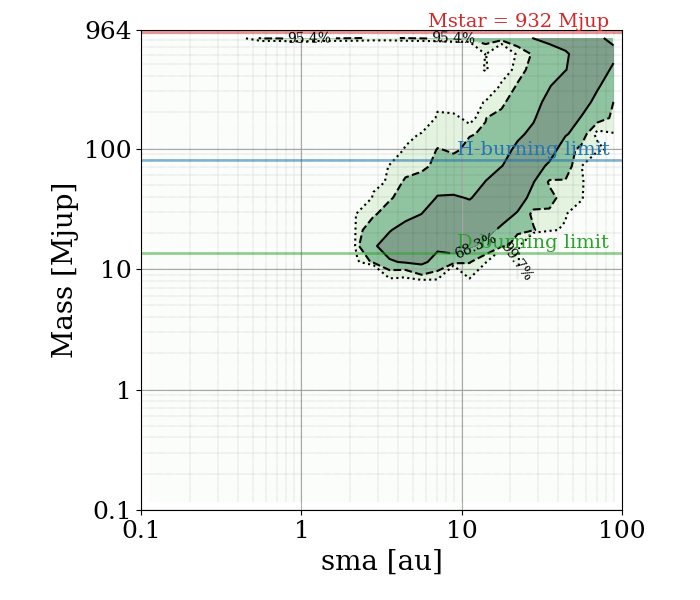} 
\includegraphics[width=40.mm,clip=True]{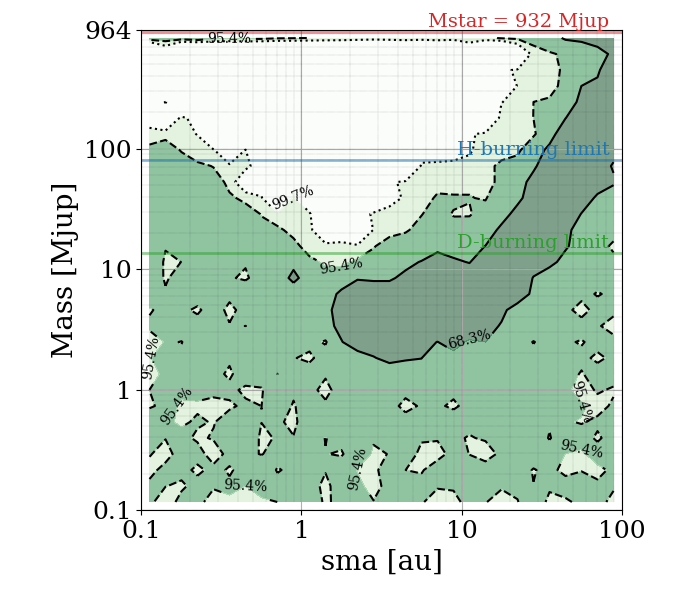} 
\includegraphics[width=40.mm,clip=True]{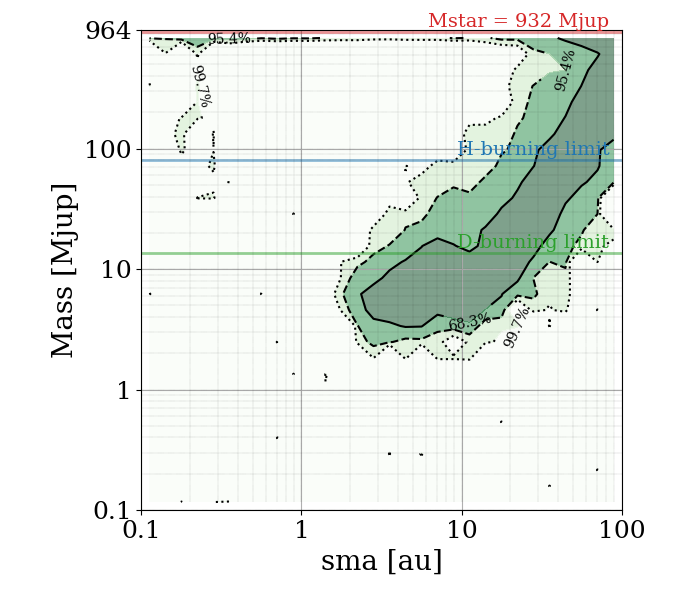} 
\includegraphics[width=40.mm,clip=True]{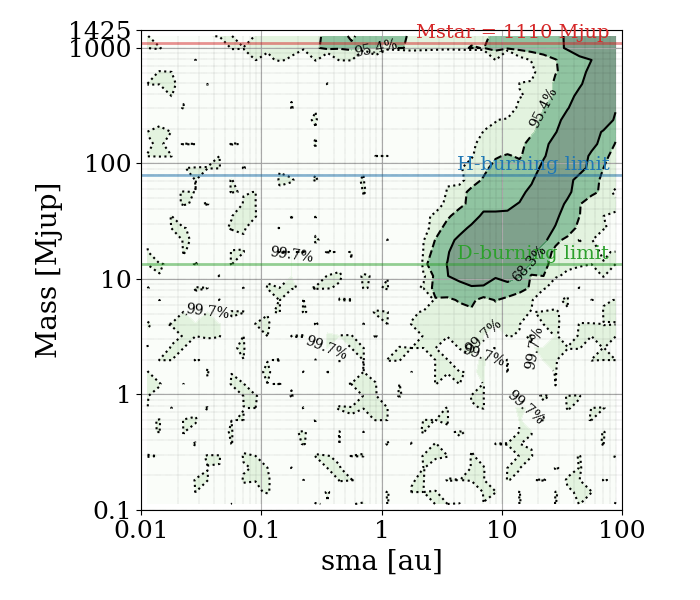} 

    \caption{Targets having companions with masses possibly in the  planetary regimes (see text). From top to bottom and left to right: HD3221, G80-21, HD40216, HD 178085, HD 222259, CD-561032, HD 53842,   PZ Tel, AF Lep, HD 41071, HD 23208, AB Pic, and  HD 14082B.}
    \label{fig:SBdowntoBDP}
\end{figure*}

Among the  binaries identified in all associations but TWA and ECH, GaiaPMEX, based on either \ruwe{} or PMa astrometric signatures, as well as combining \ruwe{} + PMa, identified 13 targets with companions with masses possibly in the planetary regime. Two companions have sma less than $\simeq $ 1 au, and the others have sma greater than $\simeq $ 1 au. We show their GaiaPMEX solutions in Figure \ref{fig:SBdowntoBDP}, and describe them below.

\begin{itemize}
    \item HD 3221 (HIP 2729, THA SACY, THA BAN; ? in ZF21) is a K4 star, reported both as a  binary with a proj. sep. of 20 mas from high contrast imaging \citep{Bonavita22}, and an RV binary \citep{Grandjean20}. The 20 mas separation is consistent with the  Frac of 0 and the very low GoF. 
    All GaiaPMEX solutions have sma less than 1 au (short period (SP) solutions). The solutions are degenerate with  L-shaped solutions. Only the equal mass binary solutions are compatible with the observed projected separation (1 au), and the flux ratio of about 1 reported by \citep{Bonavita22}.

 \item HD 40216 (HIP 28036, COL SACY, COL BAN; ? in ZF21) was classified as a binary by \citep{Bonavita22} based on data from the Spectro-Polarimetric High-contrast Exoplanet Research (SPHERE) instrument \citep{beuzit19}. It was previously detected by \citet{Galicher16} and \citet{Song03}. The mass 0.12 \Msun~and projected separation (110 au) are compatible  with the  low Frac  and GoF, and with the GaiaPMEX solutions.
 
\item HD 178085 (HIP 94235, AB Dor SACY, AB Dor BAN; ? in ZF21) was  classified as binary thanks to NACO data \citep{Chauvin15}. Interestingly, a short period (7.7 d) mini Neptune was recently reported around this star by \citet{Zhou23}, who also further constrained the properties of the stellar companion. With an sma of about 56 au (about 0.9") and mass of about 0.26 \Msun, the companion may explain the observed Frac (68) and almost zero GoF. They are also compatible with
characteristics are compatible with the GaiaPMEX solutions.

\item HD 222259 (HIP 116748, DS Tuc, THA SACY, THA BAN; ? in ZF21) has a 8.1 d period giant planet detected by TESS  \citep{Newton19} and a binary companion at 5" i.e. more than 100 au \citep{Torres06, Pierce20}. The sma and mass \cite[about 175 au (3.8") and 0.8 \Msun,][]{Pierce20} of the  binary companion, are consistent with  the low Frac  and GoF, since it must be outside Gaia's window aperture. They are also compatible with the GaiaPMEX solutions.

\item CD -561032 (HIP 22738, AB Dor SACY, AB Dor BAN; SB1 in ZF21) is classified as a SB in SIMBAD. It is member of a binary system made of two M-type stars with q=0.7 and separated by about 90 au (8")  \citep{Tokovinin16-MNRAS,Winters19}. These values are consistent with  the low  Frac  and GoF, as the binary is resolved by Gaia. They are also compatible with the GaiaPMEX solutions, even though these results should be taken with care given the angular separation and low constrast between both components (see Section 2).

\item HD 53842 (HIP 32435, THA SACY, THA BAN; ? in ZF21) is reported as a member of a binary system, with a secondary M-dwarf at a projected separation of 82 au (1.3") by \citep{Dahlqvist22}. These values are marginally compatible with the Frac value (6), given the binary separation and the window size. 
The GaiaPMEX solutions are in agreement with the properties of the companion.

\item HD 41071 (HIP 28474, RT Pic, COL SACY, COL BAN; ? in ZF21) is an eclipsing binary \citep{Malkov06} with a not yet determined period. It has a known low mass stellar companion \citep{Chauvin10,Bonavita22, Tokovinin20} with a 0.7" projected separation and a 4.6 mag contrast \citep{Tokovinin20}. Such values are consistent with  the low  Frac and GoF, and compatible with the GaiaPMEX solutions.


\item V* PZ Tel (HIP 92680, BPC SACY, BPC BAN; SB? in ZF21) is known to host a BD which is well characterized through direct imaging and RV data. The most recent estimates using also the star PMa \citep{Franson23_pztel} of its mass (27 \Mjup) and sma (27 au, 0.6") fit well with the GaiaPMEX solutions. Note that the high contrast in magnitude  explains the low Frac and GoF despite a separation of 0.6". 

\item V* AF Lep (HIP 25486, BPC SACY, BPC BAN; ? in ZF21) has been recently shown to host a planet whose properties \citep{Mesa2023} are compatible with Gaia data. Note that the high contrast in magnitude explains the low Frac and GoF despite a separation of 0.4".

\item HD 23208 (HIP 17338, OCT SACY, OCT BAN; ? in ZF21) is a close (DR3 parallax = 17.6 mas) G8V type star.  It was identified to have a PMa by \citet{Kervella19} and \citet{Brandt21}. The low Frac and GoF do not suggest the presence of a close pair, in agreement with the GaiaPMEX results. 
Palomar observations rule out stellar companion beyond about 1" (56 au) \citep{Metchev09}. This leaves us with solutions most probably in the BD or planet domain. We further characterize the companion in the next section.  We note that the smallest sma, down to 1-2 au, would require higher spatial resolution instruments on the ELT, while planetary mass companions at 10 au could be detected with VLT high contrast instruments. 

\item G80-21 (HIP 67659, AB Dor SACY, AB Dor BAN;  ? in ZF21), HD 44627 (aka AB Pic or HIP 30034,  COL SACY, CAR BANYAN, and HD 14082 B (HIP 10679, BPC SACY, BPC BAN) are found to have light companions. we discuss and characterize them  in a dedicated section below.

\end{itemize}

In summary, among the 13 candidates for planetary-mass companions, 7 are in fact stellar companions, one is a well known and characterized BD, one is a well known and characterized planet. Another one, HD 23208 is known in this paper to have a sub-stellar companion orbiting below typically 60 au, which requires additional data for further characterization. The last three targets, G80-21, AB Pic and HD14082B are shown to have sub-stellar companions that we characterize and discuss in   Section~\ref{sec:companions} thanks to high contrast imaging and RV data in addition to Gaia and PMa data. 

\subsection{Planetary candidate companions derived from \ruwe{}  analysis only}
GaiaPMEX, based on the \ruwe{}  astrometric signature only, finds 19 companions with  masses ranging from the stellar regime to the planetary regime in these associations. We wish to vet, whenever possible, between stellar companions and planetary mass companions, using available data. 
\begin{itemize}
 \item HD 217379A (AB Dor, SACY), BD +30397B (BPC, SACY), BD -211074B (BPC, SACY), HD 139084B (BPC, SACY), HD 164249 B (BPC, SACY) are members of known close bound binary systems. Hence, no planets are needed to explain the Gaia results.

  \item 2MASS J10010873-7913074 (THA SACY, THA BAN)  is also classified as a T Tauri, but is classified as a THA member by BANYAN. Its status needs further clarification. We note low values of Frac and GoF.
  \item HD 319139 (BPC SACY, BPC BAN) is reported as a binary in the WDS and as a SB2 by ZF21, and as a member of a K7+K2 system in SIMBAD. This can explain the Gaia results.   
    \item TYC 8083-45-5 (THA SACY, THA BAN) is reported as a visual binary (sep. 1"; delta K = 1.2 in the near IR) by the WDS. The high Frac  (50) and low GoF is consistent with a binary system. The Gaia data are compatible with a bound binary provided the orbit of the secondary is elliptical, but as seen in section 2, the GaiaPMEX results have to be taken with care when the Frac is high.
    \item CD -29 2531 (COL SACY, COL BAN). It is reported as a member of a  visual binary system (0.8") in the WDS. Given the star distance (more than 100 pc), the visual companion, if gravitationally bound to the star, can account for  the Gaia results  only if orbiting on a very eccentric orbit.
         \item 2MASS J08500540-7554380 (THA SACY, LCC BAN) is reported as a T Tauri star in SIMBAD, and classified as a LCC member by BANYAN. If its T Tauri status is confirmed, the Gaia results should be taken with caution. 
\item  BD -13 1328 (AB Dor BAN; ? SACY) is reported as a dubious member of a visual binary system with a 2" separation by the WDS, based on the Tycho Double Star catalog (additional DD list). It is also reported as a member of a bound binary system with a separation of 20". This remote companion cannot be responsible for Gaia results, as the maximum sma of the  solution is less than 30 au.  More observations are needed to further constrain the status of the dubious companion candidate reported in the WDS. Note that the Frac and GoF do not suggest the presence of a close stellar companion. 
    \item  EXO 02352-5216 (COL BAN, ? SACY) is not reported as a binary in the literature. The (sma, mass) solutions go well in the planetary mass regime and the maximum sma is 20 au. It is reported as a SB ? by ZF21. More observations are needed to clarify the SB status of this object. 
      \item Barta 16112 (BPC SACY, ? BAN), LP 776-25 (AB Dor SACY, AB Dor BAN), HW Cet (AB Dor SACY, AB Dor BAN), HD 309851  (ARG SACY, ARG BAN), UCAC361-3820 (THA SACY, THA BAN), CD-292360 (ARG, BANYAN, ? SACY) are not reported as members of binary systems or spectroscopic binaries. HD 309851 is not reported as a binary by \citet{Gratton24}. At this stage, these targets are then promising candidates and deserve dedicated  observations to further constrain the companions responsible for the Gaia signature. Note that all but CD -29 2360 have low Frac and GoF, while CD -29 2360 has a large Frac (64) and a low GoF, suggesting the possible presence of a stellar companion.
      
       \item RXJ 05200+0612 (THOR BAN, ? SACY) is reported to be a member of a bound binary system, with a very remote companion, which   cannot account for the \ruwe{} . This target  is then also a promising candidate, and more observations are needed to further constrain its companion. Note that Frac and GoF are low. 
 \end{itemize}

 In summary, among the 19 stars found as hosting a companion with possible mass in the planetary regime using \ruwe{}  only, eight are known binaries, and two need to be confirmed as non-accreting stars before further interpretation. Additional data are needed to further constrain the companions of the remaining nine stars.

\section{Characterization of the low mass companions around G80-21, AB Pic and HD 14082B}\label{sec:companions}

\subsection{G80-21}\label{sec:G80-21}
\subsubsection{A close sub-stellar companion to G80-21 }
\begin{figure*}[hbt]
    \centering
  \includegraphics[width=60.mm,clip=True]{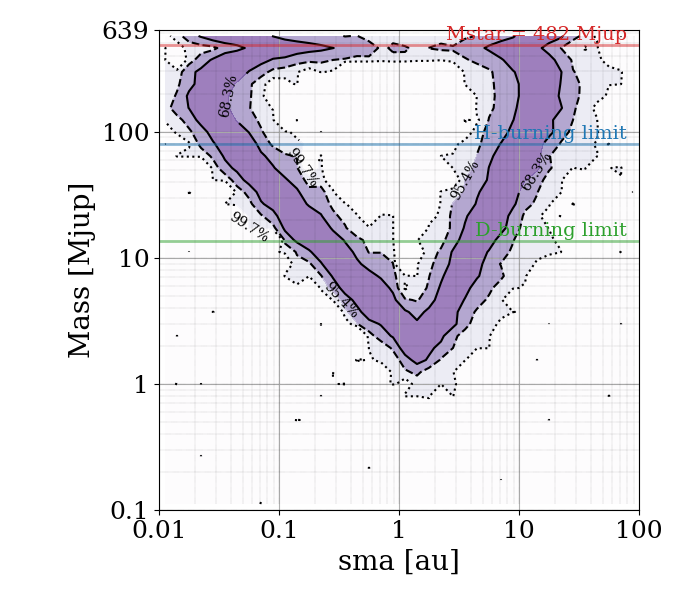}
    \includegraphics[width=60.mm,clip=True]{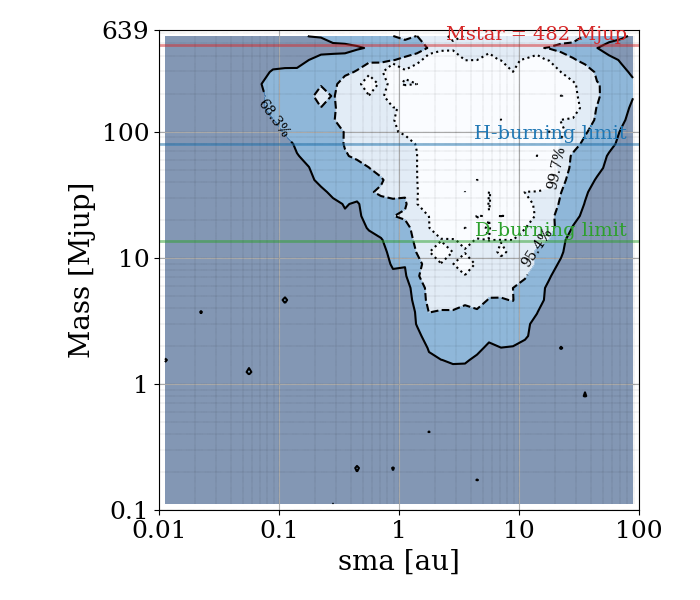} 
    \includegraphics[width=60.mm,clip=True]{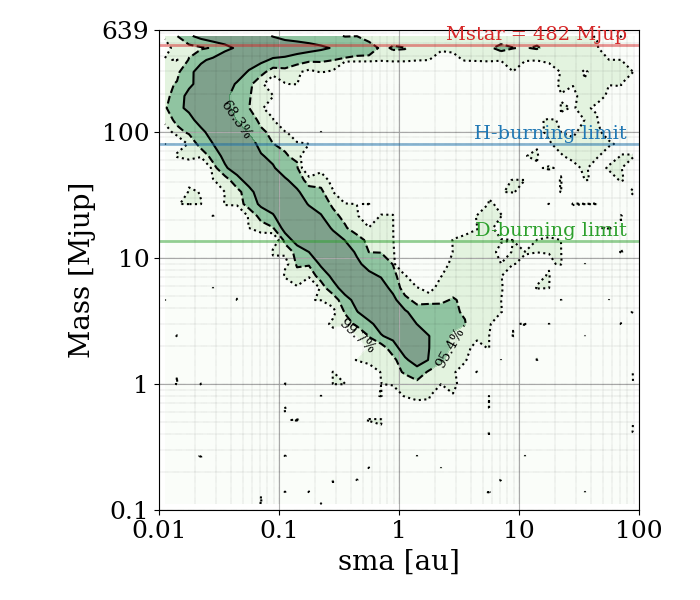} 
    \caption{From Left to Right: Gaia excess, PMa and combined maps for G80-21. }
    \label{fig:HIP17659_GaiaPMEX}
\end{figure*}

G80-21 (HIP 17659, AB Dor SACY, AB Dor BAN;  ? in ZF21) is classified as an eruptive variable, 0.46\Msun, M3-type star.  
Figure~\ref{fig:HIP17659_GaiaPMEX} shows the (sma, mass) solutions compatible with the \ruwe{}  only (Left), the PMa (Middle), and both quantities combined (Right).
 G80-21 shows a significant \aastroruwe{}  with \nsigmaruwe{} = 3.5, and no significant PMa excess at\footnote{Note that it was flagged as a binary by \citet{Kervella2019} on the basis of Gaia DR2 PMa, but  GaiaPMEX does not confirm this classification. \nsigmapma{} = 0.7}, indicative, like in the case of HD 3221, of a close (less than 1-3 au) companion. 
  The low Frac and GoF also suggest either a very close pair, as for HD 3221, or a high contrast companion BD/planetary mass companion if below 1".

ZF21 do not report a SB status for G80-21 on the basis of a few UVES data spread over 100 days. The analysis of higher RV accuracy HARPS data (described in Appendix~\ref{sec:Log_obs}) spread over about 110 days  reveal 0.4 km/s peak to peak variations (compatible with the amplitude of the UVES data). These variations occur over a few days period of time, a timescale compatible with the estimated rotation period of 3.8 days for this young and active \cite[log(R'HK) = -3.98][]{Shan24} M-star. Hence these high amplitude variations are probably linked with stellar activity. Note that, unlike the case of AU Mic, the bisector velocity spans are not fully correlated to the RVs, which might indicate either a complex spots pattern or a very close companion in addition to the spot-induced variations. The TESS light curve clearly shows a periodic signal  with a $\simeq$ 4d period, compatible with two features. Furthermore, we note that the minimum mass of the companion that would be needed to explain the RV variations would not exceed 2 Jupiter masses, and would therefore not account for the absolute astrometry  signal, unless orbiting on a very improbable pole-on orbit. We conclude that most if not all the RV variations are due to stellar activity.

Finally, G80-21 was observed by many teams in high-contrast imaging, but no companion has been found yet  \cite[see e.g.][who got a 7.76 contrast at 0.5'' at L']{Lannier16},  corresponding to roughly 10 \Mjup.

\begin{figure}[hbt]
    \centering
 \includegraphics[width=80.mm,clip=True]{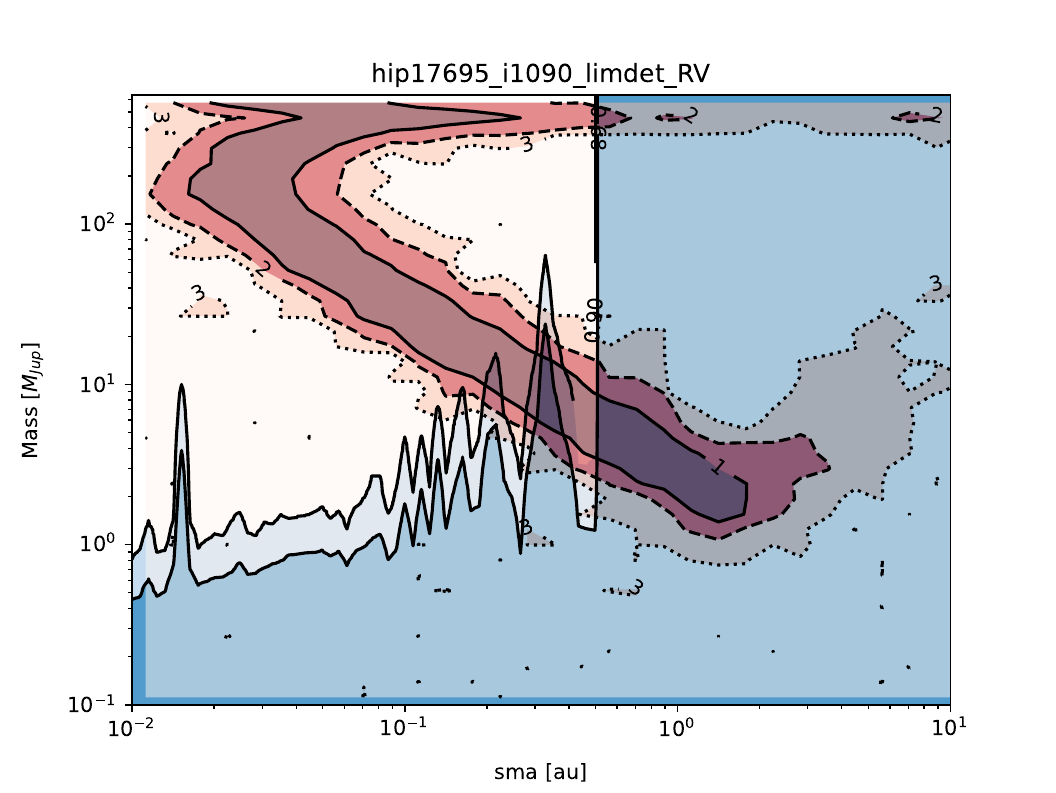}\\ 
   \includegraphics[width=80.mm,clip=True]{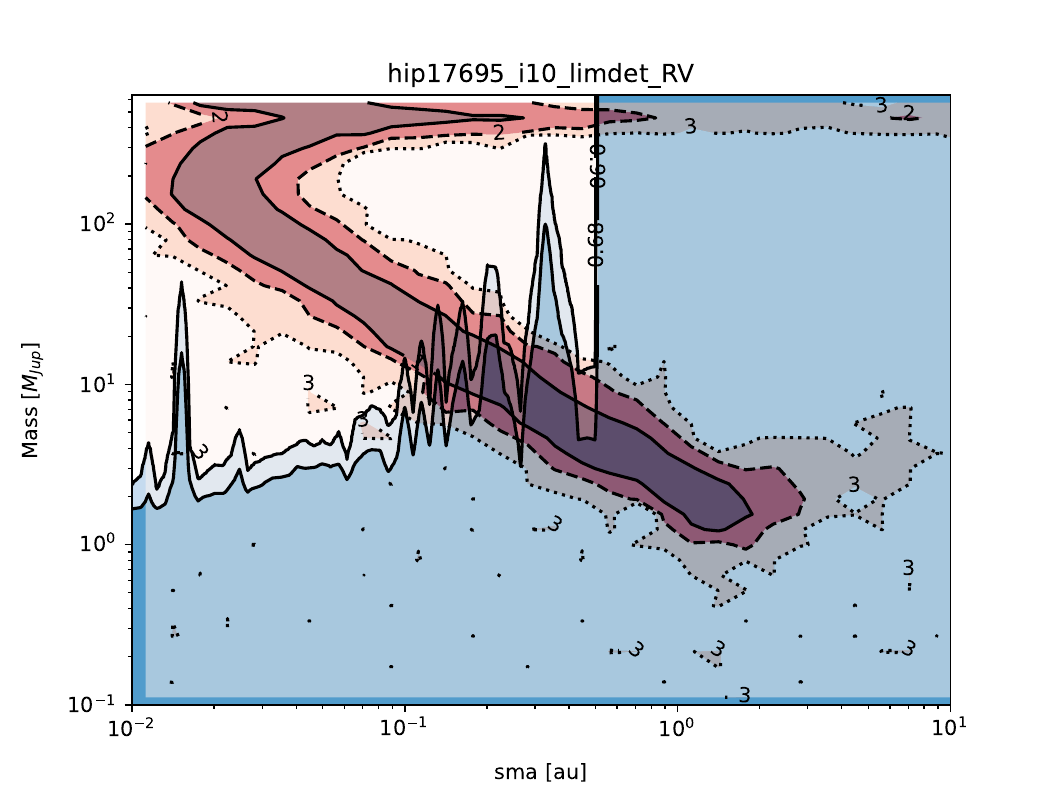}\\ 
  \includegraphics[width=80.mm,clip=True]{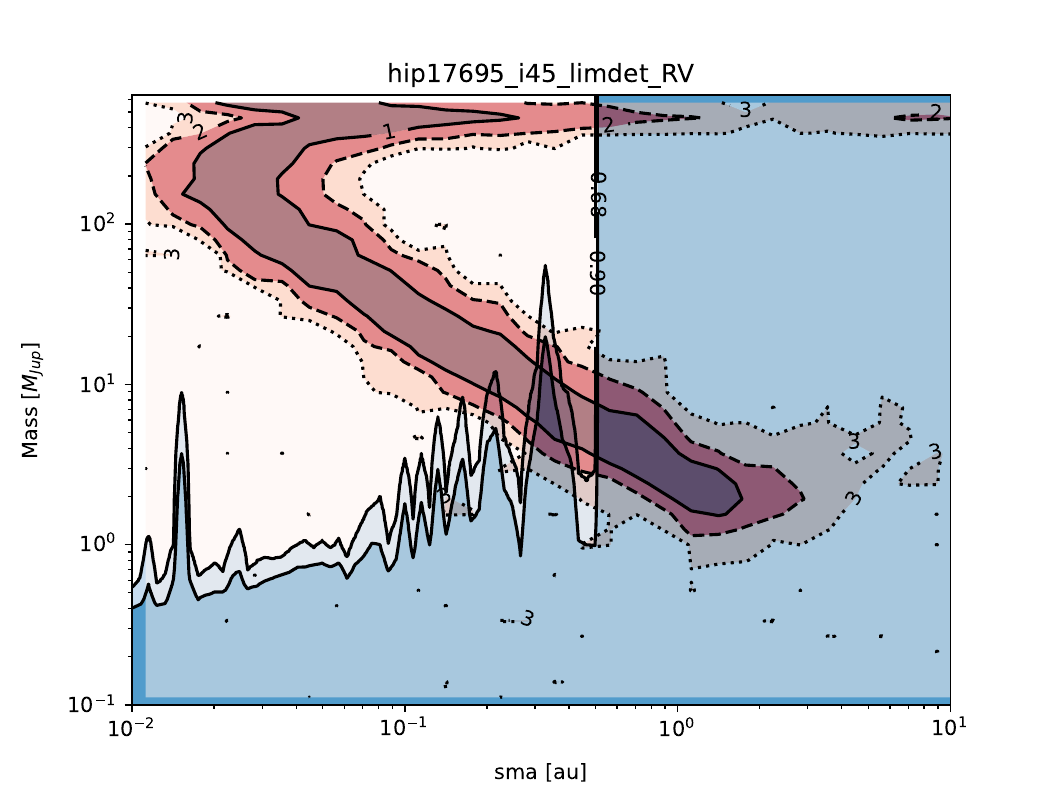}\\ 
  \includegraphics[width=80.mm,clip=True]{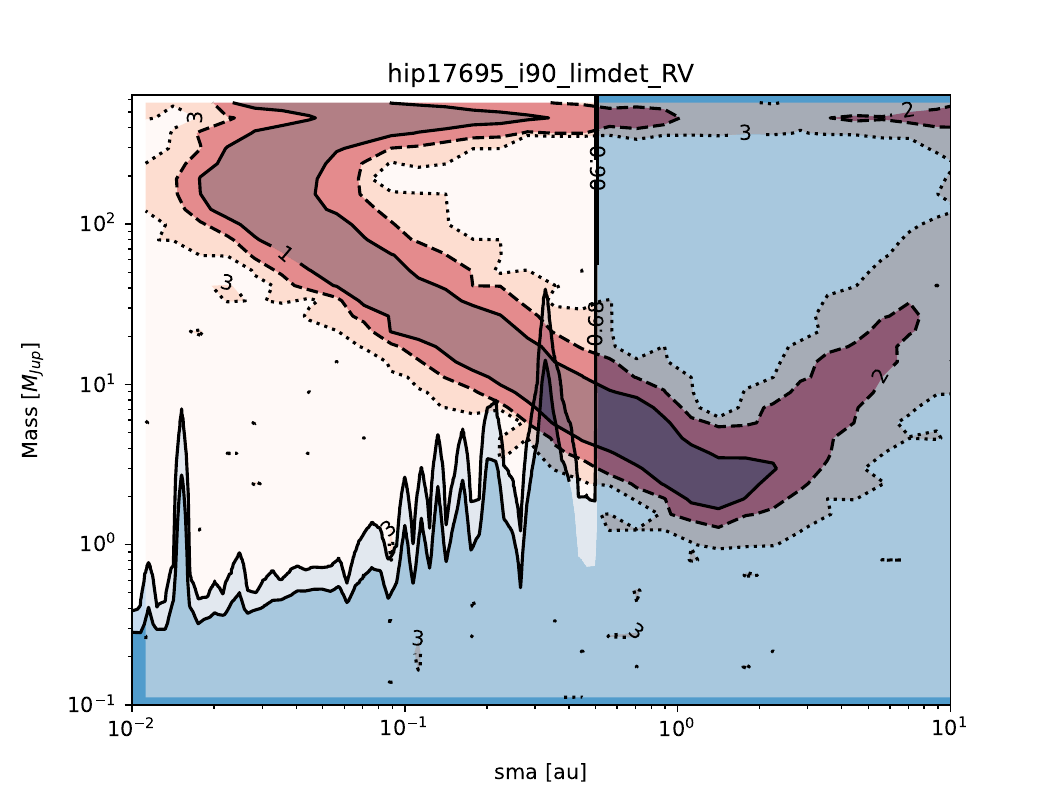}\\ 
    \caption{Possible sma and mass of G80-21 b. Superimposition of  the RV detection limits and the GaiaPMEX (sma,mass) solutions. The blacklines limiting the blue regions correspond to the RV detection limits (90 and 98 $\% $ probabilities). From Top to Bottom: an inclination between 10 and 90 $\deg $ is assumed, then inclinations of 10, 45, and 90 degrees.}
    \label{fig:HIP17659_mass_sma_overlap}
\end{figure}

We then constrained the properties of the  "astrometric" companion using these HARPS RV in addition to the absolute astrometry. We first computed the detection limits set by the RV data and overplotted the GaiaPMEX solutions. To get a global overview, we considered sma up to 10 au, an  eccentricity between 0 and 0.9, and an inclination of the orbital plane between 10 and 90\degree{} (10\degree{} were chosen instead of 0 to avoid any divergence when, at $i$\,\!$\approx$0\degree{}, $\sin i$\,\!$\approx$0). The results are showed in Figure~\ref{fig:HIP17659_mass_sma_overlap}. We see that the sma solutions are mainly in the range 0.5-2 au, and the mass is lower than 8-10 \Mjup. A peak is present at 0.3 au, which corresponds to the gap in the available RV temporal coverage. Additional smaller peaks, correspond to harmonics of this peak. These peaks are probably not physical.
We therefore conclude that unless the orbit of the companion is seen pole-on, the companion to G80-21 is a planet orbiting between 0.5 and 2 au. As the RV limits are very sensitive to the inclination of the planet, we then repeated this approach fixing the inclination to 10, 45 and 90 degrees. The results are also showed in Figure~\ref{fig:HIP17659_mass_sma_overlap}. As expected, the higher the inclination, the lower the RV detection limits, and hence the lower peaks below 0.5 au.

\begin{figure}[hbt]
    \centering
  \includegraphics[width=80.mm,clip=True]{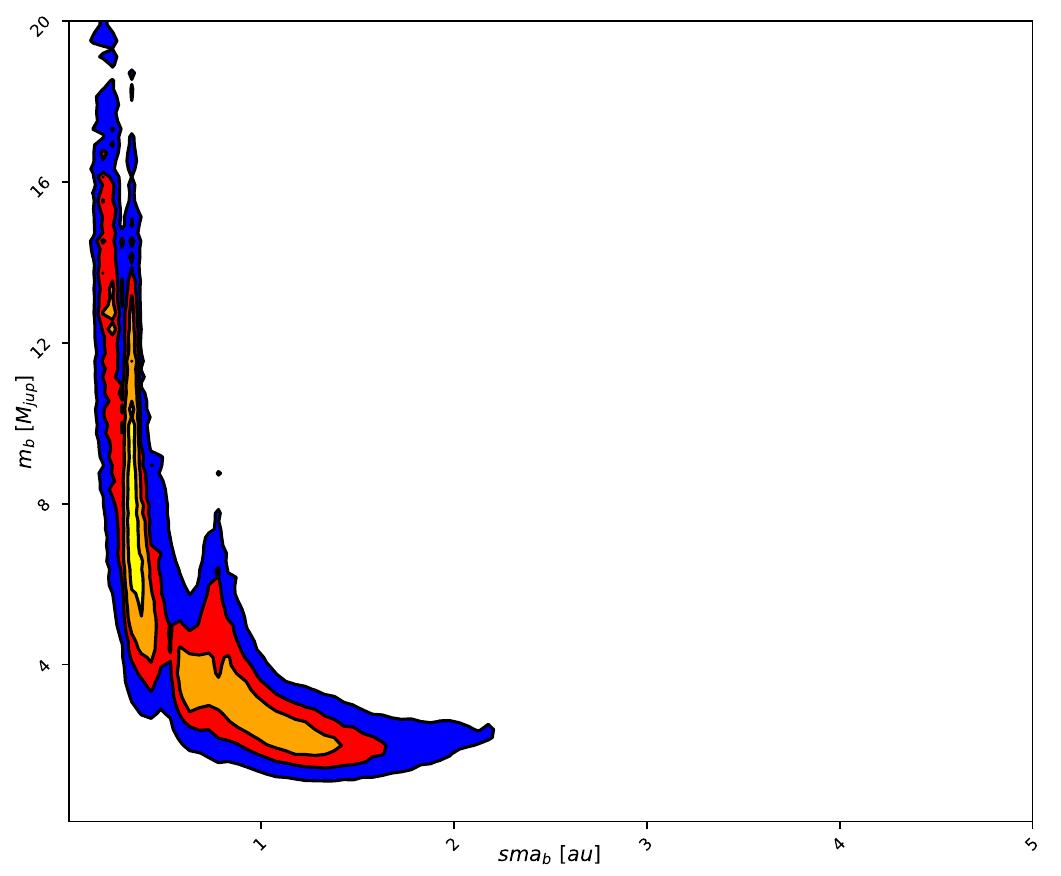} 
    \caption{(sma,mass) solutions for G80-21 b, derived from the MCMC analysis, and taking into  account the available RV data, and the Gaia and PMA data.}
    \label{fig:HIP17659_mass_sma_emcee}
\end{figure}

We then used our MCMC code described in Appendix~\ref{sec:Processing} to check these results in the general case (no assumptions on the inclination). The priors and resulting corner plot are provided in Appendix~\ref{sec:Processing}.  As expected, given the limited RV data available, most parameters are not well constrained, except the sma and the companion mass. We show in  Figure~\ref{fig:HIP17659_mass_sma_emcee} the obtained (sma, mass) solutions. We first note that the peak centered at about 0.33 au is also present, and due to the window function of the RV observations. The companion sma is therefore less in the range 0.5-2 au, with a maximum below 1 au, and its mass in the 2-6 \Mjup  range, with a maximum probability between 3 and 4 \Mjup. Additional RV data are needed to improve the temporal coverage of the data and therefore the window function, and thus further constrain its physical and orbital parameters. 

\subsubsection{Additional companions to G80-21 }\label{sec:add_comp_G80-21}
Finally, we constrain the properties of any additional companion, using at the same time the RV and absolute astrometry (Gaia and PMa). Figure~\ref{fig:MESS3_G80-21} provides these  detection limits. 
The  tool used to do so is described in Appendix~\ref{sec:Processing}.  

 \begin{figure}[hbt]
   \centering
    \includegraphics[width=90.mm,clip=True]{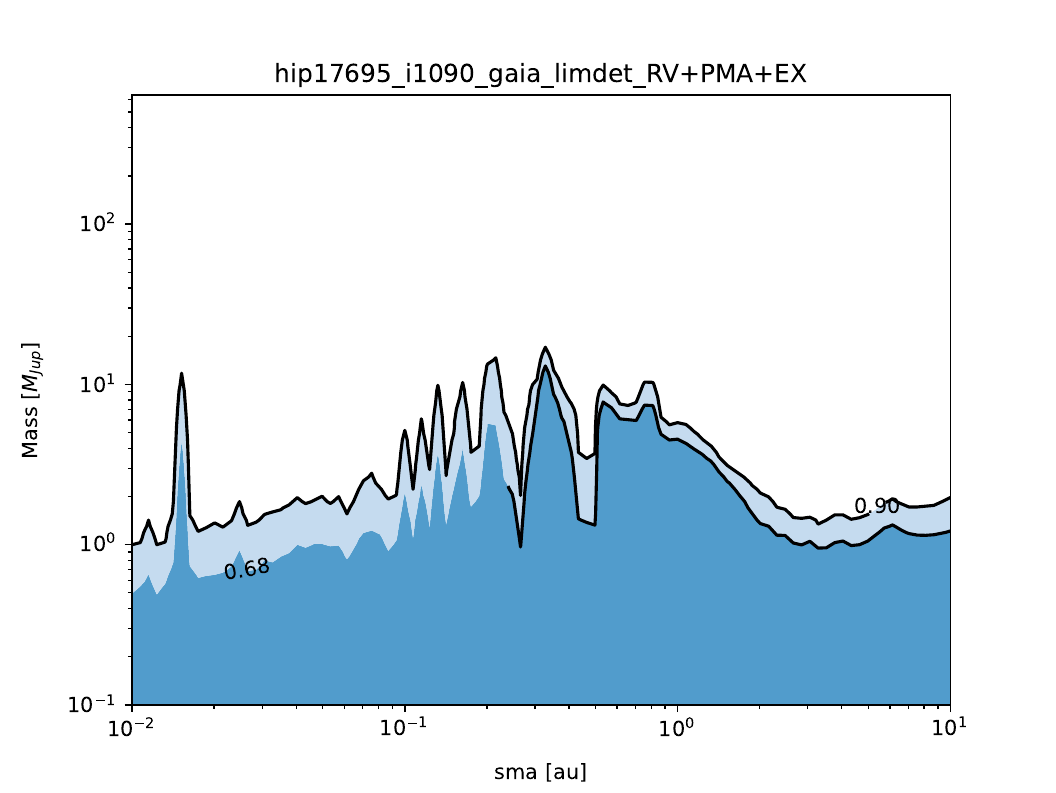} 
  \caption{ Limits on the masses of the potential planets around G80-21, using RV and GAIPMEX.}   
 \label{fig:MESS3_G80-21} 
\end{figure}

\subsubsection{Discussion }
To our knowledge, G80-21 b is the only young planet among the 11 RV planets with masses in the range 1-20 \Mjup  orbiting  stars with masses in the range 0.2-0.5 Msun (exoplanet.eu) and with a known sma between 0.017 and 5 au.  As such, it offers the possibility to study the early stages of planetary systems around low mass stars. Additional RV data are needed to further constrain its actual sma, and to precise its location with respect to  the snow line. Also, detailed RV monitoring of such a system  with an outer giant planet (if its sma is indeed in the range 1-2 au) would be particularly interesting to search for inner terrestrial planets, forming thus a young solar system-like "scaled" to a low mass star, with the possibility to image the outer giant planet in the future with Gravity+ or with the ELT.

\subsection{AB Pic}\label{sec:AB Pic}

With a $\simeq $ 13 \Mjup~companion (AB Pic b) imaged at about 260 au in the mid 2000 \citep{Chauvin05}, the solar-type star AB Pic (K1V) has become an iconic system. This was indeed the first low-mass substellar object (at the boundary between planets and BD mass) imaged around a main sequence star. The star was recently classified as a member of the Carina association (CAR) with therefore an age of about 13 Myr instead of 20 Myr as previously adopted \citep{Booth21}. Such an age would lead to a planet mass of $10 \pm 1$ \Mjup, lower than previously estimated, and well into the planetary mass domain. Due to its large projected distance to the star (about 190 au), its orbital parameters are still rather poorly constrained, though there is evidence of a non-negligible eccentricity and an almost edge-on orbit \cite[][herefater PB23]{Palma-Bifani23}. The formation process of this remote planet is still unclear. In situ formation via core accretion is not an option because of a lack of material and inappropriate timescales, but a formation closer to the star followed by an ejection due to the dynamical interaction with a third body is plausible (a scenario also considered to explain the presence of HD 106906 b orbiting currently at 650 au from the star \citep{Rodet17,Rodet19}. Alternatively, the planet could have formed by gravitational instability in an extended disk or by gravo-turbulent fragmentation. 

 \subsubsection{AB Pic c properties: previous works}
 Using the PMa estimated using  Gaia DR3 and Hipparcos data, PB23 concluded that an additional, inner planet (AB Pic c) could be present in the system. Comparing the (sma, mass) solutions derived on the one hand from the PMa constraints using \citet{Kervella19} approach, and on the other hand, the detection limits obtained using SPHERE data obtained in Dec. 2015, PB23 estimated a sma of 2 to 10 au, and a mass of  6 \Mjup~for this companion. 
These values need further confirmation for three reasons. First, PB23 did not consider the possibility of companions closer than typically 1-2 au, while such companions would have been compatible with the PMa constraints only. Second, the (sma, mass) constraints derived from the PMa based on the approach proposed by \citet{Kervella19} are approximate as they take into account the impact of inclination and eccentricity only statistically. Finally, \citet{Kervella19} does not take into account any instrumental noise  in this approach. These points were left for a more detailed study (the present one).

Recently, \citet{Gratton24} published a very narrow solution range of solutions for the mass and sma of AB Pic c: 8.4 $\pm 4.2$ \Mjup and $3.571 \pm 1.167$ au. However, no detail is provided on the way the sma and masses are computed and how the  compatibility of each simulated orbit with the various measurements (Gaia and Hipparcos astrometry, RV, high contrast imaging) is quantified, and which errors are taken into account.  

 \subsubsection{AB Pic c properties: more robust estimates} 
\begin{figure*}[hbt]
    \centering
  \includegraphics[width=60.mm,clip=True]{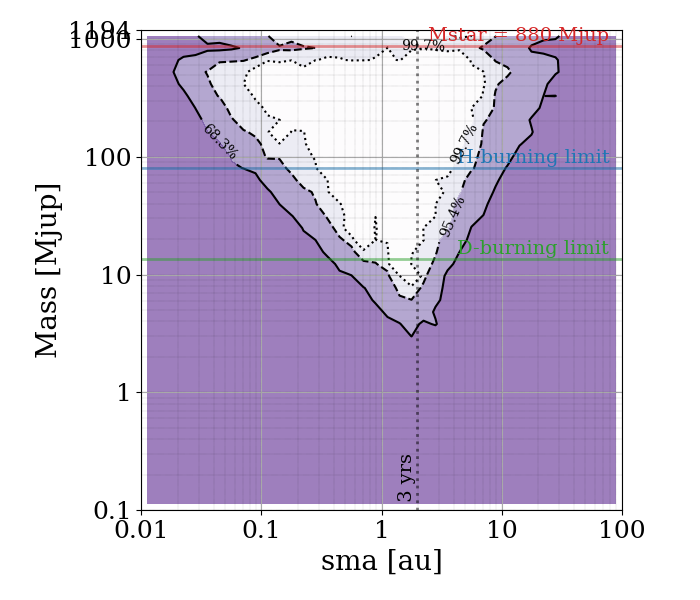} 
    \includegraphics[width=60.mm,clip=True]{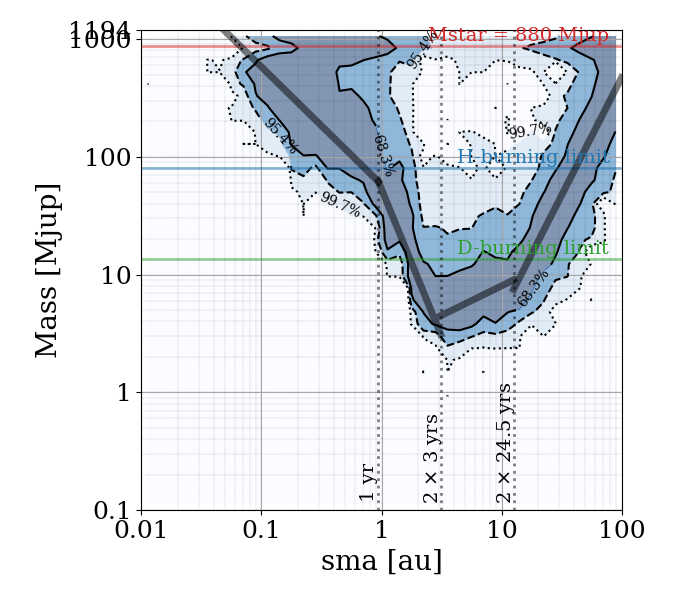} 
    \includegraphics[width=60.mm,clip=True]{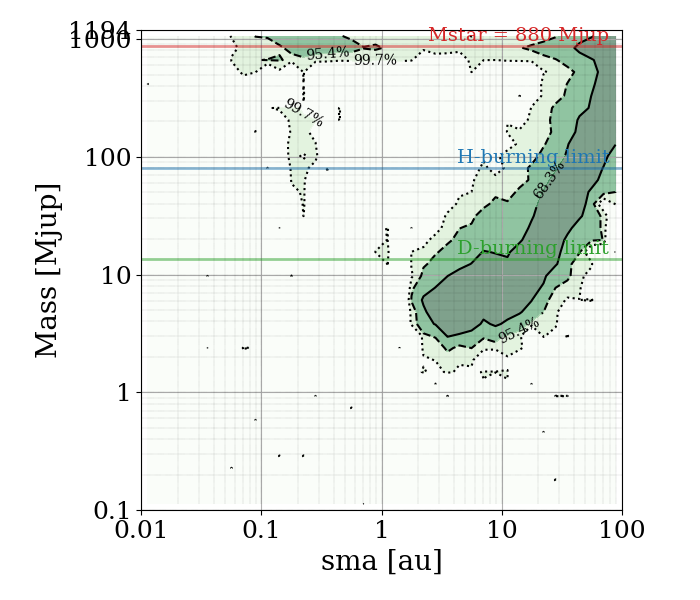} 
    \caption{From Left to Right: Gaia excess, PMa and combined maps for AB Pic. No constraint on the inclination of the orbital plane of companions.}
    \label{fig:ABPic_GaiaPMEX}
\end{figure*}

AB Pic shows a clear PMa (\nsigmapma{} = 4.2) and no significant \aastroruwe{} at \nsigmaruwe{} = 0.6, indicative of a companion beyond a few au. Figure~\ref{fig:ABPic_GaiaPMEX} shows the (sma, mass) solutions compatible with the Gaia astrometric signature only (Left), the PMa (Middle), and both quantities combined (Right). We see that the solutions found with the PMa data alone are very degenerate, while the combined Gaia signature + PMa constraints lift a significant part of the degeneracies.
A mass of $0.84 \pm 0.1$ \Msun~ was assumed for AB Pic \citep{Stassun19}. AB Pic b  (190 au, 9-11 \Mjup) is well out of the domain of possible solutions. Hence, an additional  companion is necessarily present in the system.

\begin{figure*}[hbt]
    \centering
     \includegraphics[width=60.mm,clip=True]{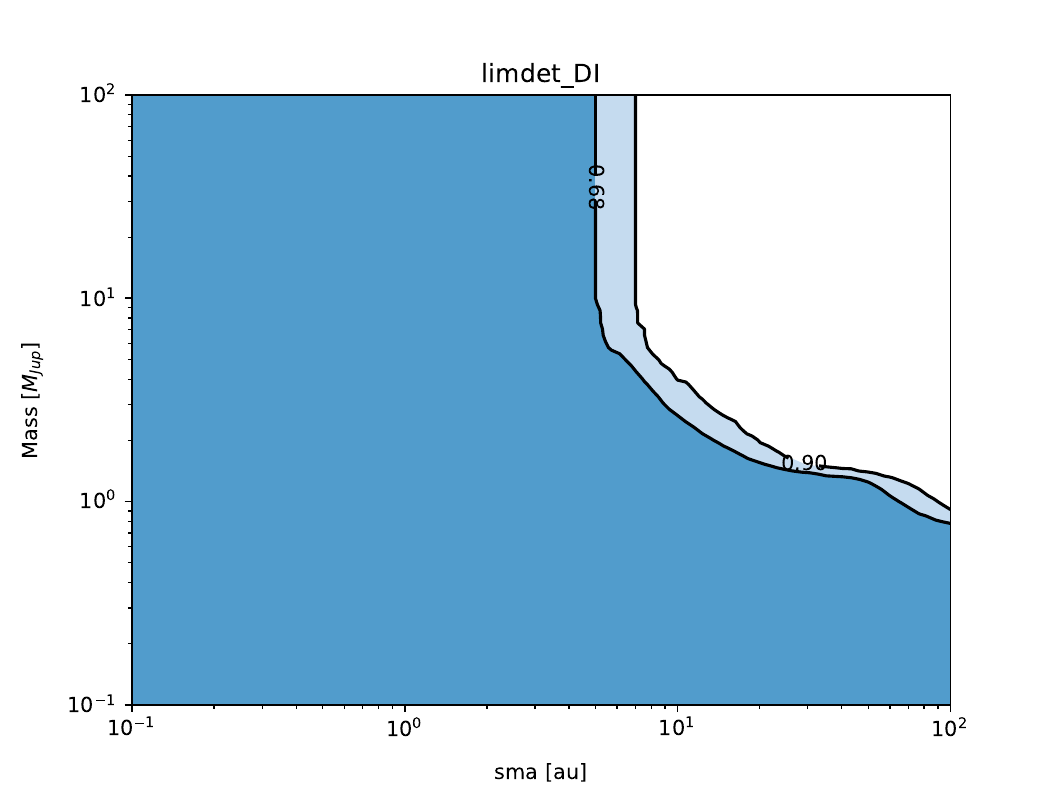} 
    \includegraphics[width=60.mm,clip=True]{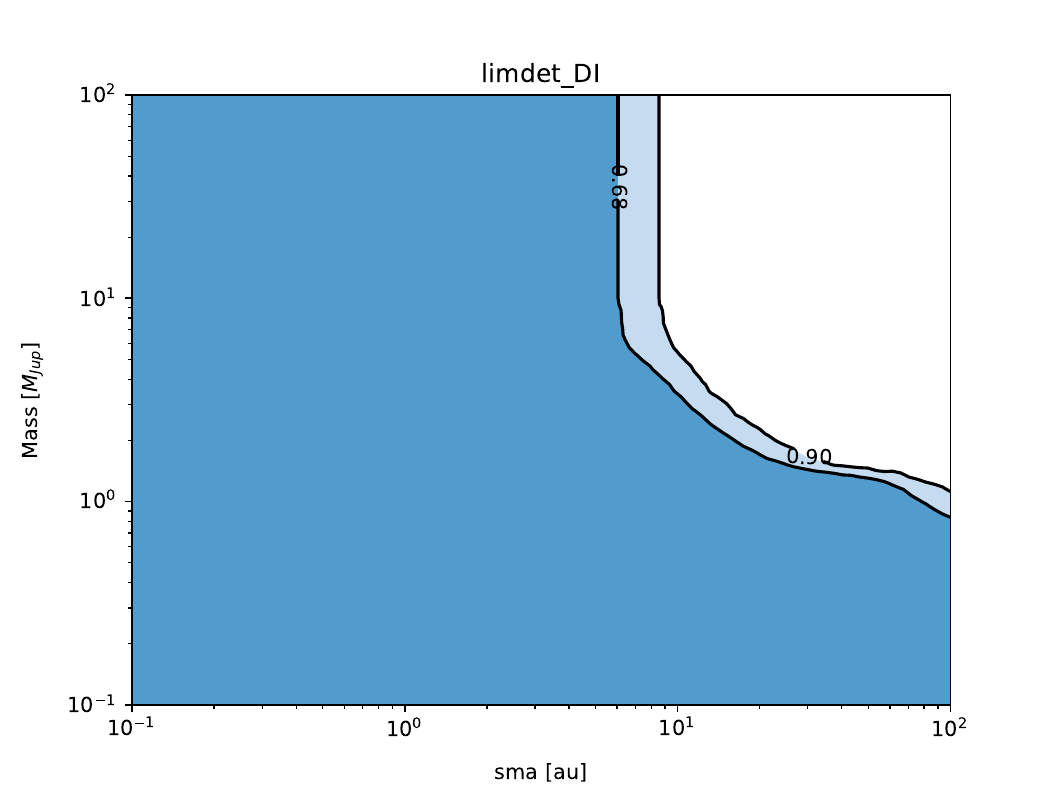} 
    \includegraphics[width=60.mm,clip=True]{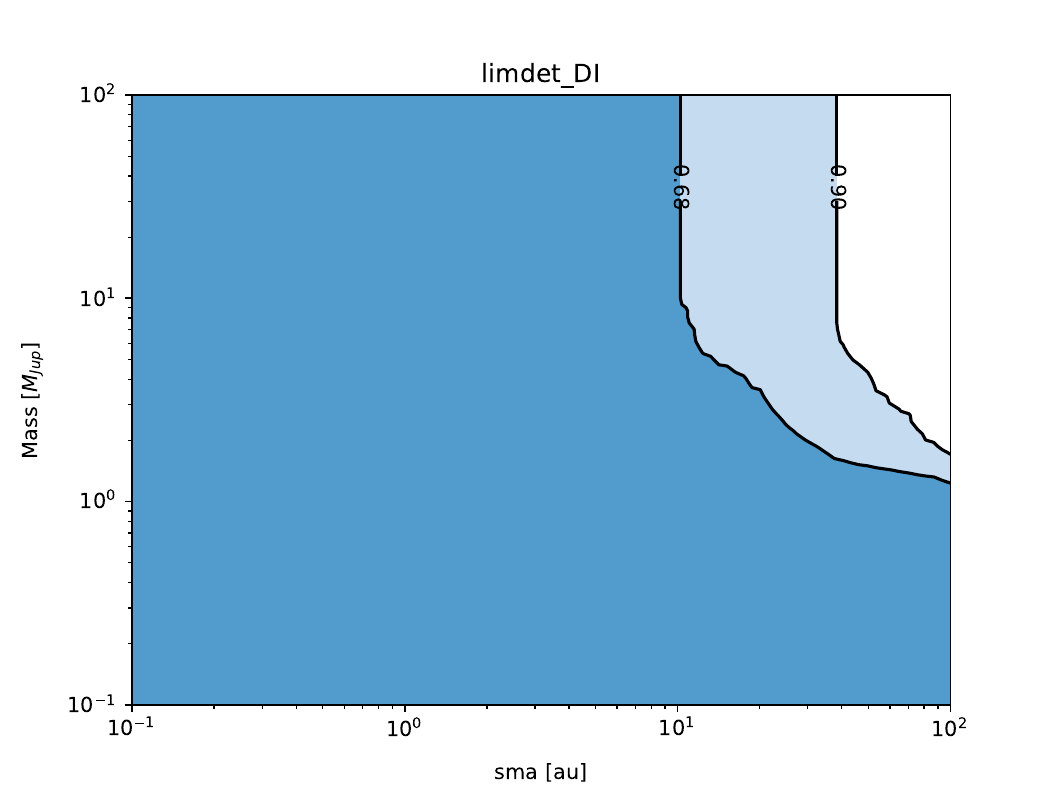}\\ 
  \includegraphics[width=60.mm,clip=True]{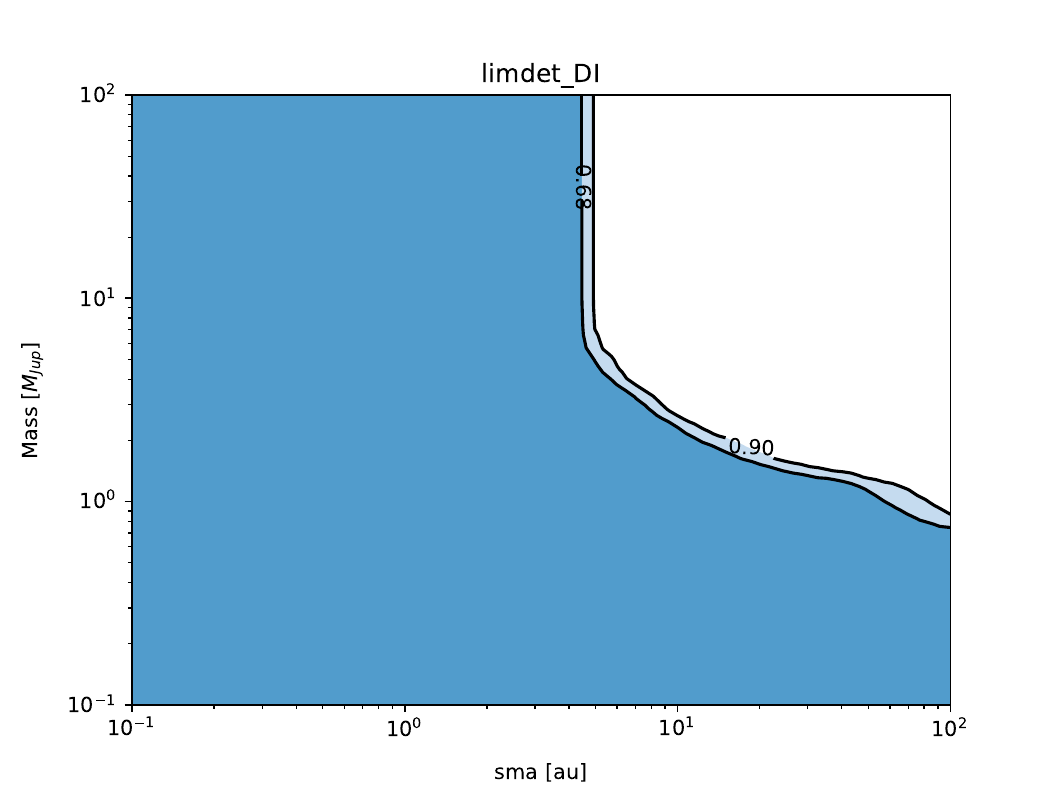} 
    \includegraphics[width=60.mm,clip=True]{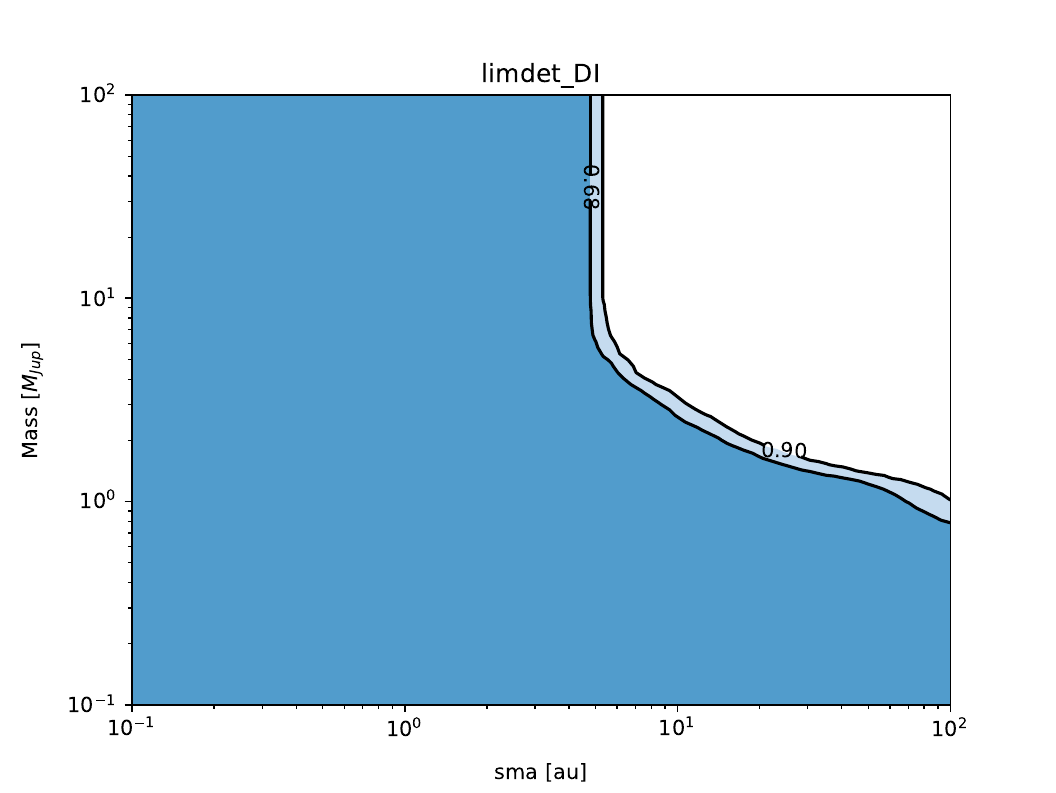} 
    \includegraphics[width=60.mm,clip=True]{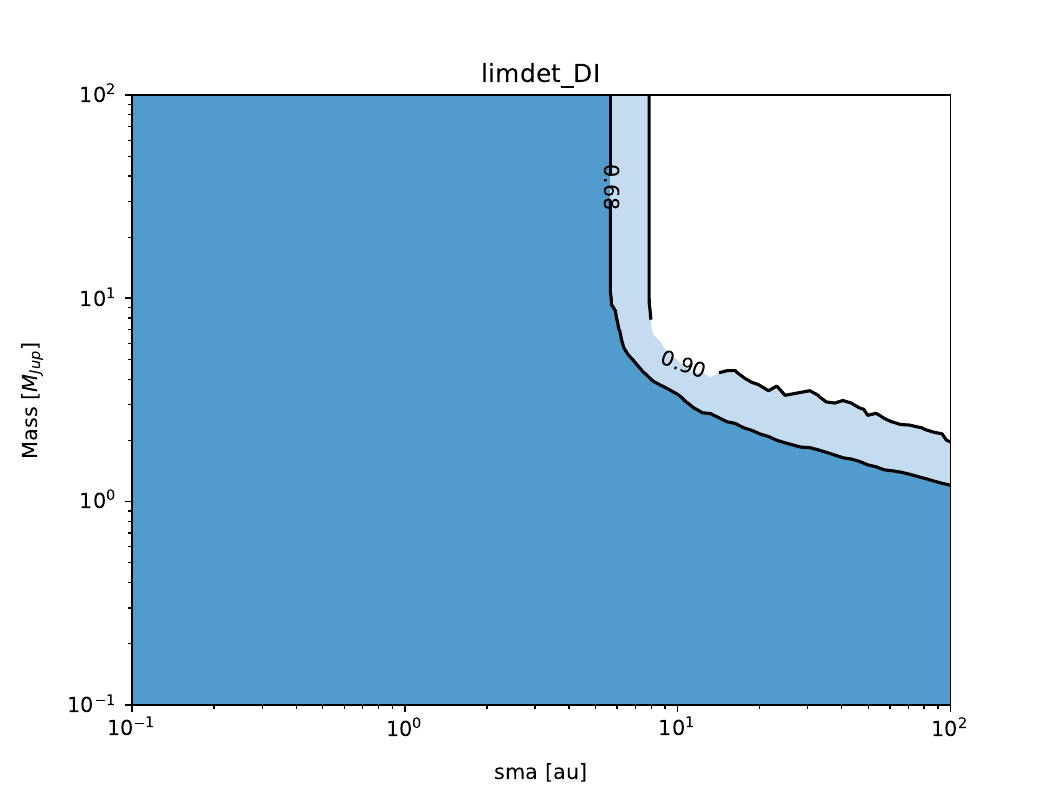} 
    \caption{Detection limits for AB Pic obtained using SPHERE high contrast imaging data obtained in Dec. 2015 (First row) and the same data plus SPHERE data obtained in Oct. 2023, as well as GPI high contrast imaging data obtained in 2018 (see Appendix). We assume, from Left to Right, inclinations of $0^\circ$, $45^\circ$ and $90^\circ$. The iso-contours correspond to  90 and 68 $\% $ detection probabilities. }
    \label{fig:ABPic_impact_inclination_all} 
\end{figure*}

We use in the following SPHERE, GPI \citep{macintosh14} and RV data to robustly constrain the properties of this additional companion. The data, as well as their reduction, are described in Appendix~\ref{sec:Log_obs}  and \ref{sec:Processing}. 
Noticeably, compared to PB23 and to \citet{Gratton24}, we use, in addition to the Dec. 2015 SPHERE data, a GPI data set, as well as an additional SPHERE data obtained in Oct. 2023 (see Appendix for a description of the data and their reduction). These two additional data sets allow to significantly improve the detection limits in the case of edge-on orbits. This can be seen comparing Figure~\ref{fig:ABPic_impact_inclination_all}, which provides the detection limits obtained respectively with the Dec 2015 set of data and the limits obtained using in addition both Dec 2015 and Oct. 2023 data (see Appendix for a description of the  computation of the detection limits). While companions can be detected (68 $\% $ probability) beyond 11 au when using the Dec 2015 data only, they can be detected beyond 5.5 au when using these additional data (case $i=90^\circ$). This significant  improvement is due to the fact that the data sets were obtained several years apart and companions hidden behind the mask in the first data set (and hence not detectable) could be outside the mask during the second observation (and hence detectable). The effect is, expectedly, not as high when considering less inclined orbits.

\begin{figure}[hbt]
    \centering
    \includegraphics[width=90.mm,clip=True]{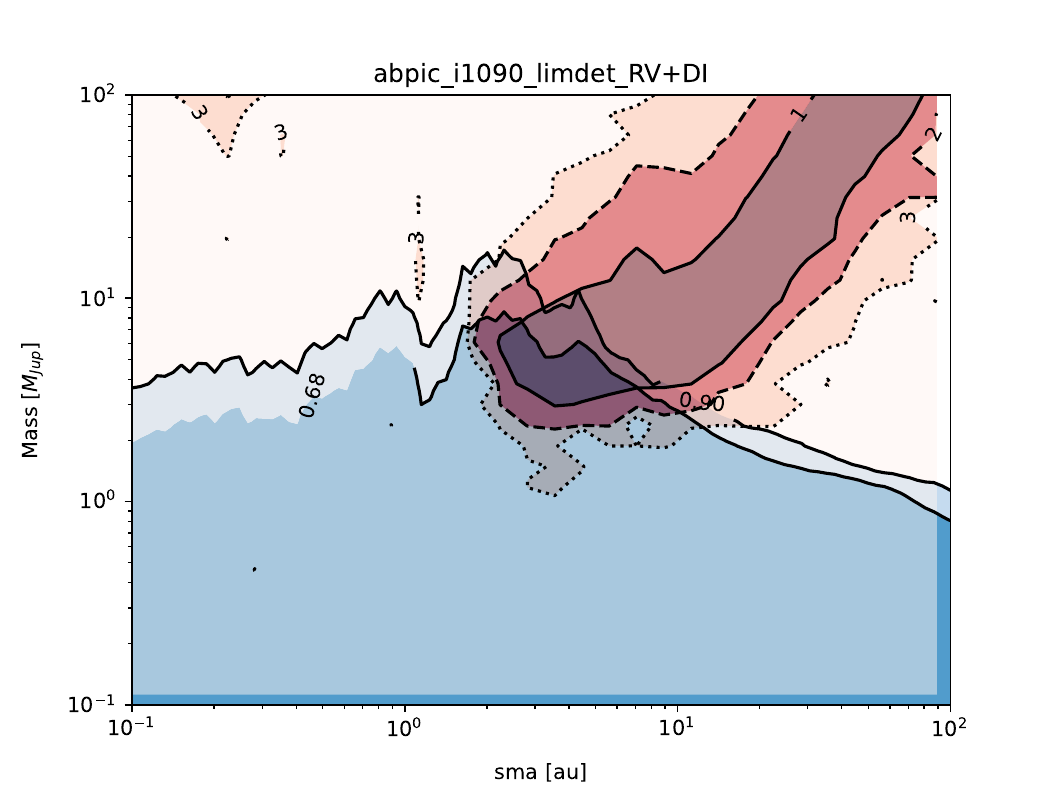} 
\includegraphics[width=90.mm,clip=True]{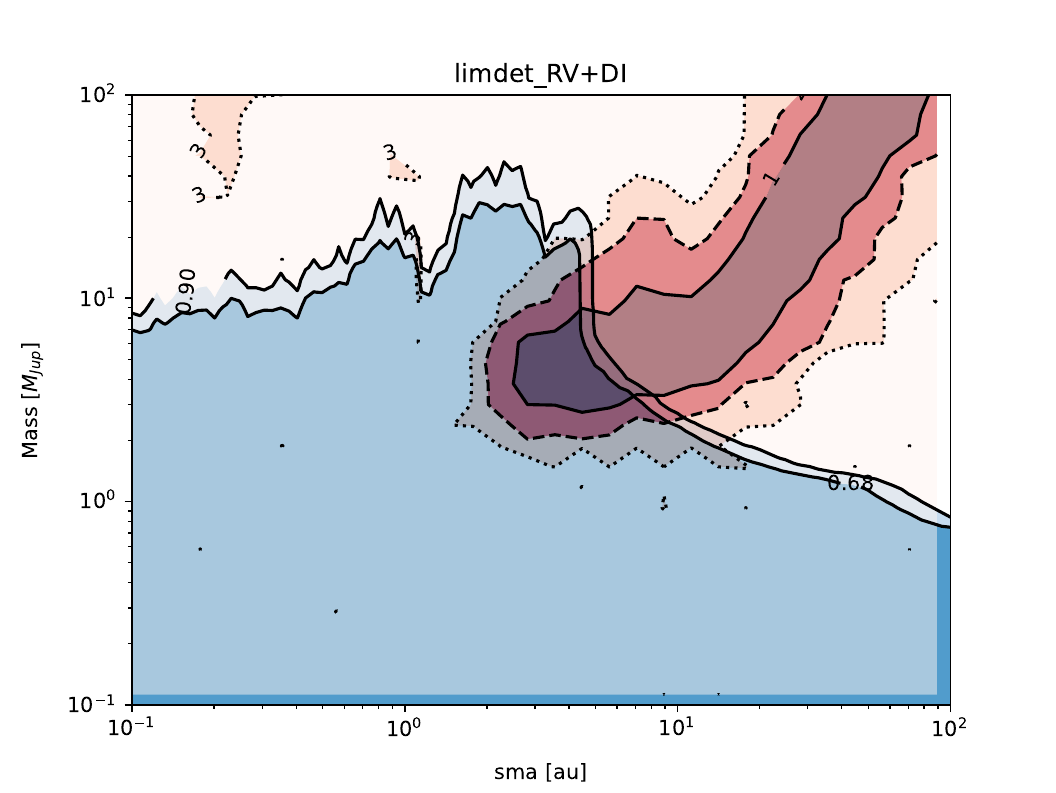} 
\includegraphics[width=90.mm,clip=True]{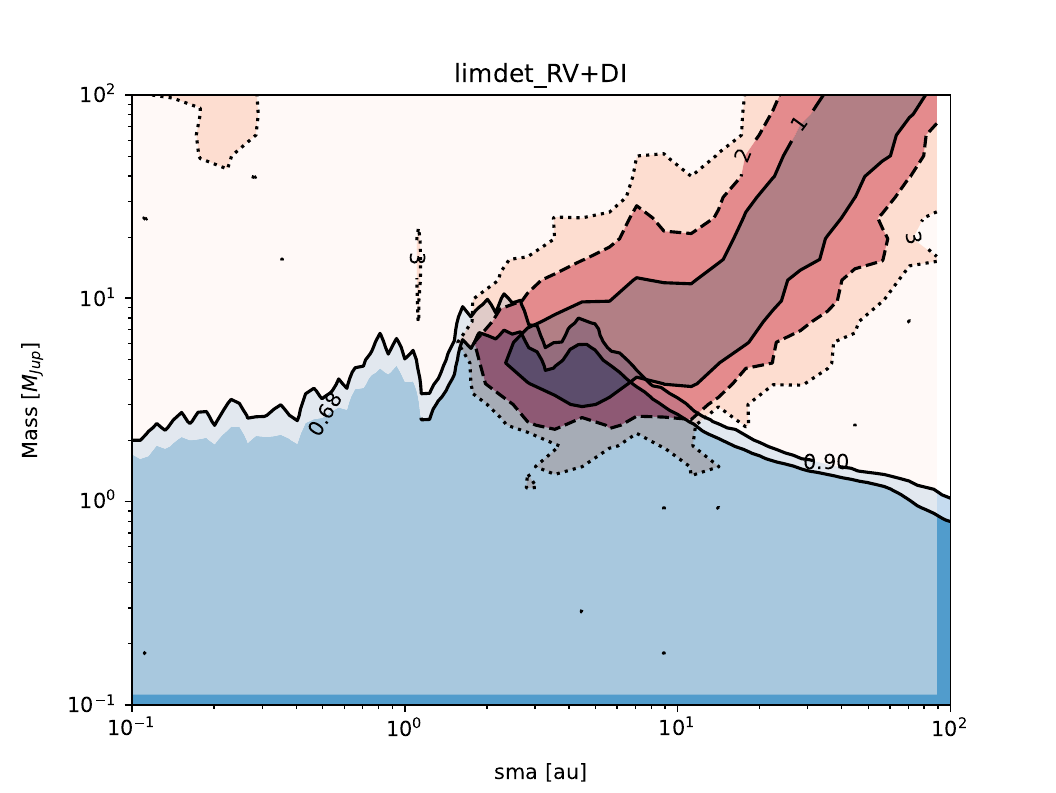} 
\includegraphics[width=90.mm,clip=True]{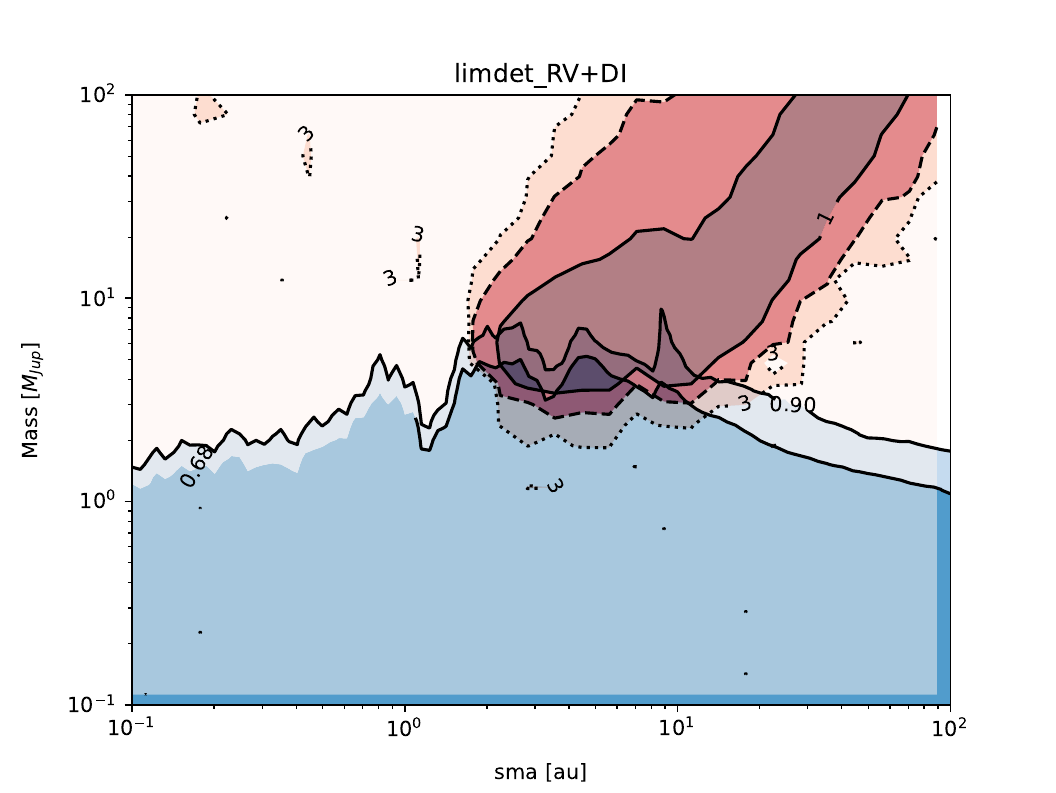} 
    \caption{AB Pic. Superimposition of the (sma,mass) solutions from the MCMC considering the PMEX and RV constraints, and the GPI+SPHERE detection limits.The blacklines limiting the blue regions correspond to the RV detection limits (90 and 98 $\% $ probabilities). From Top to Bottom: an inclination between 10 and 90 $\deg $ is assumed, then inclinations of 10, 45, and 90 degrees. }
    \label{fig:ABPic_sup_GaiaPMEX_RV_DI_inclinaison}
\end{figure}

To estimate the possible (sma, mass) of the companion, we superimpose GaiaPMEX maps with the detection limit maps obtained using all available direct imaging data (that is the two SPHERE epochs and one GPI set of data, see Appendix~\ref{sec:Log_obs}) and radial velocity data, assuming different inclinations. We show in Figure~\ref{fig:ABPic_sup_GaiaPMEX_RV_DI_inclinaison} the results obtained assuming inclinations of $90^\circ$, $45^\circ$ and\footnote{We used 10\degree{} instead of 0 to avoid any divergence when, at $i$\,\!$\approx$0\degree{}, $\sin i$\,\!$\approx$0} $10^\circ$. We see that the (sma, mass) space is wider in case of pole-on orbits, and more reduced in case of edge-on ones. For pole-on orbits, most solutions ($ 68 \%$ probability) are between 2.5 and 9 \Mjup, and sma between 2.5 and 6 au. For edge-on orbits, the solutions in terms of mass are much more reduced, between 3 and 5 \Mjup.

Given the sma of AB Pic c, and the distance of the system, a direct detection of AB Pic c is not within the reach of current high-contrast imagers on 10-m class telescopes, but should be easily detected by imagers on the ELTs. Alternatively, it might within the reach of the VLT/GRAVITY instrument if its mass and sma are on the higher sides of the current possible ranges. This would imply a pole-on orbit, hence a configuration different from the almost edge-on orbit of AB Pic b. Note that, in such a case, the detection would be more demanding in terms of performances than the detection of AF Lep b ($\sim 3$ \Mjup, $\sim 300$ mas).

  \subsubsection{Additional companions around AB Pic}
  \label{sec:add_ABPic}
\begin{figure*}[hbt]
    \centering
    \includegraphics[width=60.mm,clip=True]{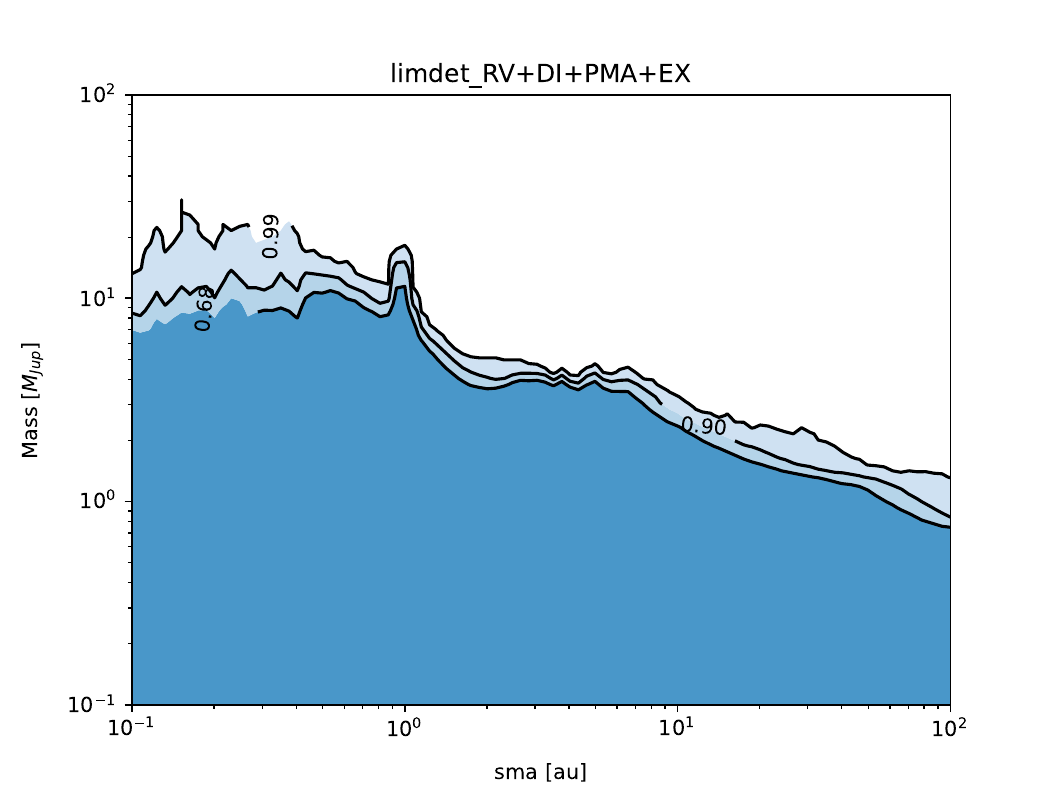} 
    \includegraphics[width=60.mm,clip=True]{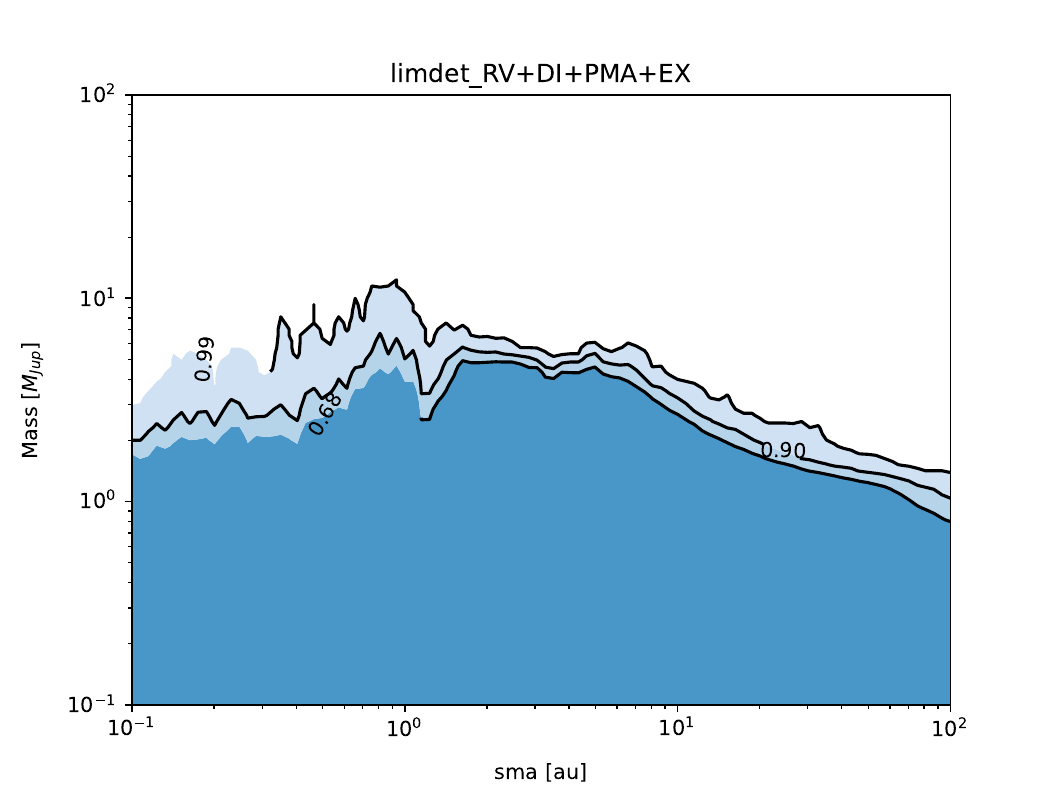} 
    \includegraphics[width=60.mm,clip=True]{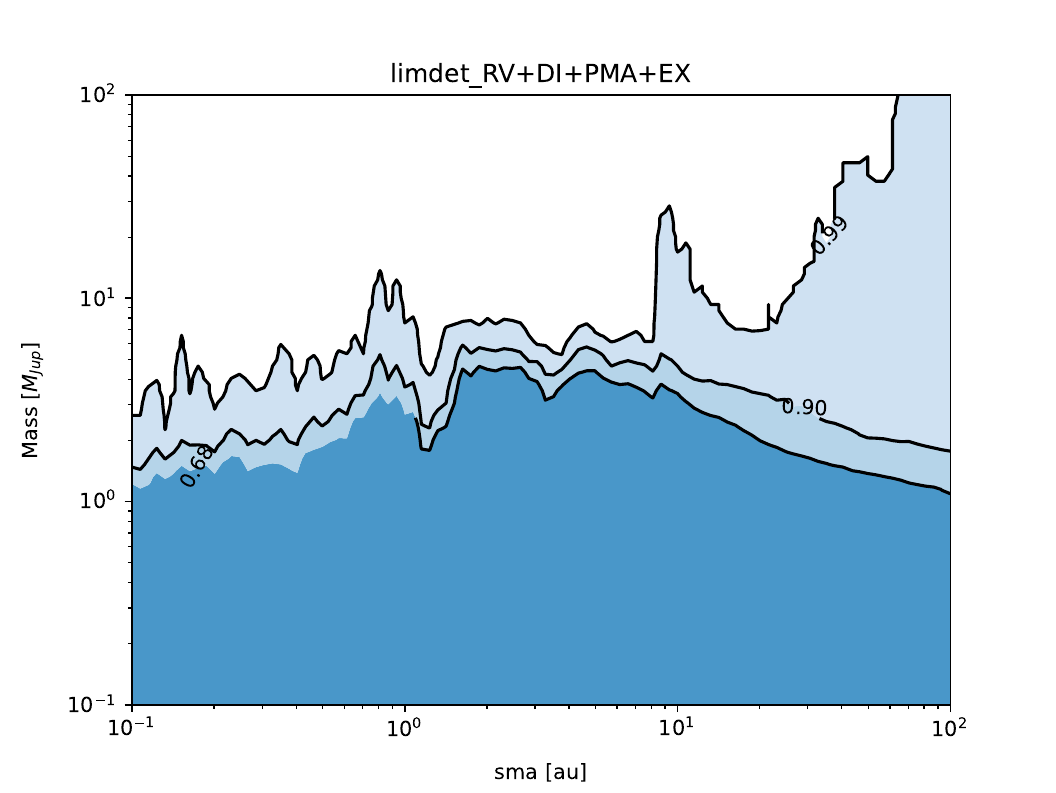} 
    \caption{From Left to Right: Detection limits for AB Pic combining RV, GaiaPMEX, and HCI data. From Left to Right: i = 0, 45, and 90 degrees.}
  \label{fig:MESS3_ABPic} 
\end{figure*}

  Finally, we constrain the properties of any additional companion, using at the same time the RV, absolute astrometry (Gaia and PMa), and the direct imaging data from SPHERE and GPI described in Table~\ref{tab:obs_cond_table}, and the detection tool described in Appendix~\ref{sec:Processing}. 
  The results  are shown in Fig.\ref{fig:MESS3_ABPic} assuming three inclinations: 90, 45, and 10 degrees. As expected, the detection limits depend on the inclination of the companion,  especially in the regions where they are dominated by the RV (sin(i) effect) or direct imaging  (reduced limits for inclined orbit with respect to pole-on one, as long period planets may happen to be located in projection behind the mask at all observation epochs). 
  
  In any case, the data at hand allow, for the first time, a quantitative estimate of the properties of any possible additional planet (planet d) around the star from less than 0.01 au to more than 100 au. We can exclude additional planets more massive than AB Pic b or c in the system.

 \subsubsection{Consequences on the system}
  This work shows that AB Pic c is definitely lighter than AB Pic b, and that there is no companion more massive than AB Pic b at all separations, up to AB Pic b distance. This severely questions any scenario of ejection of AB Pic b by a third companion close to the star to explain its present location.  This is for instance in contrast with the case of the HD 106906 system, for which the binary companion was shown to possibly be responsible for the ejection of HD 106906 b. Here any gravitational interaction between AB Pic b and AB Pic c, whether direct of resonant, would inevitably result in a stronger perturbation of the lighter body's orbit, i.e., AB Pic c. 

 Among the other possible explanations of AB Pic b's large separation, a fly-by acting on an originally much closer-in AB Pic b cannot be ruled out but is rather improbable as the perturbing star should have come very close to the initial position of AB Pic b. Subsequently, AB Pic c should have been impacted as well by this process. Hence the geometry of this encounter should be able to explain why only AB Pic b was ejected. The suitable parameter space is presumably reduced, but our still poor knowledge of AB Pic b and c's current orbital properties prevents us from initiating any detailed dynamical investigation.  

Migration within the protoplanetary disk is not favored because of the high mass of AB Pic b, and  because this would require a very massive disk up to 200 au at least. A more complex possible scenario could be that AB Pic was formerly a member of a wide (> 1000 au separation), now disrupted binary system with a low-mass star with a highly inclined orbit with respect to the orbital plane of AB Pic b. Such a configuration would inevitably lead to strong coupled eccentricity and inclination oscillations of AB Pic b due to Kozai-Lidov resonance \citep{Kozai1962} (hereafter KL). Under this process, the apoastron of AB Pic b could have been increased up to its present position. However such a scenario requires a rapid stabilization of AB Pic b around 200 au by a direct interaction at apoastron. As a matter of fact, a similar extra interaction at apoastron after ejection was also required in the HD 106906 system to freeze the present-day configuration \citep{Rodet17}. Here again also, KL oscillations triggered by the putative binary companion are also expected to affect AB Pic c as well, and one should explain why only AB Pic b was affected. The answer to this issue could be a matter of timescales. KL cycles in triple systems are characterized be a oscillation period basically equal to
\begin{equation}
P_\mathrm{KL}=\frac{M_\mathrm{tot}}{M_\mathrm{in}}\frac{P_\mathrm{out}^2}{P_\mathrm{in}}
\end{equation}
within a factor of order unity \citep{Beust2006,Kry1999,Ford2000}, where $M_\mathrm{tot}$ is the total mass, $M_\mathrm{in}$ is the inner mass, $P_\mathrm{out}$ is the outer orbital period (here the outer binary companion), and $P_\mathrm{in}$ is the inner orbital period (here AB Pic b or c). Now, if we assume that the ratio $M_\mathrm{tot}/M_\mathrm{in}$ is $\sim1.5$ (the outer companion should be less massive than AB Pic), that the outer companion's sma is $\sim 1000\,$au, and that the initial sma of AB Pic b is $\sim 40\,$au, then we derive $P_\mathrm{KL}\simeq4\,$Myr. Comparing to the estimated age of 13\,Myr for AB Pic, this gives enough time to KL to have been able to significantly affect AB Pic b. Redoing the same calculation for AB Pic c assuming now a sma of $\sim 4\,$au leads to $P_\mathrm{KL}>100\,$Myr, which is far above the age of AB Pic. KL resonance should not have had enough time to perturb AB Pic c. This way we may explain why only AB Pic b could have been ejected. However, a stabilization process at large distance is still required.  

Other possible scenarios include AB Pic b having been captured from another system or being a captured free-floating planet. Note that a capture from another system necessarily implies a similar ejection at large distance within that other system prior to capture by AB Pic, which drives us back to our previous issue. But that other system could have hold an inner binary like HD 106906 and more easily trigger the ejection.

The last, probably simpler scenarios are either in situ gravitational instability within a disk, or a gravitational collapse. In such a case, AB Pic b would be a so far unique system with two planets definitely formed by different processes. Detailed studies of the planets atmospheric compositions could further help test this hypothesis.

\subsection{HD 14082 B }\label{sec:HD14082 B}
HD 14082 B is a close (40 pc) young solar-type star, member of the BPMG. Like many members of this moving group, the star shows IR excess in the mid- and far-infrared \cite[e.g. ][]{Riviere-Marichalar14}, indicative of the presence of a yet unresolved dust disk. Noticeably, it shares a common proper motion with the F-type star HD 14082 A, orbiting 557 au away, i.e. 14" on the plane of the sky \citep{Mamajek14,Elliott-Bayo16}. 

\begin{figure*}[hbt]
    \centering
    \includegraphics[width=60.mm,clip=True]{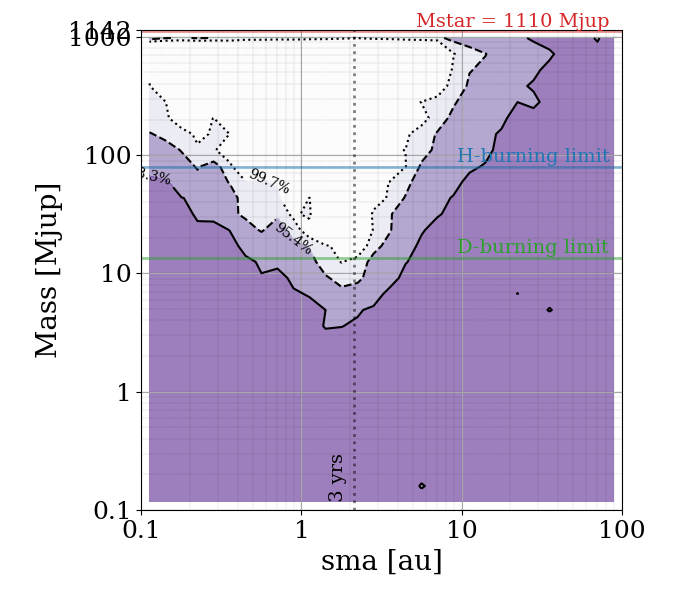} 
    \includegraphics[width=60.mm,clip=True]{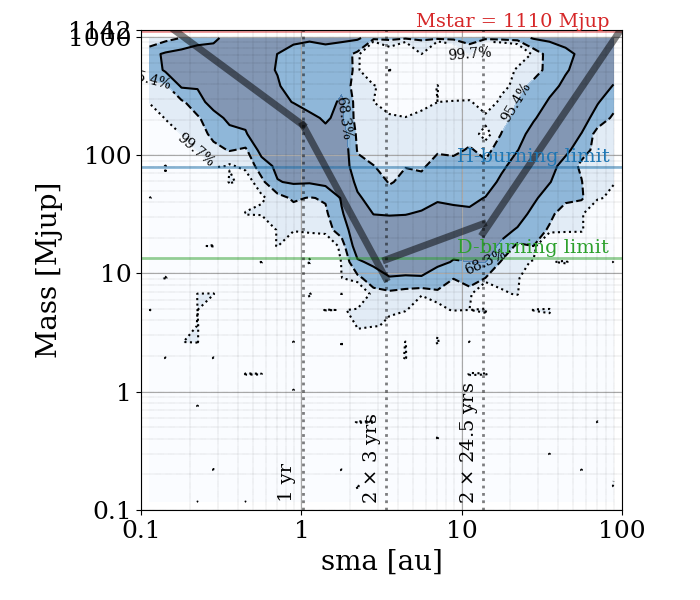} 
    \includegraphics[width=60.mm,clip=True]{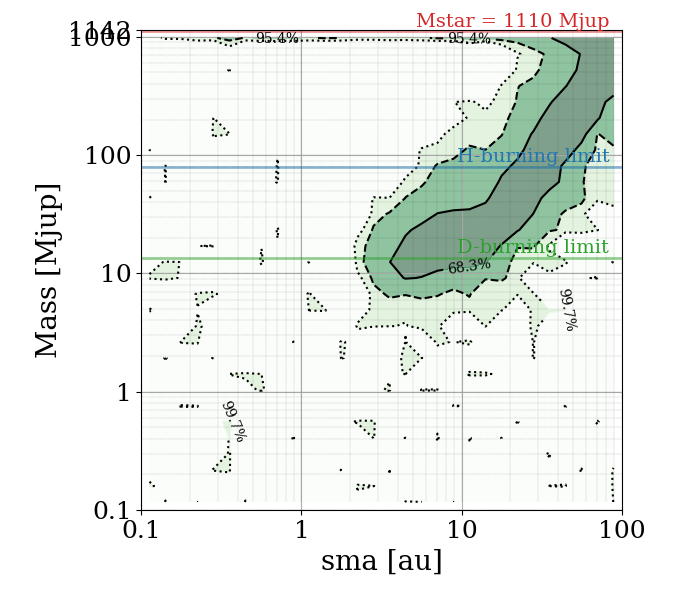} 
    \caption{From Left to Right: Gaia excess, PMa, and combined maps for HD 14082 B. No constraints on the orbital plane inclination.}
    \label{fig:HD14082B}
\end{figure*}

\subsubsection{A sub-stellar companion close to the ice-line of  HD 14082 B }
  
Like AB Pic, HD14082B shows a clear PMa (\nsigmapma{} = 5.0 and no significant \aastroruwe{} at \nsigmaruwe{} = 0.5, indicative of a companion beyond a few au. The \ruwe{} , the PMa, and the combined \ruwe{}  + PMa  maps are shown in Figure~\ref{fig:HD14082B}. They were computed considering inclinations between 0 and 90 degrees. The PMa map reveals the presence of a companion, with a still unconstrained PMa between 0.1 and about 100 au. Adding the \ruwe{}  constraints allows us to rule out short-period companions. HD 14082 A is too far to account for the observed GaiaPMEX excess. We note that the GaiaPMEX map is reminiscent of the AF Lep or AB Pic maps, where possible companion masses reach the planetary mass, albeit slightly higher. 

At the time of writing this paper, a candidate companion with a mass of $0.0054 \pm 0.0030$ \Msun~($=5.7\pm3.1$ \Mjup) and a distance of $5.74 \pm 2.96$ au was reported by \citet{Gratton24} around HD14082B. Yet, as in the case of AB Pic b, detailed information on the way these parameters and their associated uncertainties were derived is lacking. Given the limited amount of data available on this system (no RV data, less imaging data; hence less data than in the case of AB Pic), and given the fact that the PMa used does not take into account the instrumental noises, the robustness of the derived parameters is not assessed.

\begin{figure}[hbt]
\includegraphics[width=100.mm,clip=True]{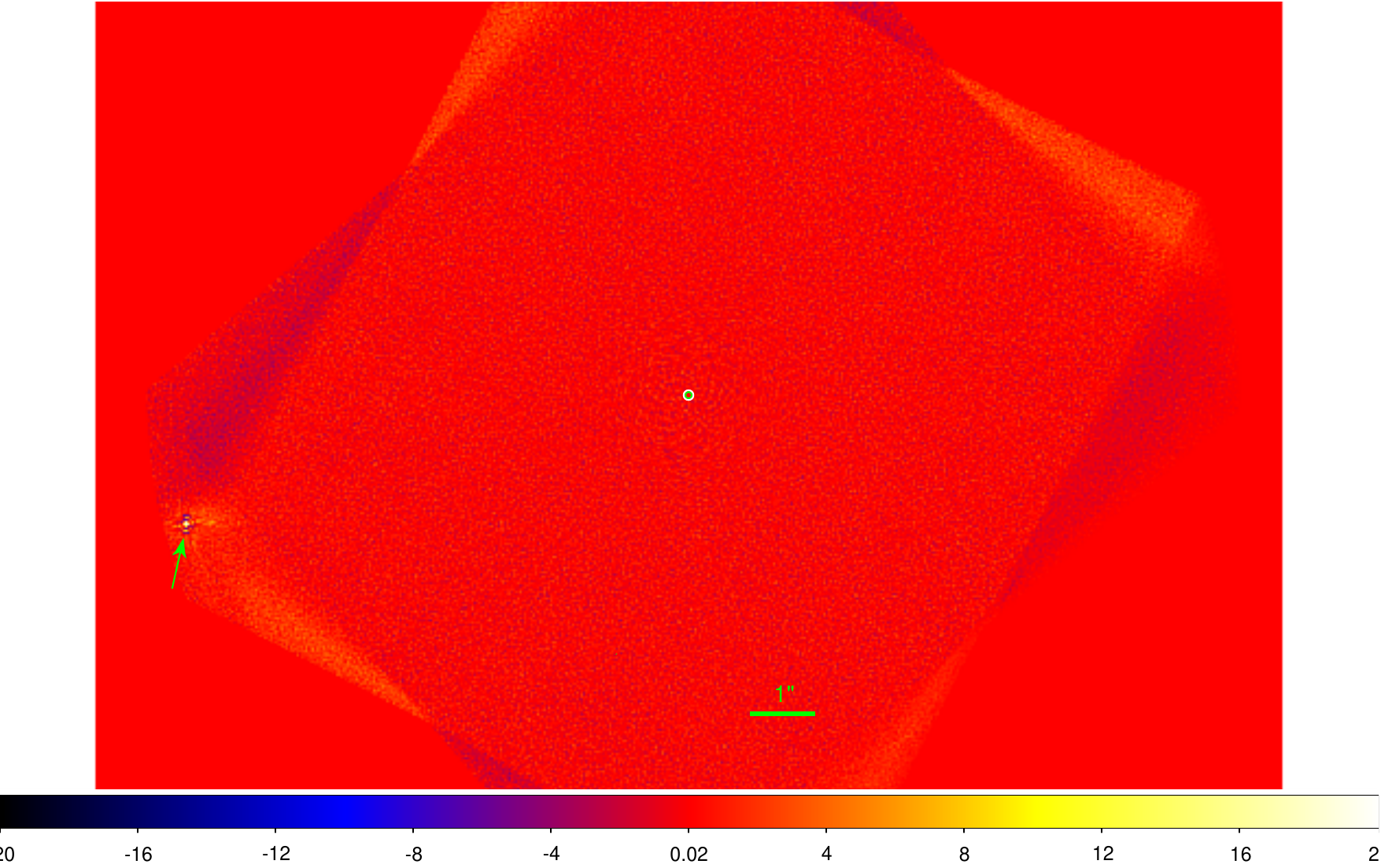} 
\label{fig:HD14082B_SPHERE}
        \caption{ SPHERE image of HD 14082B as reduced with PACO. A faint source (indicated by an arrow) is observed at 7.7'' from the star (indicated by a yellow circle, at the center of the image).}
\end{figure}

We used SPHERE data to further constrain the companion characteristics. They are described in Appendix~\ref{sec:Log_obs}. The data 
were both reduced with the Patch Covariance (PACO) tool \citep{flasseur_paco_2020}, as the AB Pic data, and with the SPEckle CALibration (SPECAL) tool 
\citep{Galicher18}. Both clearly (SNR $ > 40 $ with PACO) detect a companion close to the edge of the FoV (see Figure~\ref{fig:HD14082B_SPHERE}). Checking the Gaia DR3 database shows that this companion is a faint, M star, not connected to HD 14082B.

\begin{figure}[hbt]
    \centering
\includegraphics[width=90.mm,clip=True]{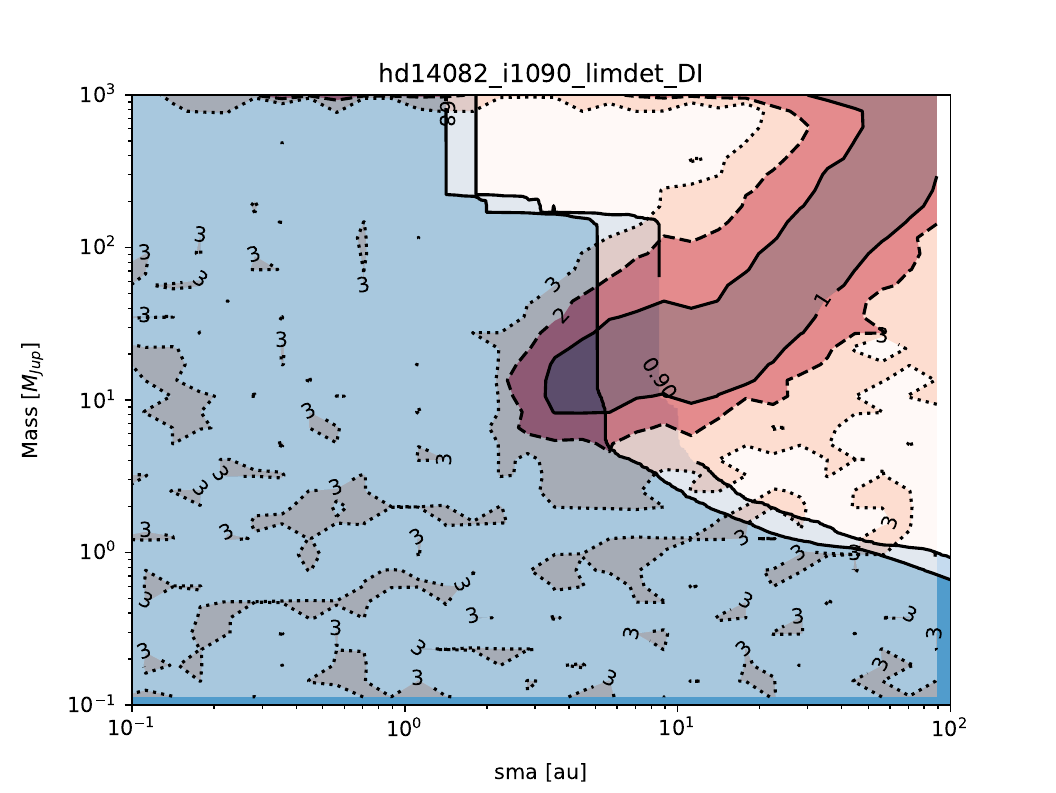} 
\includegraphics[width=90.mm,clip=True]{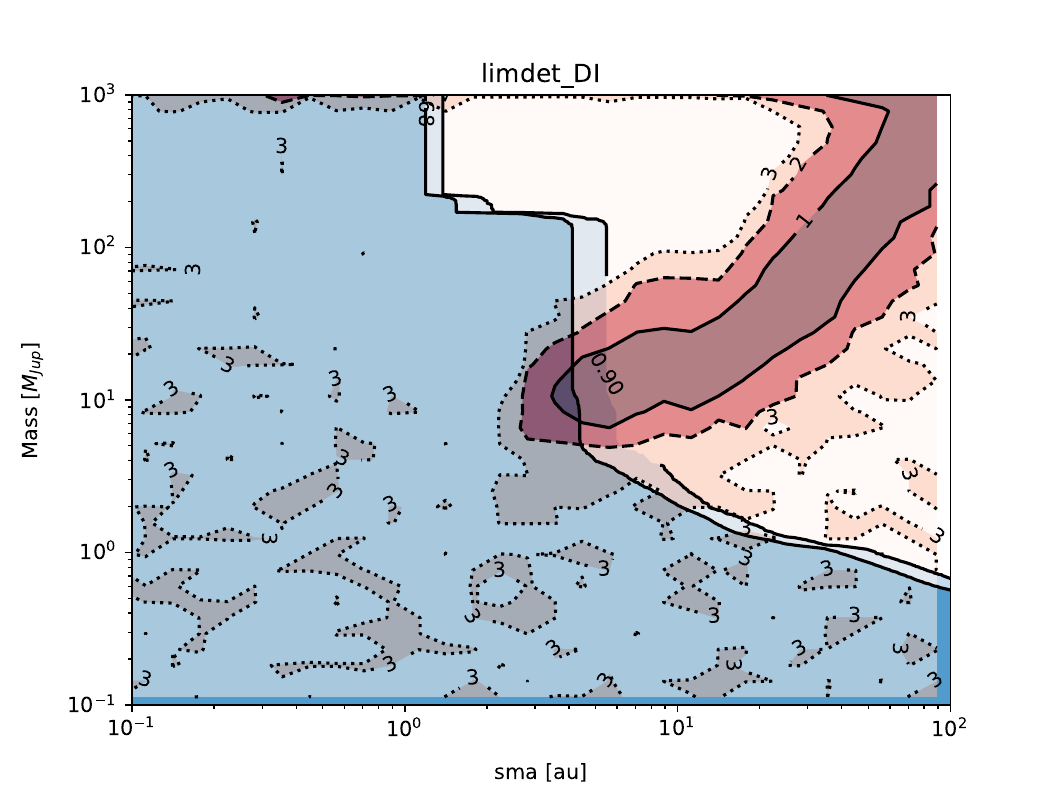} 
\includegraphics[width=90.mm,clip=True]{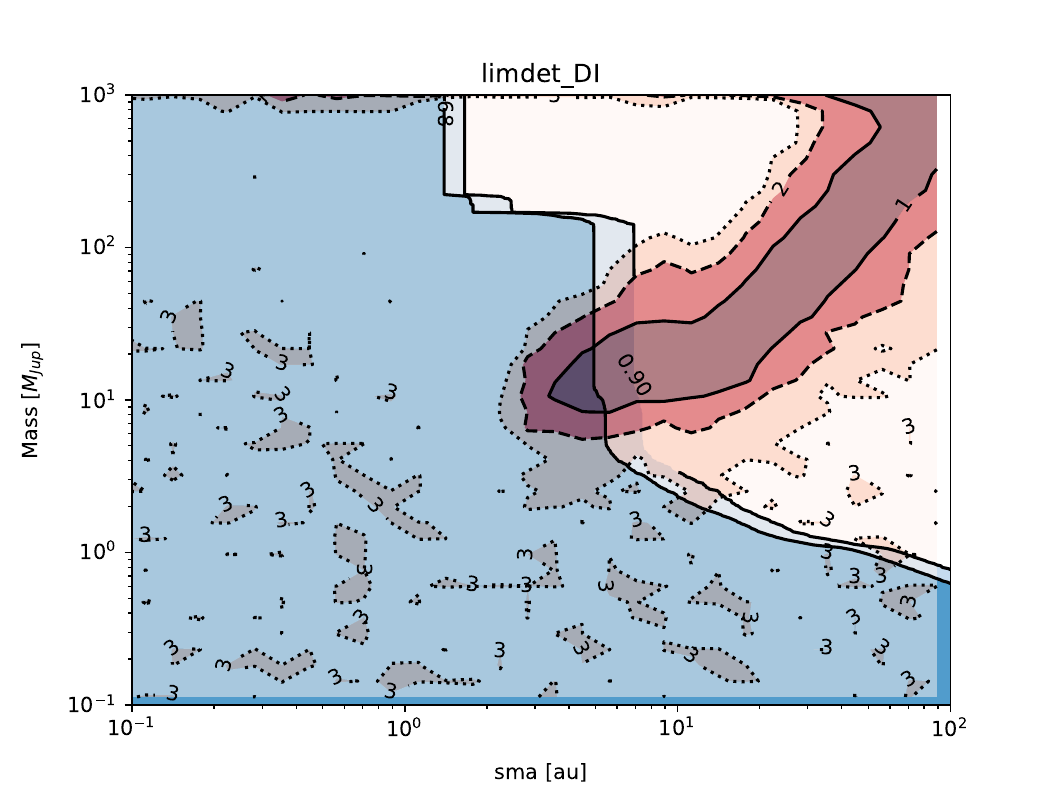} 
\includegraphics[width=90.mm,clip=True]{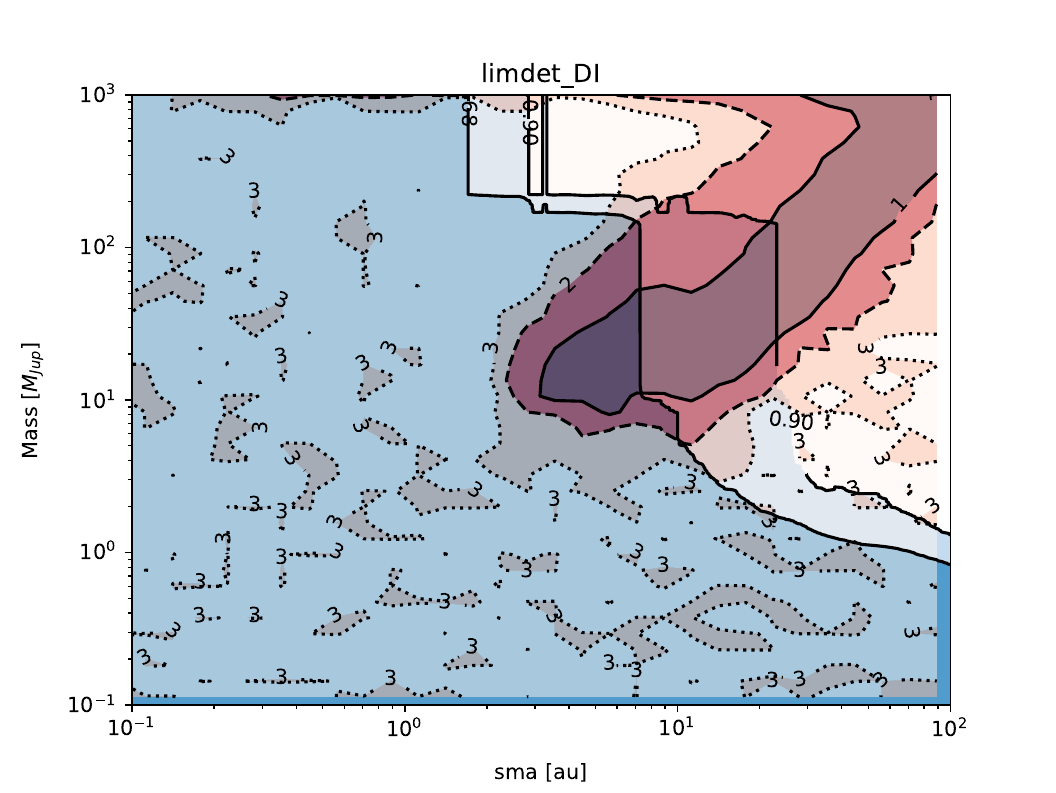} 
    \caption{HD 14082B. Superimposition of the GaiaPMEX (sma,mass) solutions and the SPHERE detection limits. The blue region corresponds to  sma and masses that are not excluded by the high contrast imaging data. From Top to Bottom: an inclination between 10 and 90 $\deg $ is assumed, then inclinations of 0, 45, and 90 degrees. }
    \label{fig:HD14082B_sup_GaiaPMEX_DI_inclinaison}
\end{figure}

As for AB Pic, we combined the direct imaging detection limits with the GaiaPMEX solutions to find the possible (sma, mass) solutions for the companion. We considered three different inclinations: 0, 45 and 90 degrees. The results are shown in Fig. \ref{fig:HD14082B_sup_GaiaPMEX_DI_inclinaison}. As expected, the solutions are much more constrained when the inclination is moderate, with a mass ranging between 8 and 20 \Mjup~and a sma between 3 and 5 au ($1\sigma$ values). For highly inclined orbits, the mass ranges between 8 and 40 \Mjup, and the sma between 3 and 7 au ($1\sigma$ values).
Given the distance of the star, the companion is well suited for direct detection with the VLTI/Gravity instrument.

\subsubsection{Additional companions around HD 14082 B}
\begin{figure*}[hbt]
    \centering
\includegraphics[width=60.mm,clip=True]{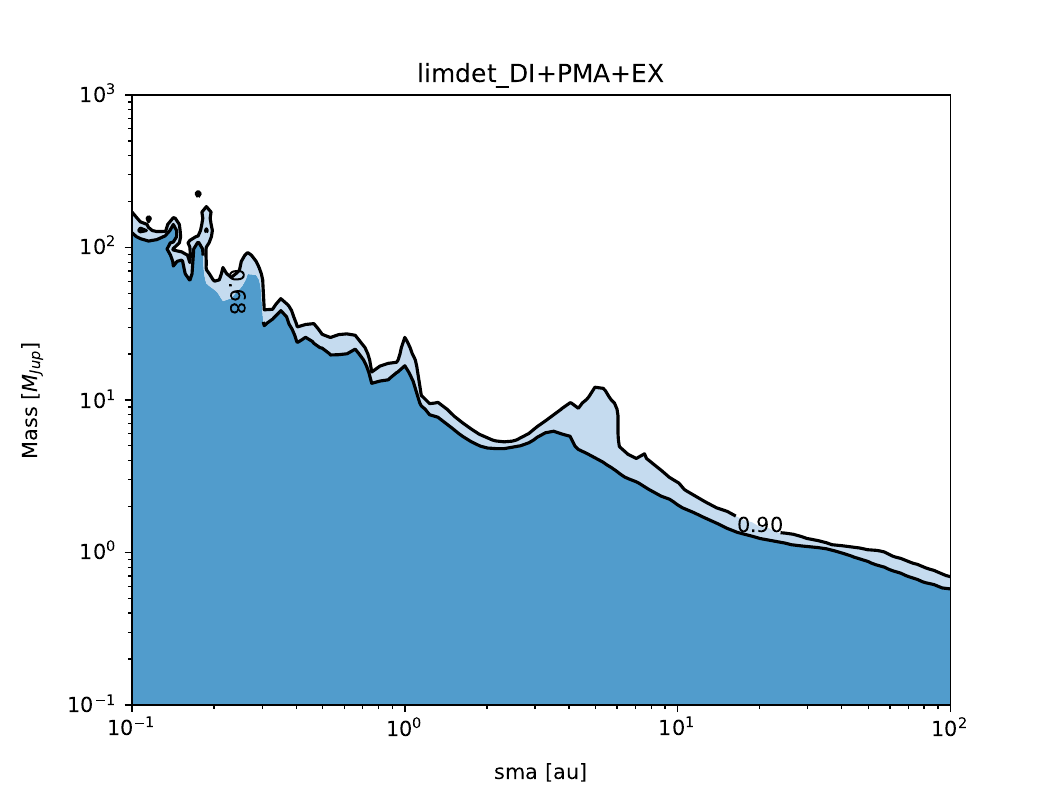} 
\includegraphics[width=60.mm,clip=True]{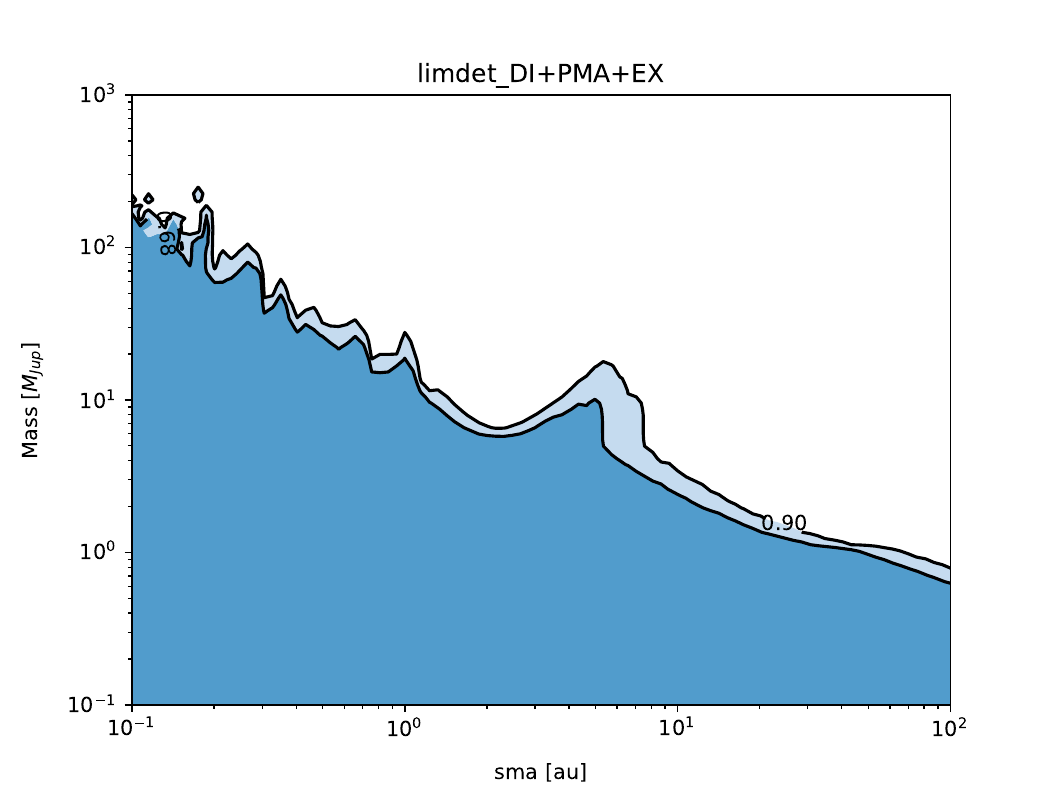} 
\includegraphics[width=60.mm,clip=True]{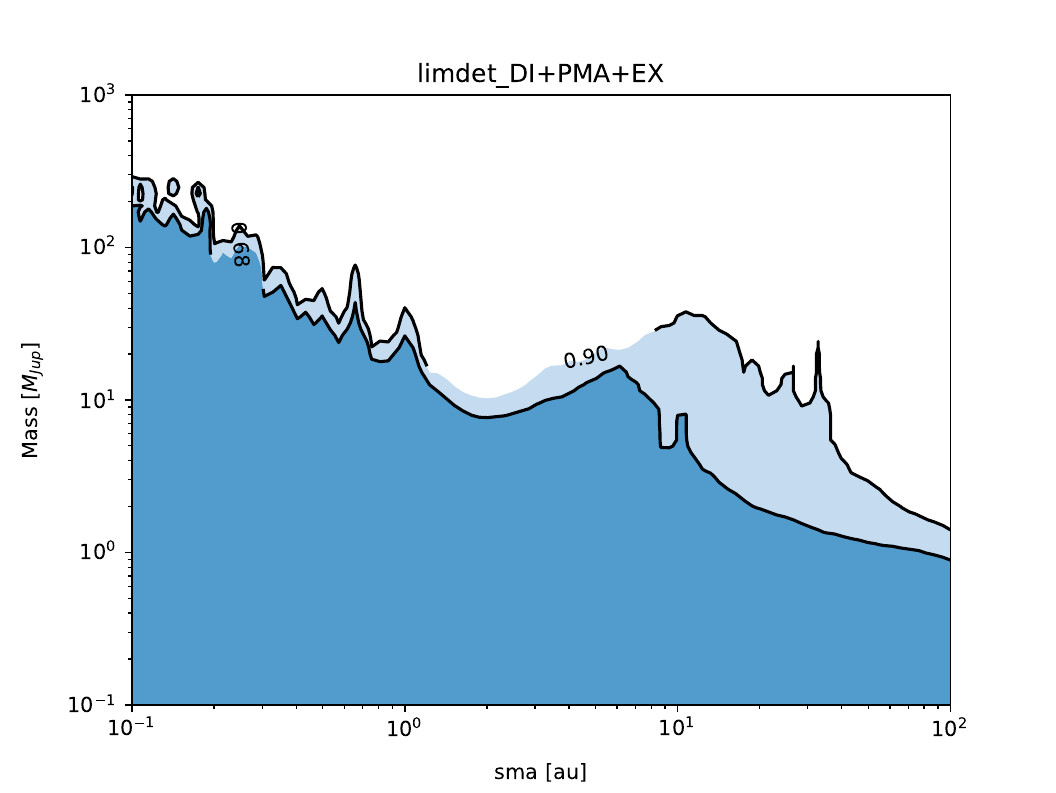} 
    \caption{From Left to Right: Detection limits for HD14082B combining GaiaPMEX and HCI data. From Left to Right: $i$ = 0, 45, and 90 degrees.}
  \label{fig:MESS3_HD14082B} 
\end{figure*}

Finally, we combine the \ruwe{} + PMa data with the direct imaging data to compute the upper limits in (sma, mass) of any additional companion to HD 14082B. We consider three inclinations: 0, 45, and 90 degrees. The results are provided in Fig. \ref{fig:MESS3_HD14082B}. As we do not have RV data, the (sma, mass) constraints are rather poor below 
a fraction of an au. At 1 au, we exclude additional companions more massive than 10 \Mjup~beyond 1 au and more massive than 2 \Mjup~at 10 au, except for edge-on orbits. For edge-on orbits, the corresponding masses are higher: 5 \Mjup~(68 $\%$
probability) and 30 \Mjup~(90 $\%$
probability) at 10 au. Adding high-contrast imaging data and RV data would help further constraining the properties of additional companions.

\subsubsection{Notes on the HD 14082 system}

Among the  confirmed exoplanets with semi-major axes in the ~3-10 au range, only three have been directly detected (HCI and/or interferometry): $\beta$ Pictoris b and c (parent star A5), and AF Lep b (parent star F8). Given its angular separation, HD 14082B b could be with SPHERE+ and with forthcoming imagers on the ELT. It may also be within the reach of the Gravity interferometer. Gravity data would in addition allow spectroscopic studies of its atmosphere.

Interestingly, HD14082B forms a binary system with HD14082A, located $\sim 500$ au away, opening up the opportunity to study planetary formation in binary systems. Finally, HD 14082B also hosts a debris disk detected through its Spitzer/MIPS infrared excess \citep{Sierchio14}, suggesting two belts at respectively 1 and
18 au \citep{Cotten16}, not detected in scattered light yet. The putative planet may therefore orbit between the
two belts, similarly to the giant planets between the asteroid and Kuiper belt in our Solar System, opening also the
perspective to study disk/planet interactions.

\section{Unconfirmed astrometric detections of planets}\label{sec:noplanets}	
\begin{figure*}[hbt]
    \centering
    \includegraphics[width=60.mm,clip=True]{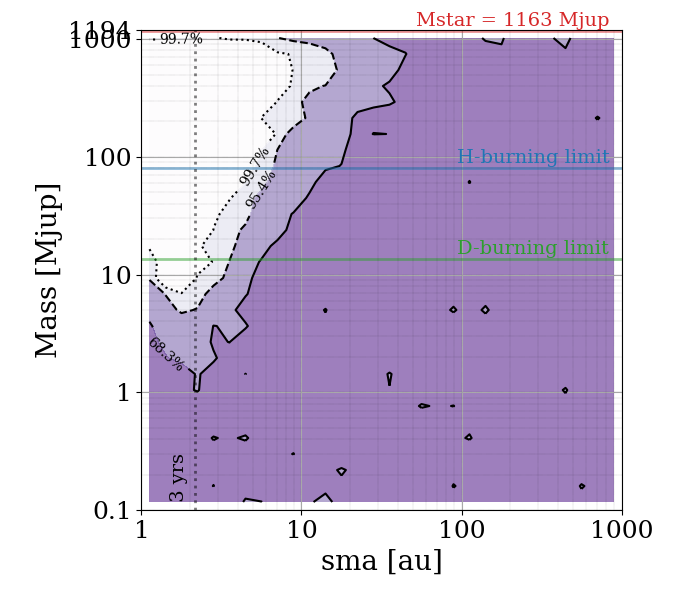} 
\includegraphics[width=60.mm,clip=True]{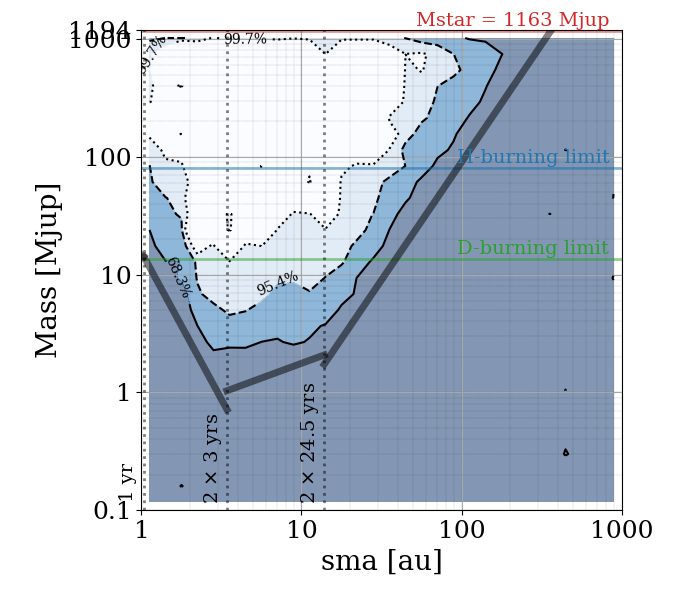} 
\includegraphics[width=60.mm,clip=True]{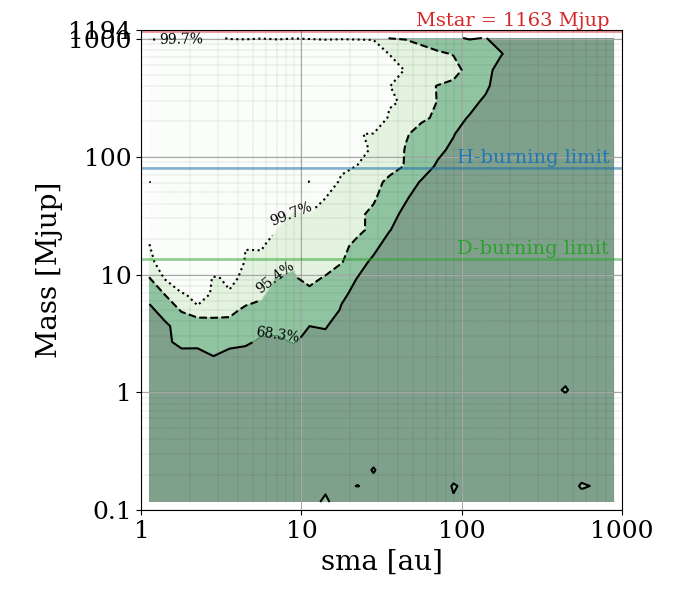} 
\includegraphics[width=60.mm,clip=True]{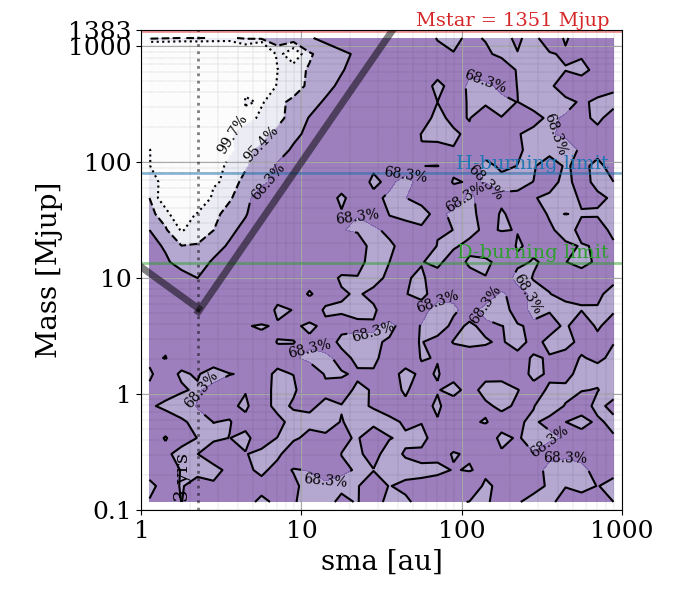} 
\includegraphics[width=60.mm,clip=True]{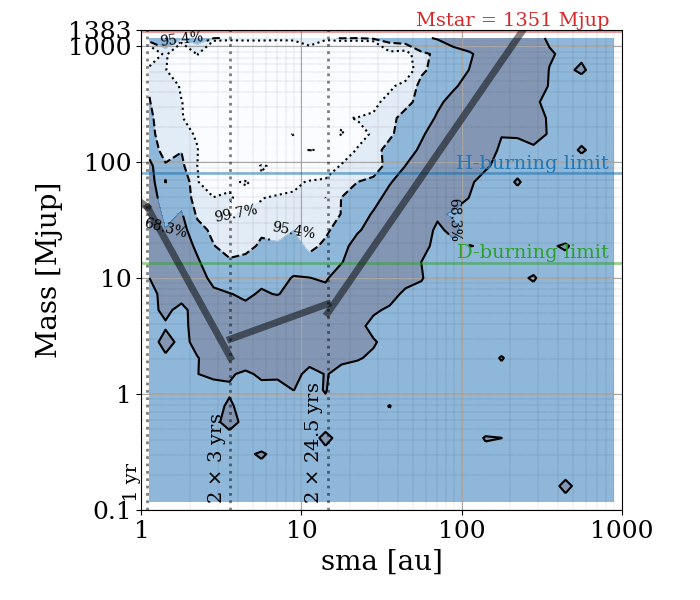} 
\includegraphics[width=60.mm,clip=True]{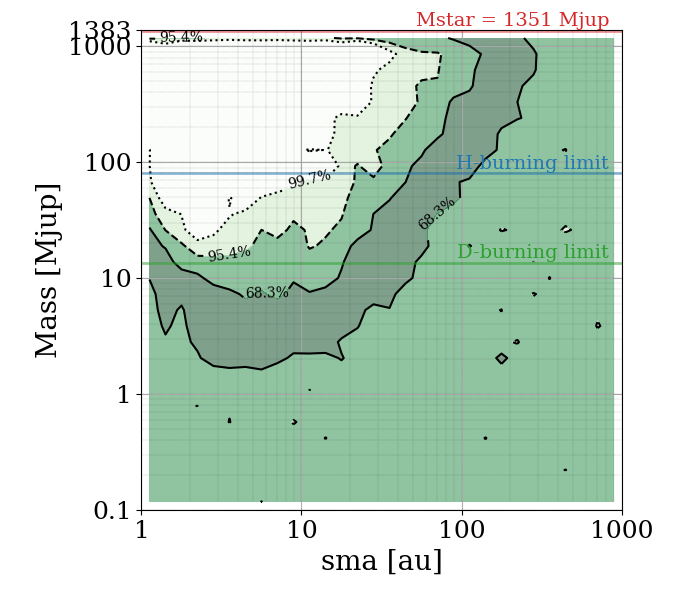} 
    \caption{GaiaPMEX results on  HD 45270 (top), and HD 164249 (bottom). From left to right: Gaia, PMA, and Gaia+PMa.}
  \label{fig:add_Gratton24} 
\end{figure*}

Recently, a list of 17 candidate planets based on a PMa computed by \citep{Kervella19} has been proposed, together with associated (sma, mass) solutions \citep{Gratton24}. In addition to AB Pic and HD 14082B that were discussed above, two candidates belong to the present sample: HIP 30314 (HD 45270) and HIP 88399 (HD 164249). 
The reported sma and masses are 
$13.28 \pm 13.93$ au and $0.0011 \pm 0.0004$
\Msun~(= $1.2 \pm 0.4$ \Mjup) for HIP 30314b, and $7.02 \pm 3.39$ au and $0.0043 \pm 0.0026$ \Msun~(= $4.5 \pm 2.7$ \Mjup) for HIP 88399b. No quantitative information is available, though, on how these numbers were determined. 
Based on our estimation of the noise budget in the measured PMa and RUWE, we rule out  any evidence of companion in the HIP 30314 system, as the \nsigmaruwe{} and \nsigmapma{}
are respectively 0.4 and 1.2, hence below 2$\sigma$. For HIP 88399, we find a  \nsigmaruwe{} of 1.4 and a \nsigmapma{}
 of 1.7, hence below 2$\sigma$, hence again below our detection threshold. Gaia and Hipparcos data do not show sufficient evidence for a companion in this system. We show in Figure~\ref{fig:add_Gratton24} the detection limits  obtained with GaiaPMEX on these targets. We attribute the discrepancy between our negative detection results and the positive detection of \citep{Gratton24} to the fact that unlike in \citet{Kervella19}, we take into account the instrumental noise when computing the PMa. These examples highlight the importance of estimating and taking into account instrumental noise when dealing with the PMa or the Gaia excess. Note that this need will remain in the context of Gaia DR4.

\section{Conclusion}\label{sec:conclusion}
We have searched for companions around a sample of more than 300  stars members of young close associations or moving groups (BPC, OCT, ECH, TWA, ARG, ABDor, COL, THA), based on available absolute astrometry data: Gaia alone (\ruwe{}), or Gaia and Hipparcos data (\ruwe    + PMa). We showed that large \ruwe{}    cannot be straightforwardly attributed to the presence of companions for some members of the two youngest associations, ECH and TWA. These specific targets were not analyzed with GaiaPMEX. 

Detection limits in terms of (mass, sma) have been derived for each target. This information will allow for optimizing the target samples in future surveys dedicated to the search for low mass companions, and avoiding spending observing time on targets for which absolute astrometry already tells that no companion is present within the instrument sensitivity domain. 

We computed the (sma, mass) solutions for all binaries. For each target with a positive detection, degeneracies are observed, so that the companions may be either stars, or stars or BDs, or stars or BD or planets (referred to as "planetary mass candidates"). 

The sample was built for a spectroscopic binary search. It allows therefore to compare the respective merits of spectroscopy and absolute astrometry for stellar binary searches. It turns out that absolute astrometry sensitivity outperforms spectroscopy for separations greater than 0.1 au, while below, spectroscopy outperforms absolute astrometry. Such techniques are then very complementary to get full and (almost) unbiased census of stellar binarity below 10 au.
We thus derive minimum unbiased binary rates for sma of different values.

Ten of the 13 planetary-mass candidates found using both \ruwe{}    and PMa in BPC, OCT, ARG, ABDor, COL, THA, happen to be in fact stellar companions or known BD/planet companions, based on  SPHERE imaging data, and/or literature. One target need additional observations to conclude on the nature of its companion. For the last three  systems, G80-21, AB Pic and HD 14082B, we show that the companions are definitely planetary/BD objects. We provide quantitative information on their possible (sma, mass). G80-21 has a 4-8 \Mjup companion with an sma in the range 0.5-1 au. Noticeably, conversely to AB Pic and HD 14082B, which show no significant \aastroruwe{}, but a significant PMa, G80-21 does not show a significant PMa, but a significant \aastroruwe{}. AB Pic c has a mass ranging between 2.5 and 9 \Mjup, and a sma between 2.5 and 6 au. The mass range can be refined according to the inclination of the orbital plane with respect to the line of sight. AB Pic c being less massive and the outer AB Pic b tend to favor a gravitational instability or gravitational collapse for the origin of AB Pic b. For HD 14082B, the mass varies between 8 and 20 \Mjup, and the sma between 3 and 5 au in case of moderate inclinations. Higher masses and sma are however possible in case of close-to-edge-on configurations.
With separations below 100 mas, the companions to G80-21, AB Pic, and HD 14082B cannot be detected with current high contrast imagers, but they could be within the reach of the future ELT imagers. In addition, AB Pic c and HD 14082B b could be within the reach of the VLTI/Gravity instrument.

From a methodological point of view, this work highlights the impact of the use of Gaia data to precise the (sma, mass)  parameters, compared to the sole use of Gaia-Hipparcos PMa, and the need for additional data (RV, imaging), to lift the degeneracies in the absolute astrometry solutions. We anticipate that the need for such additional data will remain even with the DR4 outputs, in case of long-period companions. This work also shows the important impact of the inclination of the orbit of the companion on the (sma, mass) solutions. This has to be taken into account to optimize the direct detection of these companions. 

Finally, we do not confirm the presence of a planetary-mass companion recently announced around HD 45270, and show that the companion announced around HD 164249 is below the $2\sigma$ detection threshold of GaiaPMEX. These examples highlight the importance of the instrumental noise in the analysis of the Gaia-Hipparcos PMa.

\begin{acknowledgements}
We thank the anonymous referee for her/his comments. This work has made use of data from the European Space Agency (ESA)
mission {\it Gaia} (\url{https://www.cosmos.esa.int/gaia}), processed by
the {\it Gaia} Data Processing and Analysis Consortium (DPAC,
\url{https://www.cosmos.esa.int/web/gaia/dpac/consortium}). Funding
for the DPAC has been provided by national institutions, in particular
the institutions participating in the {\it Gaia} Multilateral Agreement. This project has received funding from the European Research Council (ERC) under the European Union's Horizon 2020 research and innovation programme (COBREX; grant agreement 885593). A.Z. acknowledges support from ANID -- Millennium Science Initiative Program -- Center Code NCN2021\_080. 
This work has made use of the High Contrast Data Centre, jointly operated by OSUG/IPAG (Grenoble), PYTHEAS/LAM/CeSAM (Marseille), OCA/Lagrange (Nice), Observatoire de Paris/LESIA (Paris), and Observatoire de Lyon/CRAL, and supported by a grant from Labex OSUG@2020 (Investissements d’avenir – ANR10 LABX56). This work made use of  the European Space Agency space mission Gaia, and of the SIMBAD and VizieR CDS facilities.  
Last, AML thanks D. Segransan,  C. Mordasini and O. Schid for fruitful discussions. 
\end{acknowledgements}

\bibliographystyle{aa}
\bibliography{main_biblio}

\begin{thebibliography}{113}
\expandafter\ifx\csname natexlab\endcsname\relax\def\natexlab#1{#1}\fi

\bibitem[{Fra(2023)}]{Franson23_pztel}
 2023, \aj, 165, 246

\bibitem[{{Alcal{\'a}} {et~al.}(2021){Alcal{\'a}}, {Gangi}, {Biazzo},
  {Antoniucci}, {Frasca}, {Giannini}, {Munari}, {Nisini}, {Harutyunyan},
  {Manara}, \& {Vitali}}]{Alcala21}
{Alcal{\'a}}, J.~M., {Gangi}, M., {Biazzo}, K., {et~al.} 2021, \aap, 652, A72

\bibitem[{{Bell} {et~al.}(2015){Bell}, {Mamajek}, \& {Naylor}}]{Bell15}
{Bell}, C. P.~M., {Mamajek}, E.~E., \& {Naylor}, T. 2015, \mnras, 454, 593

\bibitem[{{Berdyugina} \& {J{\"a}rvinen}(2005)}]{Berdyugina05}
{Berdyugina}, S.~V. \& {J{\"a}rvinen}, S.~P. 2005, Astronomische Nachrichten,
  326, 283

\bibitem[{{Beust} \& {Dutrey}(2006)}]{Beust2006}
{Beust}, H. \& {Dutrey}, A. 2006, \aap, 446, 137

\bibitem[{{Beuzit} {et~al.}(2019){Beuzit}, {Vigan}, {Mouillet}, {Dohlen},
  {Gratton}, {Boccaletti}, {Sauvage}, {Schmid}, {Langlois}, {Petit},
  {Baruffolo}, {Feldt}, {Milli}, {Wahhaj}, {Abe}, {Anselmi}, {Antichi},
  {Barette}, {Baudrand}, {Baudoz}, {Bazzon}, {Bernardi}, {Blanchard}, {Brast},
  {Bruno}, {Buey}, {Carbillet}, {Carle}, {Cascone}, {Chapron}, {Charton},
  {Chauvin}, {Claudi}, {Costille}, {De Caprio}, {de Boer}, {Delboulb{\'e}},
  {Desidera}, {Dominik}, {Downing}, {Dupuis}, {Fabron}, {Fantinel}, {Farisato},
  {Feautrier}, {Fedrigo}, {Fusco}, {Gigan}, {Ginski}, {Girard}, {Giro},
  {Gisler}, {Gluck}, {Gry}, {Henning}, {Hubin}, {Hugot}, {Incorvaia}, {Jaquet},
  {Kasper}, {Lagadec}, {Lagrange}, {Le Coroller}, {Le Mignant}, {Le Ruyet},
  {Lessio}, {Lizon}, {Llored}, {Lundin}, {Madec}, {Magnard}, {Marteaud},
  {Martinez}, {Maurel}, {M{\'e}nard}, {Mesa}, {M{\"o}ller-Nilsson}, {Moulin},
  {Moutou}, {Orign{\'e}}, {Parisot}, {Pavlov}, {Perret}, {Pragt}, {Puget},
  {Rabou}, {Ramos}, {Reess}, {Rigal}, {Rochat}, {Roelfsema}, {Rousset}, {Roux},
  {Saisse}, {Salasnich}, {Santambrogio}, {Scuderi}, {Segransan}, {Sevin},
  {Siebenmorgen}, {Soenke}, {Stadler}, {Suarez}, {Tiph{\`e}ne}, {Turatto},
  {Udry}, {Vakili}, {Waters}, {Weber}, {Wildi}, {Zins}, \& {Zurlo}}]{beuzit19}
{Beuzit}, J.~L., {Vigan}, A., {Mouillet}, D., {et~al.} 2019, \aap, 631, A155

\bibitem[{{Biller} {et~al.}(2010){Biller}, {Liu}, {Wahhaj}, {Nielsen}, {Close},
  {Dupuy}, {Hayward}, {Burrows}, {Chun}, {Ftaclas}, {Clarke}, {Hartung},
  {Males}, {Reid}, {Shkolnik}, {Skemer}, {Tecza}, {Thatte}, {Alencar},
  {Artymowicz}, {Boss}, {de Gouveia Dal Pino}, {Gregorio-Hetem}, {Ida},
  {Kuchner}, {Lin}, \& {Toomey}}]{Biller10}
{Biller}, B.~A., {Liu}, M.~C., {Wahhaj}, Z., {et~al.} 2010, \apjl, 720, L82

\bibitem[{{Bonavita} {et~al.}(2022){Bonavita}, {Gratton}, {Desidera},
  {Squicciarini}, {D'Orazi}, {Zurlo}, {Biller}, {Chauvin}, {Fontanive},
  {Janson}, {Messina}, {Menard}, {Meyer}, {Vigan}, {Avenhaus}, {Asensio
  Torres}, {Beuzit}, {Boccaletti}, {Bonnefoy}, {Brandner}, {Cantalloube},
  {Cheetham}, {Cudel}, {Daemgen}, {Delorme}, {Desgrange}, {Dominik}, {Engler},
  {Feautrier}, {Feldt}, {Galicher}, {Garufi}, {Gasparri}, {Ginski}, {Girard},
  {Grandjean}, {Hagelberg}, {Henning}, {Hunziker}, {Kasper}, {Keppler},
  {Lagadec}, {Lagrange}, {Langlois}, {Lannier}, {Lazzoni}, {Le Coroller},
  {Ligi}, {Lombart}, {Maire}, {Mazevet}, {Mesa}, {Mouillet}, {Moutou},
  {M{\"u}ller}, {Peretti}, {Perrot}, {Petrus}, {Potier}, {Ramos}, {Rickman},
  {Rouan}, {Salter}, {Samland}, {Schmidt}, {Sissa}, {Stolker}, {Szul{\'a}gyi},
  {Turatto}, {Udry}, \& {Wildi}}]{Bonavita22}
{Bonavita}, M., {Gratton}, R., {Desidera}, S., {et~al.} 2022, \aap, 663, A144

\bibitem[{{Booth} {et~al.}(2021){Booth}, {del Burgo}, \& {Hambaryan}}]{Booth21}
{Booth}, M., {del Burgo}, C., \& {Hambaryan}, V.~V. 2021, \mnras, 500, 5552

\bibitem[{{Brandt}(2021{\natexlab{a}})}]{Brandt2021}
{Brandt}, T.~D. 2021{\natexlab{a}}, \apjs, 254, 42

\bibitem[{{Brandt}(2021{\natexlab{b}})}]{Brandt21}
{Brandt}, T.~D. 2021{\natexlab{b}}, \apjs, 254, 42

\bibitem[{{Chauvin} {et~al.}(2010){Chauvin}, {Lagrange}, {Bonavita},
  {Zuckerman}, {Dumas}, {Bessell}, {Beuzit}, {Bonnefoy}, {Desidera}, {Farihi},
  {Lowrance}, {Mouillet}, \& {Song}}]{Chauvin10}
{Chauvin}, G., {Lagrange}, A.~M., {Bonavita}, M., {et~al.} 2010, \aap, 509, A52

\bibitem[{{Chauvin} {et~al.}(2005){Chauvin}, {Lagrange}, {Zuckerman}, {Dumas},
  {Mouillet}, {Song}, {Beuzit}, {Lowrance}, \& {Bessell}}]{Chauvin05}
{Chauvin}, G., {Lagrange}, A.~M., {Zuckerman}, B., {et~al.} 2005, \aap, 438,
  L29

\bibitem[{{Chauvin} {et~al.}(2015){Chauvin}, {Vigan}, {Bonnefoy}, {Desidera},
  {Bonavita}, {Mesa}, {Boccaletti}, {Buenzli}, {Carson}, {Delorme},
  {Hagelberg}, {Montagnier}, {Mordasini}, {Quanz}, {Segransan}, {Thalmann},
  {Beuzit}, {Biller}, {Covino}, {Feldt}, {Girard}, {Gratton}, {Henning},
  {Kasper}, {Lagrange}, {Messina}, {Meyer}, {Mouillet}, {Moutou}, {Reggiani},
  {Schlieder}, \& {Zurlo}}]{Chauvin15}
{Chauvin}, G., {Vigan}, A., {Bonnefoy}, M., {et~al.} 2015, \aap, 573, A127

\bibitem[{{Chomez} {et~al.}(2023){Chomez}, {Lagrange}, {Delorme}, {Langlois},
  {Chauvin}, {Flasseur}, {Dallant}, {Philipot}, {Bergeon}, {Albert}, {Meunier},
  \& {Rubini}}]{Chomez23}
{Chomez}, A., {Lagrange}, A.~M., {Delorme}, P., {et~al.} 2023, \aap, 675, A205

\bibitem[{{Cotten} \& {Song}(2016)}]{Cotten16}
{Cotten}, T.~H. \& {Song}, I. 2016, \apjs, 225, 15

\bibitem[{{Cutispoto} {et~al.}(2002){Cutispoto}, {Pastori}, {Pasquini}, {de
  Medeiros}, {Tagliaferri}, \& {Andersen}}]{Cutispoto02}
{Cutispoto}, G., {Pastori}, L., {Pasquini}, L., {et~al.} 2002, \aap, 384, 491

\bibitem[{{Cutri} {et~al.}(2021){Cutri}, {Wright}, {Conrow}, {Fowler},
  {Eisenhardt}, {Grillmair}, {Kirkpatrick}, {Masci}, {McCallon}, {Wheelock},
  {Fajardo-Acosta}, {Yan}, {Benford}, {Harbut}, {Jarrett}, {Lake}, {Leisawitz},
  {Ressler}, {Stanford}, {Tsai}, {Liu}, {Helou}, {Mainzer}, {Gettngs},
  {Gonzalez}, {Hoffman}, {Marsh}, {Padgett}, {Skrutskie}, {Beck}, {Papin}, \&
  {Wittman}}]{wise}
{Cutri}, R.~M., {Wright}, E.~L., {Conrow}, T., {et~al.} 2021, VizieR Online
  Data Catalog, II/328

\bibitem[{{Dahlqvist} {et~al.}(2022){Dahlqvist}, {Milli}, {Absil},
  {Cantalloube}, {Matra}, {Choquet}, {del Burgo}, {Marshall}, {Wyatt}, \&
  {Ertel}}]{Dahlqvist22}
{Dahlqvist}, C.~H., {Milli}, J., {Absil}, O., {et~al.} 2022, \aap, 666, A33

\bibitem[{{De Rosa} {et~al.}(2023){De Rosa}, {Nielsen}, {Wahhaj}, {Ruffio},
  {Kalas}, {Peck}, {Hirsch}, \& {Roberson}}]{deRosa23}
{De Rosa}, R.~J., {Nielsen}, E.~L., {Wahhaj}, Z., {et~al.} 2023, \aap, 672, A94

\bibitem[{{Dickson-Vandervelde} {et~al.}(2021){Dickson-Vandervelde}, {Wilson},
  \& {Kastner}}]{Dickson-Vandervelde21}
{Dickson-Vandervelde}, D.~A., {Wilson}, E.~C., \& {Kastner}, J.~H. 2021, \aj,
  161, 87

\bibitem[{{Elliott} \& {Bayo}(2016)}]{Elliott-Bayo16}
{Elliott}, P. \& {Bayo}, A. 2016, \mnras, 459, 4499

\bibitem[{{Elliott} {et~al.}(2014){Elliott}, {Bayo}, {Melo}, {Torres},
  {Sterzik}, \& {Quast}}]{Elliott14}
{Elliott}, P., {Bayo}, A., {Melo}, C.~H.~F., {et~al.} 2014, \aap, 568, A26

\bibitem[{{Elliott} {et~al.}(2016){Elliott}, {Bayo}, {Melo}, {Torres},
  {Sterzik}, {Quast}, {Montes}, \& {Brahm}}]{Elliott16}
{Elliott}, P., {Bayo}, A., {Melo}, C.~H.~F., {et~al.} 2016, \aap, 590, A13

\bibitem[{Flasseur {et~al.}(2020{\natexlab{a}})Flasseur, Denis, Thi{\'e}baut,
  \& Langlois}]{flasseur2020robustness}
Flasseur, O., Denis, L., Thi{\'e}baut, {\'E}., \& Langlois, M.
  2020{\natexlab{a}}, Astronomy \& Astrophysics, 634, A2

\bibitem[{Flasseur {et~al.}(2020{\natexlab{b}})Flasseur, Denis, Thiébaut, \&
  Langlois}]{flasseur_paco_2020}
Flasseur, O., Denis, L., Thiébaut, E., \& Langlois, M. 2020{\natexlab{b}},
  Astronomy \& Astrophysics, 637, A9

\bibitem[{{Ford} {et~al.}(2000){Ford}, {Kozinsky}, \& {Rasio}}]{Ford2000}
{Ford}, E.~B., {Kozinsky}, B., \& {Rasio}, F.~A. 2000, \apj, 535, 385

\bibitem[{{Franson} {et~al.}(2023){Franson}, {Bowler}, {Zhou}, {Pearce},
  {Bardalez Gagliuffi}, {Biddle}, {Brandt}, {Crepp}, {Dupuy}, {Faherty},
  {Jensen-Clem}, {Morgan}, {Sanghi}, {Theissen}, {Tran}, \&
  {Wolf}}]{Franson2023_aflep}
{Franson}, K., {Bowler}, B.~P., {Zhou}, Y., {et~al.} 2023, arXiv e-prints,
  arXiv:2302.05420

\bibitem[{{Gagn{\'e}} {et~al.}(2018){Gagn{\'e}}, {Mamajek}, {Malo}, {Riedel},
  {Rodriguez}, {Lafreni{\`e}re}, {Faherty}, {Roy-Loubier}, {Pueyo}, {Robin}, \&
  {Doyon}}]{Gagne18}
{Gagn{\'e}}, J., {Mamajek}, E.~E., {Malo}, L., {et~al.} 2018, \apj, 856, 23

\bibitem[{{Gaia Collaboration} {et~al.}(2023{\natexlab{a}}){Gaia
  Collaboration}, {Arenou}, {Babusiaux}, {Barstow}, {Faigler}, {Jorissen},
  {Kervella}, {Mazeh}, {Mowlavi}, {Panuzzo}, {Sahlmann}, {Shahaf}, {Sozzetti},
  {Bauchet}, {Damerdji}, {Gavras}, {Giacobbe}, {Gosset}, {Halbwachs}, {Holl},
  {Lattanzi}, {Leclerc}, {Morel}, {Pourbaix}, {Re Fiorentin}, {Sadowski},
  {S{\'e}gransan}, {Siopis}, {Teyssier}, {Zwitter}, {Planquart}, {Brown},
  {Vallenari}, {Prusti}, {de Bruijne}, {Biermann}, {Creevey}, {Ducourant},
  {Evans}, {Eyer}, {Guerra}, {Hutton}, {Jordi}, {Klioner}, {Lammers},
  {Lindegren}, {Luri}, {Mignard}, {Panem}, {Randich}, {Sartoretti}, {Soubiran},
  {Tanga}, {Walton}, {Bailer-Jones}, {Bastian}, {Drimmel}, {Jansen}, {Katz},
  {van Leeuwen}, {Bakker}, {Cacciari}, {Casta{\~n}eda}, {De Angeli},
  {Fabricius}, {Fouesneau}, {Fr{\'e}mat}, {Galluccio}, {Guerrier}, {Heiter},
  {Masana}, {Messineo}, {Nicolas}, {Nienartowicz}, {Pailler}, {Riclet}, {Roux},
  {Seabroke}, {Sordo}, {Th{\'e}venin}, {Gracia-Abril}, {Portell}, {Altmann},
  {Andrae}, {Audard}, {Bellas-Velidis}, {Benson}, {Berthier}, {Blomme},
  {Burgess}, {Busonero}, {Busso}, {C{\'a}novas}, {Carry}, {Cellino}, {Cheek},
  {Clementini}, {Davidson}, {de Teodoro}, {Nu{\~n}ez Campos}, {Delchambre},
  {Dell'Oro}, {Esquej}, {Fern{\'a}ndez-Hern{\'a}ndez}, {Fraile}, {Garabato},
  {Garc{\'\i}a-Lario}, {Haigron}, {Hambly}, {Harrison}, {Hern{\'a}ndez},
  {Hestroffer}, {Hodgkin}, {Jan{\ss}en}, {Jevardat de Fombelle}, {Jordan},
  {Krone-Martins}, {Lanzafame}, {L{\"o}ffler}, {Marchal}, {Marrese},
  {Moitinho}, {Muinonen}, {Osborne}, {Pancino}, {Pauwels}, {Recio-Blanco},
  {Reyl{\'e}}, {Riello}, {Rimoldini}, {Roegiers}, {Rybizki}, {Sarro}, {Smith},
  {Utrilla}, {van Leeuwen}, {Abbas}, {{\'A}brah{\'a}m}, {Abreu Aramburu},
  {Aerts}, {Aguado}, {Ajaj}, {Aldea-Montero}, {Altavilla}, {{\'A}lvarez},
  {Alves}, {Anders}, {Anderson}, {Anglada Varela}, {Antoja}, {Baines}, {Baker},
  {Balaguer-N{\'u}{\~n}ez}, {Balbinot}, {Balog}, {Barache}, {Barbato},
  {Barros}, {Bartolom{\'e}}, {Bassilana}, {Becciani}, {Bellazzini},
  {Berihuete}, {Bernet}, {Bertone}, {Bianchi}, {Binnenfeld}, {Blanco-Cuaresma},
  {Blazere}, {Boch}, {Bombrun}, {Bossini}, {Bouquillon}, {Bragaglia},
  {Bramante}, {Breedt}, {Bressan}, {Brouillet}, {Brugaletta}, {Bucciarelli},
  {Burlacu}, {Butkevich}, {Buzzi}, {Caffau}, {Cancelliere}, {Cantat-Gaudin},
  {Carballo}, {Carlucci}, {Carnerero}, {Carrasco}, {Casamiquela}, {Castellani},
  {Castro-Ginard}, {Chaoul}, {Charlot}, {Chemin}, {Chiaramida}, {Chiavassa},
  {Chornay}, {Comoretto}, {Contursi}, {Cooper}, {Cornez}, {Cowell}, {Crifo},
  {Cropper}, {Crosta}, {Crowley}, {Dafonte}, {Dapergolas}, {David}, {de
  Laverny}, {De Luise}, {De March}, {De Ridder}, {de Souza}, {de Torres}, {del
  Peloso}, {del Pozo}, {Delbo}, {Delgado}, {Delisle}, {Demouchy},
  {Dharmawardena}, {Diakite}, {Diener}, {Distefano}, {Dolding}, {Enke},
  {Fabre}, {Fabrizio}, {Fedorets}, {Fernique}, {Figueras}, {Fournier},
  {Fouron}, {Fragkoudi}, {Gai}, {Garcia-Gutierrez}, {Garcia-Reinaldos},
  {Garc{\'\i}a-Torres}, {Garofalo}, {Gavel}, {Gerlach}, {Geyer}, {Gilmore},
  {Girona}, {Giuffrida}, {Gomel}, {Gomez}, {Gonz{\'a}lez-N{\'u}{\~n}ez},
  {Gonz{\'a}lez-Santamar{\'\i}a}, {Gonz{\'a}lez-Vidal}, {Granvik}, {Guillout},
  {Guiraud}, {Guti{\'e}rrez-S{\'a}nchez}, {Guy}, {Hatzidimitriou}, {Hauser},
  {Haywood}, {Helmer}, {Helmi}, {Sarmiento}, {Hidalgo}, {Hilger},
  {H{\l}adczuk}, {Hobbs}, {Holland}, {Huckle}, {Jardine}, {Jasniewicz},
  {Jean-Antoine Piccolo}, {Jim{\'e}nez-Arranz}, {Juaristi Campillo}, {Julbe},
  {Karbevska}, {Khanna}, {Kordopatis}, {Korn}, {K{\'o}sp{\'a}l},
  {Kostrzewa-Rutkowska}, {Kruszy{\'n}ska}, {Kun}, {Laizeau}, {Lambert},
  {Lanza}, {Lasne}, {Le Campion}, {Lebreton}, {Lebzelter}, {Leccia},
  {Lecoeur-Taibi}, {Liao}, {Licata}, {Lindstr{\o}m}, {Lister}, {Livanou},
  {Lobel}, {Lorca}, {Loup}, {Madrero Pardo}, {Magdaleno Romeo}, {Managau},
  {Mann}, {Manteiga}, {Marchant}, {Marconi}, {Marcos}, {Marcos Santos},
  {Mar{\'\i}n Pina}, {Marinoni}, {Marocco}, {Marshall}, {Martin Polo},
  {Mart{\'\i}n-Fleitas}, {Marton}, {Mary}, {Masip}, {Massari},
  {Mastrobuono-Battisti}, {McMillan}, {Messina}, {Michalik}, {Millar}, {Mints},
  {Molina}, {Molinaro}, {Moln{\'a}r}, {Monari}, {Mongui{\'o}}, {Montegriffo},
  {Montero}, {Mor}, {Mora}, {Morbidelli}, {Morris}, {Muraveva}, {Murphy},
  {Musella}, {Nagy}, {Noval}, {Oca{\~n}a}, {Ogden}, {Ordenovic}, {Osinde},
  {Pagani}, {Pagano}, {Palaversa}, {Palicio}, {Pallas-Quintela}, {Panahi},
  {Payne-Wardenaar}, {Pe{\~n}alosa Esteller}, {Penttil{\"a}}, {Pichon},
  {Piersimoni}, {Pineau}, {Plachy}, {Plum}, {Poggio}, {Pr{\v{s}}a}, {Pulone},
  {Racero}, {Ragaini}, {Rainer}, {Raiteri}, {Ramos}, {Ramos-Lerate}, {Regibo},
  {Richards}, {Rios Diaz}, {Ripepi}, {Riva}, {Rix}, {Rixon}, {Robichon},
  {Robin}, {Robin}, {Roelens}, {Rogues}, {Rohrbasser}, {Romero-G{\'o}mez},
  {Rowell}, {Royer}, {Ruz Mieres}, {Rybicki}, {S{\'a}ez N{\'u}{\~n}ez},
  {Sagrist{\`a} Sell{\'e}s}, {Salguero}, {Samaras}, {Sanchez Gimenez}, {Sanna},
  {Santove{\~n}a}, {Sarasso}, {Schultheis}, {Sciacca}, {Segol}, {Segovia},
  {Semeux}, {Siddiqui}, {Siebert}, {Siltala}, {Silvelo}, {Slezak}, {Slezak},
  {Smart}, {Snaith}, {Solano}, {Solitro}, {Souami}, {Souchay}, {Spagna},
  {Spina}, {Spoto}, {Steele}, {Steidelm{\"u}ller}, {Stephenson}, {S{\"u}veges},
  {Surdej}, {Szabados}, {Szegedi-Elek}, {Taris}, {Taylor}, {Teixeira},
  {Tolomei}, {Tonello}, {Torra}, {Torra}, {Torralba Elipe}, {Trabucchi},
  {Tsounis}, {Turon}, {Ulla}, {Unger}, {Vaillant}, {van Dillen}, {van Reeven},
  {Vanel}, {Vecchiato}, {Viala}, {Vicente}, {Voutsinas}, {Weiler}, {Wevers},
  {Wyrzykowski}, {Yoldas}, {Yvard}, {Zhao}, {Zorec}, \& {Zucker}}]{Arenou2023}
{Gaia Collaboration}, {Arenou}, F., {Babusiaux}, C., {et~al.}
  2023{\natexlab{a}}, \aap, 674, A34

\bibitem[{{Gaia Collaboration} {et~al.}(2023{\natexlab{b}}){Gaia
  Collaboration}, {Vallenari}, {Brown}, {Prusti}, {de Bruijne}, {Arenou},
  {Babusiaux}, {Biermann}, {Creevey}, {Ducourant}, {Evans}, {Eyer}, {Guerra},
  {Hutton}, {Jordi}, {Klioner}, {Lammers}, {Lindegren}, {Luri}, {Mignard},
  {Panem}, {Pourbaix}, {Randich}, {Sartoretti}, {Soubiran}, {Tanga}, {Walton},
  {Bailer-Jones}, {Bastian}, {Drimmel}, {Jansen}, {Katz}, {Lattanzi}, {van
  Leeuwen}, {Bakker}, {Cacciari}, {Casta{\~n}eda}, {De Angeli}, {Fabricius},
  {Fouesneau}, {Fr{\'e}mat}, {Galluccio}, {Guerrier}, {Heiter}, {Masana},
  {Messineo}, {Mowlavi}, {Nicolas}, {Nienartowicz}, {Pailler}, {Panuzzo},
  {Riclet}, {Roux}, {Seabroke}, {Sordo}, {Th{\'e}venin}, {Gracia-Abril},
  {Portell}, {Teyssier}, {Altmann}, {Andrae}, {Audard}, {Bellas-Velidis},
  {Benson}, {Berthier}, {Blomme}, {Burgess}, {Busonero}, {Busso},
  {C{\'a}novas}, {Carry}, {Cellino}, {Cheek}, {Clementini}, {Damerdji},
  {Davidson}, {de Teodoro}, {Nu{\~n}ez Campos}, {Delchambre}, {Dell'Oro},
  {Esquej}, {Fern{\'a}ndez-Hern{\'a}ndez}, {Fraile}, {Garabato},
  {Garc{\'\i}a-Lario}, {Gosset}, {Haigron}, {Halbwachs}, {Hambly}, {Harrison},
  {Hern{\'a}ndez}, {Hestroffer}, {Hodgkin}, {Holl}, {Jan{\ss}en}, {Jevardat de
  Fombelle}, {Jordan}, {Krone-Martins}, {Lanzafame}, {L{\"o}ffler}, {Marchal},
  {Marrese}, {Moitinho}, {Muinonen}, {Osborne}, {Pancino}, {Pauwels},
  {Recio-Blanco}, {Reyl{\'e}}, {Riello}, {Rimoldini}, {Roegiers}, {Rybizki},
  {Sarro}, {Siopis}, {Smith}, {Sozzetti}, {Utrilla}, {van Leeuwen}, {Abbas},
  {{\'A}brah{\'a}m}, {Abreu Aramburu}, {Aerts}, {Aguado}, {Ajaj},
  {Aldea-Montero}, {Altavilla}, {{\'A}lvarez}, {Alves}, {Anders}, {Anderson},
  {Anglada Varela}, {Antoja}, {Baines}, {Baker}, {Balaguer-N{\'u}{\~n}ez},
  {Balbinot}, {Balog}, {Barache}, {Barbato}, {Barros}, {Barstow},
  {Bartolom{\'e}}, {Bassilana}, {Bauchet}, {Becciani}, {Bellazzini},
  {Berihuete}, {Bernet}, {Bertone}, {Bianchi}, {Binnenfeld}, {Blanco-Cuaresma},
  {Blazere}, {Boch}, {Bombrun}, {Bossini}, {Bouquillon}, {Bragaglia},
  {Bramante}, {Breedt}, {Bressan}, {Brouillet}, {Brugaletta}, {Bucciarelli},
  {Burlacu}, {Butkevich}, {Buzzi}, {Caffau}, {Cancelliere}, {Cantat-Gaudin},
  {Carballo}, {Carlucci}, {Carnerero}, {Carrasco}, {Casamiquela}, {Castellani},
  {Castro-Ginard}, {Chaoul}, {Charlot}, {Chemin}, {Chiaramida}, {Chiavassa},
  {Chornay}, {Comoretto}, {Contursi}, {Cooper}, {Cornez}, {Cowell}, {Crifo},
  {Cropper}, {Crosta}, {Crowley}, {Dafonte}, {Dapergolas}, {David}, {David},
  {de Laverny}, {De Luise}, {De March}, {De Ridder}, {de Souza}, {de Torres},
  {del Peloso}, {del Pozo}, {Delbo}, {Delgado}, {Delisle}, {Demouchy},
  {Dharmawardena}, {Di Matteo}, {Diakite}, {Diener}, {Distefano}, {Dolding},
  {Edvardsson}, {Enke}, {Fabre}, {Fabrizio}, {Faigler}, {Fedorets}, {Fernique},
  {Fienga}, {Figueras}, {Fournier}, {Fouron}, {Fragkoudi}, {Gai},
  {Garcia-Gutierrez}, {Garcia-Reinaldos}, {Garc{\'\i}a-Torres}, {Garofalo},
  {Gavel}, {Gavras}, {Gerlach}, {Geyer}, {Giacobbe}, {Gilmore}, {Girona},
  {Giuffrida}, {Gomel}, {Gomez}, {Gonz{\'a}lez-N{\'u}{\~n}ez},
  {Gonz{\'a}lez-Santamar{\'\i}a}, {Gonz{\'a}lez-Vidal}, {Granvik}, {Guillout},
  {Guiraud}, {Guti{\'e}rrez-S{\'a}nchez}, {Guy}, {Hatzidimitriou}, {Hauser},
  {Haywood}, {Helmer}, {Helmi}, {Sarmiento}, {Hidalgo}, {Hilger},
  {H{\l}adczuk}, {Hobbs}, {Holland}, {Huckle}, {Jardine}, {Jasniewicz},
  {Jean-Antoine Piccolo}, {Jim{\'e}nez-Arranz}, {Jorissen}, {Juaristi
  Campillo}, {Julbe}, {Karbevska}, {Kervella}, {Khanna}, {Kontizas},
  {Kordopatis}, {Korn}, {K{\'o}sp{\'a}l}, {Kostrzewa-Rutkowska},
  {Kruszy{\'n}ska}, {Kun}, {Laizeau}, {Lambert}, {Lanza}, {Lasne}, {Le
  Campion}, {Lebreton}, {Lebzelter}, {Leccia}, {Leclerc}, {Lecoeur-Taibi},
  {Liao}, {Licata}, {Lindstr{\o}m}, {Lister}, {Livanou}, {Lobel}, {Lorca},
  {Loup}, {Madrero Pardo}, {Magdaleno Romeo}, {Managau}, {Mann}, {Manteiga},
  {Marchant}, {Marconi}, {Marcos}, {Marcos Santos}, {Mar{\'\i}n Pina},
  {Marinoni}, {Marocco}, {Marshall}, {Martin Polo}, {Mart{\'\i}n-Fleitas},
  {Marton}, {Mary}, {Masip}, {Massari}, {Mastrobuono-Battisti}, {Mazeh},
  {McMillan}, {Messina}, {Michalik}, {Millar}, {Mints}, {Molina}, {Molinaro},
  {Moln{\'a}r}, {Monari}, {Mongui{\'o}}, {Montegriffo}, {Montero}, {Mor},
  {Mora}, {Morbidelli}, {Morel}, {Morris}, {Muraveva}, {Murphy}, {Musella},
  {Nagy}, {Noval}, {Oca{\~n}a}, {Ogden}, {Ordenovic}, {Osinde}, {Pagani},
  {Pagano}, {Palaversa}, {Palicio}, {Pallas-Quintela}, {Panahi},
  {Payne-Wardenaar}, {Pe{\~n}alosa Esteller}, {Penttil{\"a}}, {Pichon},
  {Piersimoni}, {Pineau}, {Plachy}, {Plum}, {Poggio}, {Pr{\v{s}}a}, {Pulone},
  {Racero}, {Ragaini}, {Rainer}, {Raiteri}, {Rambaux}, {Ramos}, {Ramos-Lerate},
  {Re Fiorentin}, {Regibo}, {Richards}, {Rios Diaz}, {Ripepi}, {Riva}, {Rix},
  {Rixon}, {Robichon}, {Robin}, {Robin}, {Roelens}, {Rogues}, {Rohrbasser},
  {Romero-G{\'o}mez}, {Rowell}, {Royer}, {Ruz Mieres}, {Rybicki}, {Sadowski},
  {S{\'a}ez N{\'u}{\~n}ez}, {Sagrist{\`a} Sell{\'e}s}, {Sahlmann}, {Salguero},
  {Samaras}, {Sanchez Gimenez}, {Sanna}, {Santove{\~n}a}, {Sarasso},
  {Schultheis}, {Sciacca}, {Segol}, {Segovia}, {S{\'e}gransan}, {Semeux},
  {Shahaf}, {Siddiqui}, {Siebert}, {Siltala}, {Silvelo}, {Slezak}, {Slezak},
  {Smart}, {Snaith}, {Solano}, {Solitro}, {Souami}, {Souchay}, {Spagna},
  {Spina}, {Spoto}, {Steele}, {Steidelm{\"u}ller}, {Stephenson}, {S{\"u}veges},
  {Surdej}, {Szabados}, {Szegedi-Elek}, {Taris}, {Taylor}, {Teixeira},
  {Tolomei}, {Tonello}, {Torra}, {Torra}, {Torralba Elipe}, {Trabucchi},
  {Tsounis}, {Turon}, {Ulla}, {Unger}, {Vaillant}, {van Dillen}, {van Reeven},
  {Vanel}, {Vecchiato}, {Viala}, {Vicente}, {Voutsinas}, {Weiler}, {Wevers},
  {Wyrzykowski}, {Yoldas}, {Yvard}, {Zhao}, {Zorec}, {Zucker}, \&
  {Zwitter}}]{GaiaDR3}
{Gaia Collaboration}, {Vallenari}, A., {Brown}, A.~G.~A., {et~al.}
  2023{\natexlab{b}}, \aap, 674, A1

\bibitem[{{Galicher} {et~al.}(2018){Galicher}, {Boccaletti}, {Mesa}, {Delorme},
  {Gratton}, {Langlois}, {Lagrange}, {Maire}, {Le Coroller}, {Chauvin},
  {Biller}, {Cantalloube}, {Janson}, {Lagadec}, {Meunier}, {Vigan},
  {Hagelberg}, {Bonnefoy}, {Zurlo}, {Rocha}, {Maurel}, {Jaquet}, {Buey}, \&
  {Weber}}]{Galicher18}
{Galicher}, R., {Boccaletti}, A., {Mesa}, D., {et~al.} 2018, \aap, 615, A92

\bibitem[{{Galicher} {et~al.}(2016){Galicher}, {Marois}, {Macintosh},
  {Zuckerman}, {Barman}, {Konopacky}, {Song}, {Patience}, {Lafreni{\`e}re},
  {Doyon}, \& {Nielsen}}]{Galicher16}
{Galicher}, R., {Marois}, C., {Macintosh}, B., {et~al.} 2016, \aap, 594, A63

\bibitem[{{Galland} {et~al.}(2005){Galland}, {Lagrange}, {Udry}, {Chelli},
  {Pepe}, {Queloz}, {Beuzit}, \& {Mayor}}]{Galland05}
{Galland}, F., {Lagrange}, A.~M., {Udry}, S., {et~al.} 2005, \aap, 443, 337

\bibitem[{{Gangi} {et~al.}(2022){Gangi}, {Antoniucci}, {Biazzo}, {Frasca},
  {Nisini}, {Alcal{\'a}}, {Giannini}, {Manara}, {Giunta}, {Harutyunyan},
  {Munari}, \& {Vitali}}]{Gangi22}
{Gangi}, M., {Antoniucci}, S., {Biazzo}, K., {et~al.} 2022, \aap, 667, A124

\bibitem[{{Grandjean} {et~al.}(2020){Grandjean}, {Lagrange}, {Keppler},
  {Meunier}, {Mignon}, {Borgniet}, {Chauvin}, {Desidera}, {Galland}, {Messina},
  {Sterzik}, {Pantoja}, {Rodet}, \& {Zicher}}]{Grandjean20}
{Grandjean}, A., {Lagrange}, A.~M., {Keppler}, M., {et~al.} 2020, \aap, 633,
  A44

\bibitem[{{Gratton} {et~al.}(2024){Gratton}, {Bonavita}, {Mesa}, {Desidera},
  {Zurlo}, {Marino}, {D'Orazi}, {Rigliaco}, {Nascimbeni}, {Barbato}, {Columba},
  \& {Squicciarini}}]{Gratton24}
{Gratton}, R., {Bonavita}, M., {Mesa}, D., {et~al.} 2024, \aap, 685, A119

\bibitem[{{Griffin}(2010)}]{Griffin10}
{Griffin}, R.~F. 2010, The Observatory, 130, 125

\bibitem[{{Hagelberg} {et~al.}(2020){Hagelberg}, {Engler}, {Fontanive},
  {Daemgen}, {Quanz}, {K{\"u}hn}, {Reggiani}, {Meyer}, {Jayawardhana}, \&
  {Kostov}}]{Hagelberg20}
{Hagelberg}, J., {Engler}, N., {Fontanive}, C., {et~al.} 2020, \aap, 643, A98

\bibitem[{{Hales} {et~al.}(2014){Hales}, {De Gregorio-Monsalvo}, {Montesinos},
  {Casassus}, {Dent}, {Dougados}, {Eiroa}, {Hughes}, {Garay}, {Mardones},
  {M{\'e}nard}, {Palau}, {P{\'e}rez}, {Phillips}, {Torrelles}, \&
  {Wilner}}]{Hales14}
{Hales}, A.~S., {De Gregorio-Monsalvo}, I., {Montesinos}, B., {et~al.} 2014,
  \aj, 148, 47

\bibitem[{{Hall} {et~al.}(2007){Hall}, {Lockwood}, \& {Skiff}}]{Hall07}
{Hall}, J.~C., {Lockwood}, G.~W., \& {Skiff}, B.~A. 2007, \aj, 133, 862

\bibitem[{{Holl} {et~al.}(2022){Holl}, {Perryman}, {Lindegren}, {Segransan}, \&
  {Raimbault}}]{Holl2022}
{Holl}, B., {Perryman}, M., {Lindegren}, L., {Segransan}, D., \& {Raimbault},
  M. 2022, \aap, 661, A151

\bibitem[{{Holl} {et~al.}(2023){Holl}, {Sozzetti}, {Sahlmann}, {Giacobbe},
  {S{\'e}gransan}, {Unger}, {Delisle}, {Barbato}, {Lattanzi}, {Morbidelli}, \&
  {Sosnowska}}]{Holl2023}
{Holl}, B., {Sozzetti}, A., {Sahlmann}, J., {et~al.} 2023, \aap, 674, A10

\bibitem[{{James} {et~al.}(2016){James}, {Aarnio}, {Richert}, {Cargile},
  {Santos}, {Melo}, \& {Bouvier}}]{James16}
{James}, D.~J., {Aarnio}, A.~N., {Richert}, A.~J.~W., {et~al.} 2016, \mnras,
  459, 1363

\bibitem[{{James} {et~al.}(2006){James}, {Melo}, {Santos}, \&
  {Bouvier}}]{James06}
{James}, D.~J., {Melo}, C., {Santos}, N.~C., \& {Bouvier}, J. 2006, \aap, 446,
  971

\bibitem[{{Kastner} {et~al.}(2010){Kastner}, {Hily-Blant}, {Sacco},
  {Forveille}, \& {Zuckerman}}]{Kastner10}
{Kastner}, J.~H., {Hily-Blant}, P., {Sacco}, G.~G., {Forveille}, T., \&
  {Zuckerman}, B. 2010, \apjl, 723, L248

\bibitem[{{Kastner} \& {Principe}(2022)}]{Kastner22}
{Kastner}, J.~H. \& {Principe}, D.~A. 2022, in Handbook of X-ray and Gamma-ray
  Astrophysics, 49

\bibitem[{{Kervella} {et~al.}(2019{\natexlab{a}}){Kervella}, {Arenou},
  {Mignard}, \& {Th{\'e}venin}}]{Kervella2019}
{Kervella}, P., {Arenou}, F., {Mignard}, F., \& {Th{\'e}venin}, F.
  2019{\natexlab{a}}, \aap, 623, A72

\bibitem[{{Kervella} {et~al.}(2019{\natexlab{b}}){Kervella}, {Arenou},
  {Mignard}, \& {Th{\'e}venin}}]{Kervella19}
{Kervella}, P., {Arenou}, F., {Mignard}, F., \& {Th{\'e}venin}, F.
  2019{\natexlab{b}}, \aap, 623, A72

\bibitem[{{Kervella} {et~al.}(2022){Kervella}, {Arenou}, \&
  {Th{\'e}venin}}]{Kervella2022}
{Kervella}, P., {Arenou}, F., \& {Th{\'e}venin}, F. 2022, \aap, 657, A7

\bibitem[{{Kiefer} {et~al.}(2024{\natexlab{a}}){Kiefer}, {Lagrange}, {Rubini},
  \& {Philipot}}]{Kiefer2024a}
{Kiefer}, F., {Lagrange}, A.-M., {Rubini}, P., \& {Philipot}, F.
  2024{\natexlab{a}}, arXiv e-prints, A\&A, in press, arXiv:2409.16992

\bibitem[{{Kiefer} {et~al.}(2024{\natexlab{b}}){Kiefer}, {Lagrange}, {Rubini},
  \& {Philipot}}]{Kiefer2024b}
{Kiefer}, F., {Lagrange}, A.-M., {Rubini}, P., \& {Philipot}, F.
  2024{\natexlab{b}}, arXiv e-prints, A\&A, in press, arXiv:2409.16993

\bibitem[{{Kozai}(1962)}]{Kozai1962}
{Kozai}, Y. 1962, \aj, 67, 591

\bibitem[{{Kraus} {et~al.}(2014){Kraus}, {Shkolnik}, {Allers}, \&
  {Liu}}]{kraus14}
{Kraus}, A.~L., {Shkolnik}, E.~L., {Allers}, K.~N., \& {Liu}, M.~C. 2014, \aj,
  147, 146

\bibitem[{{Krymolowski} \& {Mazeh}(1999)}]{Kry1999}
{Krymolowski}, Y. \& {Mazeh}, T. 1999, \mnras, 304, 720

\bibitem[{{Lagrange} {et~al.}(2011){Lagrange}, {Meunier}, {Desort}, \&
  {Malbet}}]{Lagrange11}
{Lagrange}, A.~M., {Meunier}, N., {Desort}, M., \& {Malbet}, F. 2011, \aap,
  528, L9

\bibitem[{{Lannier} {et~al.}(2016){Lannier}, {Delorme}, {Lagrange}, {Borgniet},
  {Rameau}, {Schlieder}, {Gagn{\'e}}, {Bonavita}, {Malo}, {Chauvin},
  {Bonnefoy}, \& {Girard}}]{Lannier16}
{Lannier}, J., {Delorme}, P., {Lagrange}, A.~M., {et~al.} 2016, \aap, 596, A83

\bibitem[{{Lannier} {et~al.}(2017){Lannier}, {Lagrange}, {Bonavita},
  {Borgniet}, {Delorme}, {Meunier}, {Desidera}, {Messina}, {Chauvin}, \&
  {Keppler}}]{Lannier17}
{Lannier}, J., {Lagrange}, A.~M., {Bonavita}, M., {et~al.} 2017, \aap, 603, A54

\bibitem[{{Lockwood} {et~al.}(2007){Lockwood}, {Skiff}, {Henry}, {Henry},
  {Radick}, {Baliunas}, {Donahue}, \& {Soon}}]{Lockwood07}
{Lockwood}, G.~W., {Skiff}, B.~A., {Henry}, G.~W., {et~al.} 2007, \apjs, 171,
  260

\bibitem[{{Lopez Mart{\'\i}} {et~al.}(2013){Lopez Mart{\'\i}}, {Jimenez
  Esteban}, {Bayo}, {Barrado}, {Solano}, \& {Rodrigo}}]{Lopez-Marti13}
{Lopez Mart{\'\i}}, B., {Jimenez Esteban}, F., {Bayo}, A., {et~al.} 2013, \aap,
  551, A46

\bibitem[{{Luhman}(2023)}]{luhman23}
{Luhman}, K.~L. 2023, \aj, 165, 269

\bibitem[{{Macintosh} {et~al.}(2014){Macintosh}, {Graham}, {Ingraham},
  {Konopacky}, {Marois}, {Perrin}, {Poyneer}, {Bauman}, {Barman}, {Burrows},
  {Cardwell}, {Chilcote}, {De Rosa}, {Dillon}, {Doyon}, {Dunn}, {Erikson},
  {Fitzgerald}, {Gavel}, {Goodsell}, {Hartung}, {Hibon}, {Kalas}, {Larkin},
  {Maire}, {Marchis}, {Marley}, {McBride}, {Millar-Blanchaer}, {Morzinski},
  {Norton}, {Oppenheimer}, {Palmer}, {Patience}, {Pueyo}, {Rantakyro},
  {Sadakuni}, {Saddlemyer}, {Savransky}, {Serio}, {Soummer},
  {Sivaramakrishnan}, {Song}, {Thomas}, {Wallace}, {Wiktorowicz}, \&
  {Wolff}}]{macintosh14}
{Macintosh}, B., {Graham}, J.~R., {Ingraham}, P., {et~al.} 2014, Proceedings of
  the National Academy of Science, 111, 12661

\bibitem[{{Malkov} {et~al.}(2006){Malkov}, {Oblak}, {Snegireva}, \&
  {Torra}}]{Malkov06}
{Malkov}, O.~Y., {Oblak}, E., {Snegireva}, E.~A., \& {Torra}, J. 2006, \aap,
  446, 785

\bibitem[{{Mamajek} \& {Bell}(2014)}]{Mamajek14}
{Mamajek}, E.~E. \& {Bell}, C. P.~M. 2014, \mnras, 445, 2169

\bibitem[{{Martioli} {et~al.}(2021){Martioli}, {H{\'e}brard}, {Correia},
  {Laskar}, \& {Lecavelier des Etangs}}]{Martioli21}
{Martioli}, E., {H{\'e}brard}, G., {Correia}, A.~C.~M., {Laskar}, J., \&
  {Lecavelier des Etangs}, A. 2021, \aap, 649, A177

\bibitem[{{McGinnis} {et~al.}(2015){McGinnis}, {Alencar}, {Guimar{\~a}es},
  {Sousa}, {Stauffer}, {Bouvier}, {Rebull}, {Fonseca}, {Venuti}, {Hillenbrand},
  {Cody}, {Teixeira}, {Aigrain}, {Favata}, {F{\H{u}}r{\'e}sz}, {Vrba},
  {Flaccomio}, {Turner}, {Gameiro}, {Dougados}, {Herbst},
  {Morales-Calder{\'o}n}, \& {Micela}}]{McGinnis2015}
{McGinnis}, P.~T., {Alencar}, S.~H.~P., {Guimar{\~a}es}, M.~M., {et~al.} 2015,
  \aap, 577, A11

\bibitem[{{Mesa} {et~al.}(2023{\natexlab{a}}){Mesa}, {Gratton}, {Kervella},
  {Bonavita}, {Desidera}, {D'Orazi}, {Marino}, {Zurlo}, \& {Rigliaco}}]{Mesa23}
{Mesa}, D., {Gratton}, R., {Kervella}, P., {et~al.} 2023{\natexlab{a}}, \aap,
  672, A93

\bibitem[{{Mesa} {et~al.}(2023{\natexlab{b}}){Mesa}, {Gratton}, {Kervella},
  {Bonavita}, {Desidera}, {D'Orazi}, {Marino}, {Zurlo}, \&
  {Rigliaco}}]{Mesa2023}
{Mesa}, D., {Gratton}, R., {Kervella}, P., {et~al.} 2023{\natexlab{b}}, \aap,
  672, A93

\bibitem[{{Metchev} \& {Hillenbrand}(2009)}]{Metchev09}
{Metchev}, S.~A. \& {Hillenbrand}, L.~A. 2009, \apjs, 181, 62

\bibitem[{{Meunier}(2021)}]{Meunier21}
{Meunier}, N. 2021, arXiv e-prints, arXiv:2104.06072

\bibitem[{{Meunier} \& {Lagrange}(2022)}]{Meunier22}
{Meunier}, N. \& {Lagrange}, A.~M. 2022, \aap, 659, A104

\bibitem[{{Mignon} {et~al.}(2023){Mignon}, {Meunier}, {Delfosse}, {Bonfils},
  {Santos}, {Forveille}, {Gaisn{\'e}}, {Astudillo-Defru}, {Lovis}, \&
  {Udry}}]{Mignon23}
{Mignon}, L., {Meunier}, N., {Delfosse}, X., {et~al.} 2023, \aap, 675, A168

\bibitem[{{Mugrauer} {et~al.}(2010){Mugrauer}, {Vogt}, {Neuh{\"a}user}, \&
  {Schmidt}}]{Mugrauer10}
{Mugrauer}, M., {Vogt}, N., {Neuh{\"a}user}, R., \& {Schmidt}, T.~O.~B. 2010,
  \aap, 523, L1

\bibitem[{{Murphy} \& {Lawson}(2015)}]{murphy15}
{Murphy}, S.~J. \& {Lawson}, W.~A. 2015, \mnras, 447, 1267

\bibitem[{{Newton} {et~al.}(2019){Newton}, {Mann}, {Tofflemire}, {Pearce},
  {Rizzuto}, {Vanderburg}, {Martinez}, {Wang}, {Ruffio}, {Kraus}, {Johnson},
  {Thao}, {Wood}, {Rampalli}, {Nielsen}, {Collins}, {Dragomir}, {Hellier},
  {Anderson}, {Barclay}, {Brown}, {Feiden}, {Hart}, {Isopi}, {Kielkopf},
  {Mallia}, {Nelson}, {Rodriguez}, {Stockdale}, {Waite}, {Wright}, {Lissauer},
  {Ricker}, {Vanderspek}, {Latham}, {Seager}, {Winn}, {Jenkins}, {Bouma},
  {Burke}, {Davies}, {Fausnaugh}, {Li}, {Morris}, {Mukai}, {Villase{\~n}or},
  {Villeneuva}, {De Rosa}, {Macintosh}, {Mengel}, {Okumura}, \&
  {Wittenmyer}}]{Newton19}
{Newton}, E.~R., {Mann}, A.~W., {Tofflemire}, B.~M., {et~al.} 2019, \apjl, 880,
  L17

\bibitem[{{Nguyen} {et~al.}(2022){Nguyen}, {Costa}, {Girardi}, {Volpato},
  {Bressan}, {Chen}, {Marigo}, {Fu}, \& {Goudfrooij}}]{parsec}
{Nguyen}, C.~T., {Costa}, G., {Girardi}, L., {et~al.} 2022, \aap, 665, A126

\bibitem[{{Offner} {et~al.}(2023){Offner}, {Moe}, {Kratter}, {Sadavoy},
  {Jensen}, \& {Tobin}}]{Offner22}
{Offner}, S.~S.~R., {Moe}, M., {Kratter}, K.~M., {et~al.} 2023, in Astronomical
  Society of the Pacific Conference Series, Vol. 534, Protostars and Planets
  VII, ed. S.~{Inutsuka}, Y.~{Aikawa}, T.~{Muto}, K.~{Tomida}, \& M.~{Tamura},
  275

\bibitem[{{Palma-Bifani} {et~al.}(2023){Palma-Bifani}, {Chauvin}, {Bonnefoy},
  {Rojo}, {Petrus}, {Rodet}, {Langlois}, {Allard}, {Charnay}, {Desgrange},
  {Homeier}, {Lagrange}, {Beuzit}, {Baudoz}, {Boccaletti}, {Chomez}, {Delorme},
  {Desidera}, {Feldt}, {Ginski}, {Gratton}, {Maire}, {Meyer}, {Samland},
  {Snellen}, {Vigan}, \& {Zhang}}]{Palma-Bifani23}
{Palma-Bifani}, P., {Chauvin}, G., {Bonnefoy}, M., {et~al.} 2023, \aap, 670,
  A90

\bibitem[{{Pearce} {et~al.}(2020){Pearce}, {Kraus}, {Dupuy}, {Mann}, {Newton},
  {Tofflemire}, \& {Vanderburg}}]{Pierce20}
{Pearce}, L.~A., {Kraus}, A.~L., {Dupuy}, T.~J., {et~al.} 2020, \apj, 894, 115

\bibitem[{Perrin {et~al.}(2016)Perrin, Ingraham, Follette, Maire, Wang,
  Savransky, Arriaga, Bailey, Bruzzone, Chilcote, De~Rosa, Draper, Fitzgerald,
  Greenbaum, Hung, Konopacky, Macintosh, Marchis, Marois, Millar-Blanchaer,
  Nielsen, Rajan, Rameau, Rantakyro, Ruffio, Ward-Duong, Wolff, \&
  Zalesky}]{evans_gemini_2016}
Perrin, M.~D., Ingraham, P., Follette, K.~B., {et~al.} 2016, in , Edinburgh,
  United Kingdom, 990837

\bibitem[{Perrin {et~al.}(2014)Perrin, Maire, Ingraham, Savransky,
  Millar-Blanchaer, Wolff, Ruffio, Wang, Draper, Sadakuni, Marois, Rajan,
  Fitzgerald, Macintosh, Graham, Doyon, Larkin, Chilcote, Goodsell, Palmer,
  Labrie, Beaulieu, De~Rosa, Greenbaum, Hartung, Hibon, Konopacky, Lafreniere,
  Lavigne, Marchis, Patience, Pueyo, Rantakyrö, Soummer, Sivaramakrishnan,
  Thomas, Ward-Duong, \& Wiktorowicz}]{perrin_gemini_2014}
Perrin, M.~D., Maire, J., Ingraham, P., {et~al.} 2014, in , 91473J,
  arXiv:1407.2301 [astro-ph]

\bibitem[{{Perryman} {et~al.}(2014){Perryman}, {Hartman}, {Bakos}, \&
  {Lindegren}}]{Perryman2014}
{Perryman}, M., {Hartman}, J., {Bakos}, G.~{\'A}., \& {Lindegren}, L. 2014,
  \apj, 797, 14

\bibitem[{{Riviere-Marichalar} {et~al.}(2014){Riviere-Marichalar}, {Barrado},
  {Montesinos}, {Duch{\^e}ne}, {Bouy}, {Pinte}, {Menard}, {Donaldson}, {Eiroa},
  {Krivov}, {Kamp}, {Mendigut{\'\i}a}, {Dent}, \&
  {Lillo-Box}}]{Riviere-Marichalar14}
{Riviere-Marichalar}, P., {Barrado}, D., {Montesinos}, B., {et~al.} 2014, \aap,
  565, A68

\bibitem[{{Rodet} {et~al.}(2019){Rodet}, {Beust}, {Bonnefoy}, {De Rosa},
  {Kalas}, \& {Lagrange}}]{Rodet19}
{Rodet}, L., {Beust}, H., {Bonnefoy}, M., {et~al.} 2019, \aap, 631, A139

\bibitem[{{Rodet} {et~al.}(2017){Rodet}, {Beust}, {Bonnefoy}, {Lagrange},
  {Galli}, {Ducourant}, \& {Teixeira}}]{Rodet17}
{Rodet}, L., {Beust}, H., {Bonnefoy}, M., {et~al.} 2017, \aap, 602, A12

\bibitem[{{Sacco} {et~al.}(2014){Sacco}, {Kastner}, {Forveille}, {Principe},
  {Montez}, {Zuckerman}, \& {Hily-Blant}}]{Sacco14}
{Sacco}, G.~G., {Kastner}, J.~H., {Forveille}, T., {et~al.} 2014, \aap, 561,
  A42

\bibitem[{{Sahlmann} {et~al.}(2015){Sahlmann}, {Triaud}, \&
  {Martin}}]{Sahlmann2015}
{Sahlmann}, J., {Triaud}, A.~H.~M.~J., \& {Martin}, D.~V. 2015, \mnras, 447,
  287

\bibitem[{{Shan} {et~al.}(2024){Shan}, {Revilla}, {Skrzypinski}, {Dreizler},
  {B{\'e}jar}, {Caballero}, {Cardona Guill{\'e}n}, {Cifuentes}, {Fuhrmeister},
  {Reiners}, {Vanaverbeke}, {Ribas}, {Quirrenbach}, {Amado}, {Aceituno},
  {Casanova}, {Cort{\'e}s-Contreras}, {Dubois}, {Gorrini}, {Henning},
  {Herrero}, {Jeffers}, {Kemmer}, {Lalitha}, {Lodieu}, {Logie}, {L{\'o}pez
  Gonz{\'a}lez}, {Mart{\'\i}n-Ruiz}, {Montes}, {Morales}, {Nagel}, {Pall{\'e}},
  {Perdelwitz}, {P{\'e}rez-Torres}, {Pollacco}, {Rau},
  {Rodr{\'\i}guez-L{\'o}pez}, {Rodr{\'\i}guez}, {Sch{\"o}fer}, {Seifert},
  {Sota}, {Zapatero Osorio}, \& {Zechmeister}}]{Shan24}
{Shan}, Y., {Revilla}, D., {Skrzypinski}, S.~L., {et~al.} 2024, \aap, 684, A9

\bibitem[{{Shkolnik} {et~al.}(2017){Shkolnik}, {Allers}, {Kraus}, {Liu}, \&
  {Flagg}}]{Shkolnik17}
{Shkolnik}, E.~L., {Allers}, K.~N., {Kraus}, A.~L., {Liu}, M.~C., \& {Flagg},
  L. 2017, \aj, 154, 69

\bibitem[{{Sierchio} {et~al.}(2014){Sierchio}, {Rieke}, {Su}, \&
  {G{\'a}sp{\'a}r}}]{Sierchio14}
{Sierchio}, J.~M., {Rieke}, G.~H., {Su}, K.~Y.~L., \& {G{\'a}sp{\'a}r}, A.
  2014, \apj, 785, 33

\bibitem[{{Skrutskie} {et~al.}(2006){Skrutskie}, {Cutri}, {Stiening},
  {Weinberg}, {Schneider}, {Carpenter}, {Beichman}, {Capps}, {Chester},
  {Elias}, {Huchra}, {Liebert}, {Lonsdale}, {Monet}, {Price}, {Seitzer},
  {Jarrett}, {Kirkpatrick}, {Gizis}, {Howard}, {Evans}, {Fowler}, {Fullmer},
  {Hurt}, {Light}, {Kopan}, {Marsh}, {McCallon}, {Tam}, {Van Dyk}, \&
  {Wheelock}}]{2mass}
{Skrutskie}, M.~F., {Cutri}, R.~M., {Stiening}, R., {et~al.} 2006, \aj, 131,
  1163

\bibitem[{{Song} {et~al.}(2003){Song}, {Zuckerman}, \& {Bessell}}]{Song03}
{Song}, I., {Zuckerman}, B., \& {Bessell}, M.~S. 2003, \apj, 599, 342

\bibitem[{{Squicciarini} \& {Bonavita}(2022)}]{squicciarini22}
{Squicciarini}, V. \& {Bonavita}, M. 2022, \aap, 666, A15

\bibitem[{{Stassun} {et~al.}(2019){Stassun}, {Oelkers}, {Paegert}, {Torres},
  {Pepper}, {De Lee}, {Collins}, {Latham}, {Muirhead}, {Chittidi},
  {Rojas-Ayala}, {Fleming}, {Rose}, {Tenenbaum}, {Ting}, {Kane}, {Barclay},
  {Bean}, {Brassuer}, {Charbonneau}, {Ge}, {Lissauer}, {Mann}, {McLean},
  {Mullally}, {Narita}, {Plavchan}, {Ricker}, {Sasselov}, {Seager}, {Sharma},
  {Shiao}, {Sozzetti}, {Stello}, {Vanderspek}, {Wallace}, \&
  {Winn}}]{Stassun19}
{Stassun}, K.~G., {Oelkers}, R.~J., {Paegert}, M., {et~al.} 2019, \aj, 158, 138

\bibitem[{{Thomas} {et~al.}(2023){Thomas}, {Nielsen}, {De Rosa}, {Peck},
  {Macintosh}, {Chilcote}, {Kalas}, {Wang}, {Blunt}, {Greenbaum}, {Konopacky},
  {Ireland}, {Tuthill}, {Ward-Duong}, {Hirsch}, {Czekala}, {Marchis}, {Marois},
  {Millar-Blanchaer}, {Roberson}, {Smith}, {Gallamore}, \&
  {Klusmeyer}}]{Thomas23}
{Thomas}, A.~D., {Nielsen}, E.~L., {De Rosa}, R.~J., {et~al.} 2023, \aj, 166,
  246

\bibitem[{{Tokovinin}(2016)}]{Tokovinin16a}
{Tokovinin}, A. 2016, \aj, 152, 11

\bibitem[{{Tokovinin} \& {Horch}(2016)}]{Tokovinin16b}
{Tokovinin}, A. \& {Horch}, E.~P. 2016, \aj, 152, 116

\bibitem[{{Tokovinin} \& {Kiyaeva}(2016)}]{Tokovinin16-MNRAS}
{Tokovinin}, A. \& {Kiyaeva}, O. 2016, VizieR Online Data Catalog,
  J/MNRAS/456/2070

\bibitem[{{Tokovinin} {et~al.}(2020){Tokovinin}, {Mason}, {Mendez}, {Costa}, \&
  {Horch}}]{Tokovinin20}
{Tokovinin}, A., {Mason}, B.~D., {Mendez}, R.~A., {Costa}, E., \& {Horch},
  E.~P. 2020, \aj, 160, 7

\bibitem[{{Torres} {et~al.}(2006){Torres}, {Quast}, {da Silva}, {de La Reza},
  {Melo}, \& {Sterzik}}]{Torres06}
{Torres}, C.~A.~O., {Quast}, G.~R., {da Silva}, L., {et~al.} 2006, \aap, 460,
  695

\bibitem[{{Torres} {et~al.}(2008){Torres}, {Quast}, {Melo}, \&
  {Sterzik}}]{Torres08}
{Torres}, C.~A.~O., {Quast}, G.~R., {Melo}, C.~H.~F., \& {Sterzik}, M.~F. 2008,
  in Handbook of Star Forming Regions, Volume II, ed. B.~{Reipurth}, Vol.~5,
  757

\bibitem[{{Trifonov} {et~al.}(2020){Trifonov}, {Tal-Or}, {Zechmeister},
  {Kaminski}, {Zucker}, \& {Mazeh}}]{Trifonov20}
{Trifonov}, T., {Tal-Or}, L., {Zechmeister}, M., {et~al.} 2020, \aap, 636, A74

\bibitem[{{van Leeuwen}(2007)}]{vanleeuwen2007}
{van Leeuwen}, F. 2007, \aap, 474, 653

\bibitem[{{Venuti} {et~al.}(2019){Venuti}, {Stelzer}, {Alcal{\'a}}, {Manara},
  {Frasca}, {Jayawardhana}, {Antoniucci}, {Argiroffi}, {Natta}, {Nisini},
  {Randich}, \& {Scholz}}]{Venuti19}
{Venuti}, L., {Stelzer}, B., {Alcal{\'a}}, J.~M., {et~al.} 2019, \aap, 632, A46

\bibitem[{{Vogt} {et~al.}(2015){Vogt}, {Mugrauer}, {Neuh{\"a}user}, {Schmidt},
  {Contreras-Quijada}, \& {Schmidt}}]{vogt15}
{Vogt}, N., {Mugrauer}, M., {Neuh{\"a}user}, R., {et~al.} 2015, Astronomische
  Nachrichten, 336, 97

\bibitem[{{Wahhaj} {et~al.}(2010){Wahhaj}, {Cieza}, {Koerner}, {Stapelfeldt},
  {Padgett}, {Case}, {Keller}, {Mer{\'\i}n}, {Evans}, {Harvey}, {Sargent}, {van
  Dishoeck}, {Allen}, {Blake}, {Brooke}, {Chapman}, {Mundy}, \&
  {Myers}}]{Wahhaj10}
{Wahhaj}, Z., {Cieza}, L., {Koerner}, D.~W., {et~al.} 2010, \apj, 724, 835

\bibitem[{Wang {et~al.}(2015)Wang, Ruffio, De~Rosa, Aguilar, Wolff, \&
  Pueyo}]{wang_pyklip_2015}
Wang, J.~J., Ruffio, J.-B., De~Rosa, R.~J., {et~al.} 2015, Astrophysics Source
  Code Library, ascl:1506.001, aDS Bibcode: 2015ascl.soft06001W

\bibitem[{{Wang} {et~al.}(2023){Wang}, {Xia}, \& {Fu}}]{Wang23-PXVir}
{Wang}, X., {Xia}, F., \& {Fu}, Y. 2023, \pasj, 75, 368

\bibitem[{{Winters} {et~al.}(2019){Winters}, {Henry}, {Jao}, {Subasavage},
  {Chatelain}, {Slatten}, {Riedel}, {Silverstein}, \& {Payne}}]{Winters19}
{Winters}, J.~G., {Henry}, T.~J., {Jao}, W.-C., {et~al.} 2019, \aj, 157, 216

\bibitem[{{Zhou} {et~al.}(2022){Zhou}, {Wirth}, {Huang}, {Venner}, {Franson},
  {Quinn}, {Bouma}, {Kraus}, {Mann}, {Newton}, {Dragomir}, {Heitzmann},
  {Lowson}, {Douglas}, {Battley}, {Gillen}, {Triaud}, {Latham}, {Howell},
  {Hartman}, {Tofflemire}, {Wittenmyer}, {Bowler}, {Horner}, {Kane},
  {Kielkopf}, {Plavchan}, {Wright}, {Addison}, {Mengel}, {Okumura}, {Ricker},
  {Vanderspek}, {Seager}, {Jenkins}, {Winn}, {Daylan}, {Fausnaugh}, \&
  {Kunimoto}}]{Zhou23}
{Zhou}, G., {Wirth}, C.~P., {Huang}, C.~X., {et~al.} 2022, \aj, 163, 289

\bibitem[{{Zicher} {et~al.}(2022){Zicher}, {Barrag{\'a}n}, {Klein}, {Aigrain},
  {Owen}, {Gandolfi}, {Lagrange}, {Serrano}, {Kaye}, {Nielsen}, {Rajpaul},
  {Grandjean}, {Goffo}, \& {Nicholson}}]{Zicher22}
{Zicher}, N., {Barrag{\'a}n}, O., {Klein}, B., {et~al.} 2022, \mnras, 512, 3060

\bibitem[{{Z{\'u}{\~n}iga-Fern{\'a}ndez}
  {et~al.}(2021){Z{\'u}{\~n}iga-Fern{\'a}ndez}, {Bayo}, {Elliott}, {Zamora},
  {Corval{\'a}n}, {Haubois}, {Corral-Santana}, {Olofsson}, {Hu{\'e}lamo},
  {Sterzik}, {Torres}, {Quast}, \& {Melo}}]{Zuniga21}
{Z{\'u}{\~n}iga-Fern{\'a}ndez}, S., {Bayo}, A., {Elliott}, P., {et~al.} 2021,
  \aap, 645, A30

\bibitem[{{Zuckerman}(2019)}]{Zuckerman19}
{Zuckerman}, B. 2019, \apj, 870, 27

\end{thebibliography}

\appendix

\newpage
\onecolumn
\section{Acronyms and symbols}
\begin{table*}[hbt]
    \caption{List and definition of relevant, non-trivial, acronyms and symbols used in the text.}
    \label{tab:acronyms}
    \centering
\begin{tabular}{l|l}
        Acronyms or Symbols & Definition \\
        \hline
        \multicolumn{2}{l}{\textit{Young associations and moving groups}} \\ 
        \multicolumn{2}{l}{} \\
        MG & Moving Group \\ 
        YMG & Young Moving Groups \\
        BPC & $\beta$\,Pictoris \\
        BPMG & $\beta$\,Pictoris moving group \\
        OCT & Octans \\
\hline
    \end{tabular}
\end{table*}
\newpage
\twocolumn

\section{Target sample}\label{sec:Sample}
\pagestyle{empty}
\onecolumn
\begin{landscape}
\begin{longtable}[c]{l@{~~}cccccccccc}
\caption{Targets sample. The star masses are from ZF21.}
\label{tab:planet_catalog}\\
\hline
Star name               & Gaia DR3 ID         & $RA_{ICRS}$   & $DEC_{ICRS}$  & SACYMG & SACYprob & BANMG  & BANprob & star mass & plx(mas) & plx err \\
\hline
\endfirsthead
\multicolumn{11}{c}%
{{\bfseries Table \thetable\ continued from previous page}} \\
\hline
Star name               & Gaia DR3 ID         & $RA_{ICRS}$   & $DEC_{ICRS}$  & SACYMG & SACYprob & BANMG  & BANprob & star mass & plx(mas) & plx err \\
\hline
\endhead
\hline
\endfoot
\endlastfoot
HD 105                  & 4995853014646189696 & 00:05:52.6847 & -41:45:12.269 & THA    & 0.997    & THA    & 1.000 & 1.11      & 25.75    & 0.02    \\
HD 987                  & 4688292536884945536 & 00:13:53.3451 & -74:41:18.636 & THA    & 0.996    & THA    & 1.000 & 0.9       & 21.8     & 0.01    \\
HD 984                  & 2431157720981843200 & 00:14:10.3654 & -07:11:57.900 & COL    & 0.843    & COL    & 1.000 & 1.26      & 21.88    & 0.02    \\
HD 1466                 & 4901229043960053248 & 00:18:26.3386 & -63:28:39.928 & THA    & 0.997    & THA    & 1.000 & 1.15      & 23.36    & 0.02    \\
Smethells 165           & 4901926409210453760 & 00:24:09.1823 & -62:11:05.115 & THA    & 0.950    & THA    & 1.000 & 0.74      & 22.8     & 0.06    \\
V* CT Tuc               & 4902014095262270336 & 00:25:14.8587 & -61:30:49.151 & THA    & 0.996    & THA    & 1.000 & 0.7       & 22.65    & 0.01    \\
HD 3221                 & 4901802507993393664 & 00:34:51.4029 & -61:54:58.972 & THA    & 0.801    & THA    & 1.000 & 0.96      & 22.51    & 0.02    \\
CD-78 24                & 4635593356881798912 & 00:42:20.7179 & -77:47:40.215 & THA    & 0.998    & THA    & 1.000 & 0.82      & 20.08    & 0.01    \\
\hline
\end{longtable}
\end{landscape}
\twocolumn

\section{Results}\label{sec:Results}
This table provides the results on the stars not known to be accreting and found to be have a companion in the present study. 

The column labeled ZF21 indicates the binary status reported by ZF21. 
WDS stands for Washington visual Double Stars catalogue \citep{Mason2020}. SB9 stands for the 
9th Catalogue of Spectroscopic Binary Orbits \citep{Pourbaix04}.
NSS refers to the Gaia DR3 Part 3 Non-single stars (Gaia Collaboration, 2022). FMP refers to $IPD\_ frac\_ multi\_ peak $. 
$\sigma_{\rm \texttt{ruwe}}$ refers to the \nsigmaruwe{} , and $\sigma_{\rm PMa}$ refers to the \nsigmapma{}  (see Text). Note that values of 10.0 indicate actual values equal or larger than 10. 
Max sma  indicates the maximum sma of the (sma, Mass) solutions for each identified binary, and Min mass gives the minimum mass in terms of class (Stellar, BD, or Planet) of the (sma, Mass) solutions.

\newpage
\onecolumn
\begin{landscape}
\begin{longtable}[c]{l@{~~}c@{~~}c@{~~}c@{~~}c@{~~}c@{~~}c@{~~}c@{~~}c@{~~}c@{~~}c@{~~}c@{~~}c@{~~}c@{~~}c@{~~}c@{~~}c@{~~}c@{~~}c}
\caption{Binary status from GaiaPMEX. For a description on the columns, see text.}
\label{tab:results}\\
\hline
Star name	&	Gaia DR3 ID	&	ZF21	&	Nfov	&	Npts	&	Dup.	&	Par.	&	RUWE	&	WDS	&	SB9	&	NSS	&	FMP	&	Phot Var.	&	$\sigma_{\rm \texttt{ruwe}}$	&	$\sigma_{\rm PMa}$	&	Max sma	&	Min mass	\\
                   &                     &    &      &      &   & solved &        & flag & flag & flag &flag &  &        &                &               & sma      & mass \\

                       &                     &                     &  &      &      &       &        &        &      &      &      &            &     &           &               & (au)     &      \\

\hline
\endhead
\hline
\endfoot
\endlastfoot

CD-29 2360	&	2905633272252337920	&	?	&	56	&	491	&	0	&	31	&	1913.0	&	0	&	0	&	0	&	12.0	&	NAV	&	8.2	&		&	20	&	BD	\\
CD-44 1173	&	4848412525451859328	&	?	&	41	&	357	&	0	&	95	&	1811.0	&	1	&	0	&	0	&	36.0	&	NAV	&	8.0	&		&	20	&	BD	\\
BD-20 951	&	2974596325173490816	&	SB2	&	68	&	602	&	0	&	31	&	1806.0	&	0	&	0	&	2	&	1.0	&	NAV	&	7.8	&		&	20	&	BD	\\
V* V376 Vel	&	5318512260971706496	&	?	&	40	&	352	&	0	&	31	&	1919.0	&	0	&	0	&	0	&	1.0	&	NAV	&	7.7	&		&	15	&	BD	\\
V* DX Cha	&	5836666564476158336	&	?	&	33	&	293	&	1	&	31	&	2071.0	&	1	&	0	&	0	&	2.0	&	NAV	&	7.7	&	3.1	&	3	&	S	\\
V* V1215 Cen	&	5375025440645600000	&	SB2	&	80	&	693	&	0	&	31	&	1727.0	&	1	&	0	&	0	&	0.0	&	NAV	&	6.9	&		&	20	&	BD	\\
$$[FLG2003]$$ eps Cha 7	&	5836666564480769280	&	?	&	39	&	344	&	0	&	95	&	1554.0	&	1	&	0	&	0	&	0.0	&	NAV	&	6.5	&		&	10	&	BD	\\
BD-21 1074B	&	2962658549474035712	&	?	&	52	&	454	&	0	&	95	&	1948.0	&	1	&	0	&	0	&	59.0	&	NAV	&	6.1	&		&	20	&	P	\\
\hline
\end{longtable}
\end{landscape}

\twocolumn

\section{Log of observations}\label{sec:Log_obs}
\subsection{G80-21 }\label{sec:Log_G80-21}

The HARPS  data were standard Echelle observations taken as part of program 097.C-0864 B (PI: Lannier). The RV and Bisector velocity spans (BVS) were computed using our SAFIR software \citep{Galland05}. They are given it Table~\ref{tab:G80-21-RV}.
Note that the RV are compatible with those found by \citet{Trifonov20}.



\begin{table}
    \centering
    \caption{RV (relative to the median RV) and BVS and associated uncertainties (epsRV, epsBVS) for G80-21. }
    \label{tab:G80-21-RV}
    \begin{tabular}{ccccc}
        \toprule 
JDB-2400000& RV& epsRV& BVS&epsBVS\\
& km/s&km/s& km/s&km/s\\
 \midrule
57666.692176&0.0269& 0.0072& 0.0035	& 0.0181  \\
57666.705881&0.0353& 0.0069& 0.0346	& 0.0172  \\
57667.695570&-0.1006&0.0072& 0.0301	& 0.0180  \\
57667.706242&-0.1161&0.0075& 0.0344	& 0.0188  \\
57668.680072&0.0136& 0.0059& -0.0311	& 0.0149  \\
57668.693221&0.0219& 0.0063& -0.0547	& 0.0158  \\
57669.745330&0.0644& 0.0053& -0.0035	& 0.0134  \\
57669.759729&0.0587& 0.0054& -0.0477	& 0.0136  \\
57670.682837&0.0216& 0.0058& 0.0153	& 0.0147  \\
57670.696391&0.0321& 0.0060& -0.0073	& 0.0151  \\
57671.687117&-0.0664&0.0056& 0.0223	& 0.0142  \\
57671.701643&-0.0762&0.0057& 0.0380	& 0.0143  \\
57771.563217&0.0283& 0.0074& -0.0136	& 0.0186  \\
57771.574096&0.0122& 0.0070& -0.0563	& 0.0177  \\

        \bottomrule
    \end{tabular}
\end{table}

\subsection{AB Pic }\label{sec:Log_ABPic}
The coronagraphic SPHERE and GPI data used are described in Table~\ref{tab:obs_cond_table}. Additional PSF were recorded prior to and after the coronagraphic observations in the case of SPHERE data. The 2015 AB Pic data were obtained as part of the SPHERE SHINE survey in H2H3 mode. The Oct 2023 data were part of a GO Program (112.25.001, P.I. Alice Zurlo), and were obtained in star-hopping mode.

\begin{table*}[t!]
    \centering
    \begin{tabular}{l|c|c|c|c|c}
          \hline
        Star &  \multicolumn{3}{c|}{AB Pic} & \multicolumn{2}{c}{HD 14082 B} \\
            \hline

        instrument & SPHERE & GPI & SPHERE & SPHERE & SPHERE\\
        OBS night & 2015-02-05 & 2018-02-05 & 2023-10-25 & 2015-12-23 & 2016-09-18\\
        DIT(s)$\times$Nframe & 64$\times$64 & 60$\times$68 & 64$\times$24 &
        16$\times$144 & 16$\times$288 \\
        $\Delta$PA ($\deg $) & 28.28 & 23.60 & 22.23 & 11.15 & 22.74 \\
        seeing (")$^a$ & 1.08 & 0.99 & 0.49 & 0.84 & 0.67\\
        airmass$^a$ & 1.20 & 1.24 & 1.21 & 1.72 & 1.70\\
        IRDIS filter & DH\_H23 & -- & DH\_H23 & BB\_H & BB\_H\\
        $\tau_0$ (ms)$^{a}$ & 5.3 & unknown & 4.9 & 2.6 & 5.7\\
        coronograph & N\_ALC\_YJH\_S & {\tiny APO+FPM\_H\_G6225+080m12\_04} & N\_ALC\_YJH\_S & N\_ALC\_YJH\_S & N\_ALC\_YJH\_S\\
        program ID & 095.C-0298(H) & GS-2017B-Q-500 & 112.25QU.001 & 096.C-0388(A) &  096.C-0388(A)\\
        \hline

    \end{tabular}
    \caption{Observation logs for the SPHERE and GPI observations considered in this work.}
    \label{tab:obs_cond_table}
    \tablefoot{ DIT = detector integration time per frame, $\Delta$PA = amplitude of the parallactic rotation, $\tau_0$ = coherence time. $^a$: for SPHERE, values extracted from the updated DIMM information and averaged over the sequence.}
\end{table*}

The RV data of AB Pic were obtained with HARPS as part of a long-term search for RV exoplanets. They are presented in \cite[][]{Grandjean20}. 

\subsection{HD 14082 B }\label{sec:Log_HD14082B}

The data were taken as part of the SHARDS survey to search for disks around IR excess stars members of Sco Cen associations (P.I. J. Milli; 096.C-0388(A)). The observing set up and conditions are described in Table~\ref{tab:obs_cond_table}.
No RV data are available for HD 14082B.

\section{Data processing and analysis}\label{sec:Processing}
\subsection{Data processing}\label{sec:processing}
The reduction and analysis of SPHERE data are presented in  \cite[][]{Chomez23}. They made use of the PACO algorithm, described in \citet{flasseur2020robustness,flasseur_paco_2020} and in \cite[][]{Chomez23}.

For the GPI data, every raw frame of the archival sequence was dark subtracted, flat-fielded, cleaned of correlated detector noise, and corrected for bad pixel and flexure effects by means of the standard GPI Data Reduction Pipeline \citep[DRP v1.6;][]{perrin_gemini_2014, evans_gemini_2016}. Individual frames were therefore re-centered and combined into a spectral datacube using Pyklip \citep{wang_pyklip_2015}, which also allowed for a more accurate wavelength and PSF re-calibration. As for SPHERE, post-processing was performed using \verb+PACO ASDI+ \citep{flasseur_paco_2020}. 

\subsection{Detection maps from direct imaging}\label{sec:Detlim_DI}
The $5 \sigma$ detection maps from the DI images expressed in terms of contrast are finally converted in units of Jupiter masses using a suitable routine of the \textsc{madys} tool \citep{squicciarini22}: the conversion assumes the age of the system and is mediated by the AMES-Cond isochrones.

\subsection{Detection limits from combined data  }\label{sec:Detlim}
Finally, (sma, mass) detection limits are derived using the direct imaging detection maps, together with the RV data once corrected from stellar activity to estimate the probability of the presence of
other planets in the system. We generate (Monte
Carlo approach) planets with various orbital properties and masses and test whether these simulated planets are detected at least for one epoch in direct imaging or are detected in RV \cite[see details in ][]{Lannier17}.

\subsection{MCMC characterization of G80-21 }\label{sec:MCMC80-21}
The MCMC tool is based on the emcee 3.0 library (Foreman-Mackey et al. 2013). It uses  a mix of custom move functions to alleviate potential multi-modality problems and the cyclicity of angular variables. We consider the RV, \ruwe{}  and PMa data. For Gaia data, the logic implemented in the aforementioned GaiaPMEX tool has been used to compute Gaia-related likelihoods.

\begin{figure*}[hbt]
\includegraphics[width=180mm,clip=True]{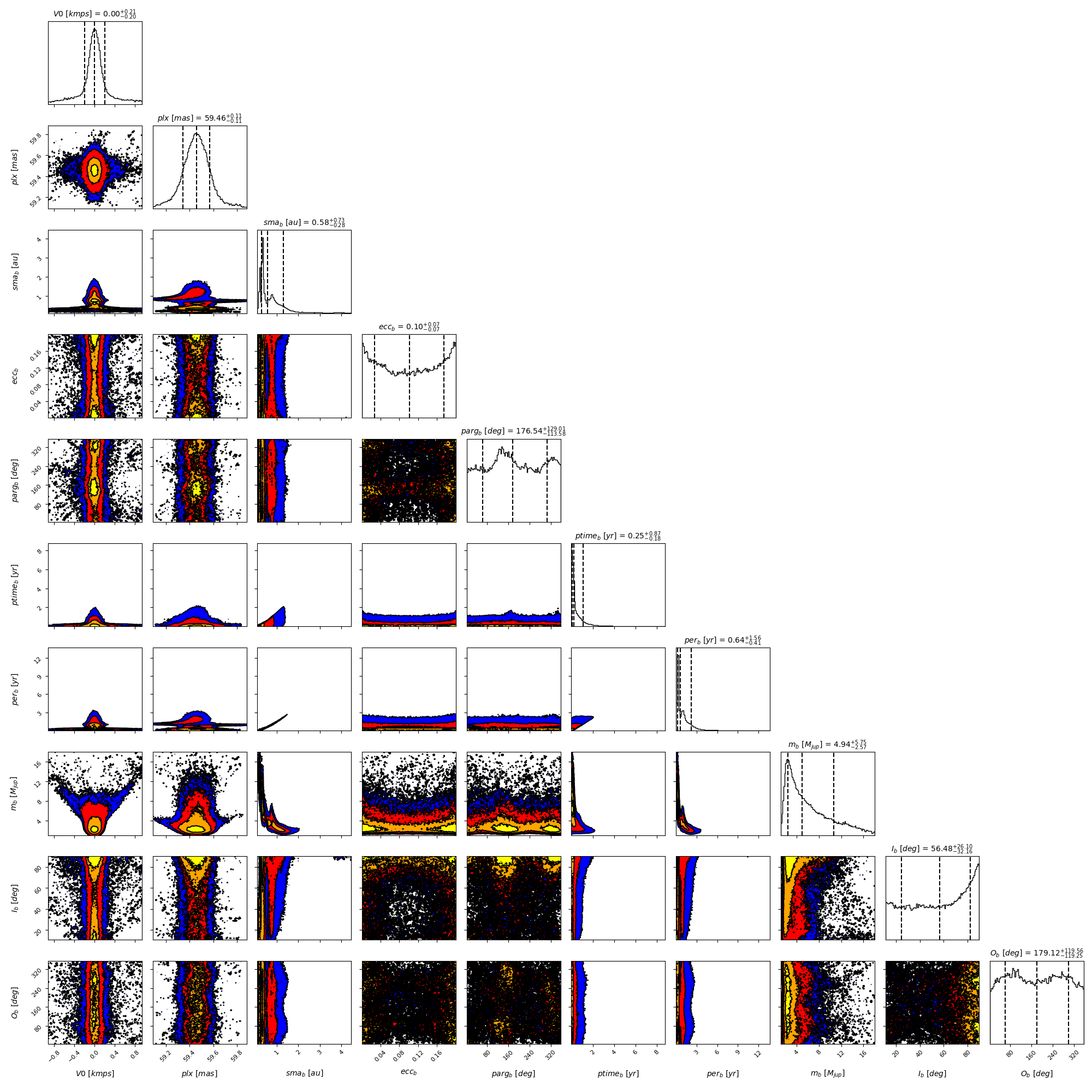}
    \caption{MCMC results for G80-21. See Text.}
    \label{fig:CornerPlot}
\end{figure*}

The free parameters considered are the orbital parameters (semi-major axis, eccentricity, argument of periastron and phase), the minimum mass(es) of
the planet(s), and the RV (global offset due to the star systemic velocity), and a stellar jitter.  A value of 200 m/s was adopted for the jitter to account for the observed short term variability. Uniform priors were considered for all parameters. The star RV was between 18 and 18.5 km/s, the range of sma was between 0.001 and 5 au, given the detection limits provided by high contrast imaging (see Section~\ref{sec:add_comp_G80-21}), the mass between 0.1 and 20 \Mjup, and the eccentricity arbitrarily taken 0 and 0.2.  
The resulting corner plot is given in Figure~\ref{fig:CornerPlot} (see discussion in Section~\ref{sec:G80-21}).

\end{document}